\documentclass[10pt,aps,pra,superscriptaddress,twocolumn,showpacs,preprintnumbers,
nofootinbib]{revtex4-1}
\usepackage{amsfonts}
\usepackage{amssymb,amsmath}
\usepackage{mathrsfs}
\usepackage{amsthm}
\usepackage{bm}
\usepackage{newlfont}
\usepackage{graphicx}

\newcommand{\usb}{\affiliation{Departamento de F\'{\i}sica, Secci\'{o}n de Fen\'{o}menos \'{O}pticos, 
Universidad Sim\'{o}n Bol\'{\i}var,Apartado Postal 89000, Caracas 1080-A, Venezuela.}}
\newcommand{\ivic}{\affiliation{Centro de F\'{\i}sica, Instituto Venezolano de Investigaciones 
Cient\'{\i}ficas, Apartado 20632 Caracas 1020-A, Venezuela.}}

\begin{document}
\title{The energy-momentum tensor of electromagnetic fields in matter} 
\author{Rodrigo Medina}\email[]{rmedina@ivic.gob.ve}\ivic
\author{J Stephany}\email[]{stephany@usb.ve}\usb

\begin{abstract}
In this paper we present a complete resolution of the Abraham-Minkowski controversy
about the energy-momentum tensor of the electromagnetic field in matter. This is done 
by introducing in our approach several new aspects which invalidate previous discussions.
These are: 1)We show that for polarized matter the center of mass theorem is no longer valid
in its usual form. A contribution related to microscopic spin should be considered.
This allows to disregard the influencing  argument done by Balacz in favor of Abraham's momentum
which discuss the motion of the center of mass of light interacting with a dielectric block.
2)The electromagnetic dipolar energy density $ -\mathbf{P}\cdot\mathbf{E}$ contributes to the inertia of matter
and should be incorporated covariantly to the the energy-momentum tensor of matter. This implies that there 
is also an electromagnetic component in matter's momentum density given by $\mathbf{P}\times\mathbf{B}$ which 
accounts for the diference between Abraham and Minkowski's momentum densities. The variation of this contribution 
explains the results of G.~B.~Walker, D.~G.~Lahoz and G.~Walker's experiment which until now was the only 
undisputed support to Abraham's force. 3)A careful averaging of microscopic Lorentz's force
results in the unambiguos expression $f^\mu_\mathrm{d}=\frac{1}{2}D_{\alpha\beta}\partial^\mu F^{\alpha\beta}$ 
for the force density which the field exerts on matter. This 
force density is different to the ones used in most of the discussions and is consistent 
with all the experimental evidence. 
4)Newton's Third law or equivalently momentum conservation 
determines the electromagnetic energy-momentum tensor as the only one consistent with Maxwell's 
equations whose divergence is minus the 
force density. It is $T_{\mathrm S}^{\mu\nu} -F^{\mu}
_{\ \alpha}D^{\nu\alpha}$, where $T_{\mathrm S}$ is the standard tensor of the field
in vacuum. This tensor is different from 
Abraham's and Minkowski's tensors, but one recovers
Minkowski's expression for the momentum density. In particular the energy density is different from
Poynting's expression but  Poynting's vector remains the same.. Our tensor is non-symmetric which allows the field to exert a
distributed torque on matter. As a result of the specific form of the
antisymmetric part of the constructed  tensor, the spin density of the electromagnetic field
decouples from the polarization of matter. To give further support to the  proposed tensor,
we also deduce its form using an alternative method  based on  direct averaging
of the microscopic equations and on imposing consistency of the dipolar coupling. We
use our results to discuss momentum and angular momentum exchange in various situations of
physical interest.  We find complete consistency of our equations in the description of the
systems considered. We also show that several alternative expressions of the field
energy-momentum tensor and force-density cannot be successfully used
in all our examples. In particular we verify in two of these examples that the center
of mass and spin introduced by us moves with constant velocity, but that the standard
center of mass does not.
\end{abstract}
\maketitle 
\thispagestyle{empty} 

\newpage
\tableofcontents
\newpage
\section{Introduction}

The definitions of the electromagnetic (EM) momentum in matter and of the
energy-momentum density tensor of the electromagnetic field in matter has been
discussed for over a century in what is known in the literature as the
Abraham-Minkowski controversy. It is among the most
prominent unsettled problems in classical physics. At the quantum level,
resolution of the Abraham-Minkowski controversy is  of paramount interest
since it answers the  fundamental  question ``what is the momentum of a
photon in matter?''  The main contenders in the dispute are the non-symmetric tensor
constructed by Minkowski in 1908 \cite{Minkowski1908} and the symmetric one proposed
by Abraham in 1909 \cite{Abraham1909,Abraham1910}, but  other options have been
discussed in the literature
 \cite{EinsteinL1908,EinsteinL1908b,Beck1953,MarxG1956,deGrootS,Peirls1976,Peirls1977}.
Each construction provides a definite  expression of the force exerted by the field
on matter.

Although hundreds of papers  investigate this issue many arguments
lack precision in  statement of the hypotheses
and definitions or present an inadequate handling of their consequences.
Various hypotheses and mechanisms that depart from the well-established
Maxwell-Lorentz  electromagnetic theory have been advanced.
For example,  proposed forms of the force density that the field exerts
on matter are in general different from the force density imposed by the microscopic
Lorentz force.  Examples of this  are the Minkowski force, the Abraham force and  the
Einstein-Laub force \cite{EinsteinL1908}. An even more puzzling hypothesis proposed in
this context is the use of the theoretical quantity known as hidden momentum.
Hidden momentum is the caloric of the XXth century. It was introduced by Shockley and
James in \cite{SJ1967} as a resource to force the Center of Mass Motion
Theorem (CMMT) \cite{Pryce1948,Hill1951} to hold, for systems which appear to
have a contribution of electromagnetic momentum at rest. A weak version of the
CMMT, states that the center of energy of an isolated system moves with a
constant velocity in a chosen reference frame. A stronger version
claims that such velocity is the total momentum of the system times $c^2$,
divided by the energy.  As we discuss below, these theorems have  only limited
validity.   The true
nature of the hidden momentum is never exhibited explicitly and the hypothesis
implies very strange consequences on the mechanical behavior of physical systems when
the value of the electromagnetic momentum changes.

An unsatisfactory situation exists because none of the sides have been able to
convince the other of the validity of their arguments.    This has led to an annoying
posture well distributed in the literature
 \cite{TangM1961,Israel1977,Dewar1977,Kranys1979,Maugin1980,Pfeifer2007,Goto2011}
which, relying on the fact that the energy-momentum tensor
of the matter-field  system may be split in many different forms, hypothesizes that
all the alternatives are equivalent if adequately interpreted. In other words, it
postulates that the problem  cannot be settled theoretically and is not amenable to
experimental elucidation.

The same unsatisfactory situation has induced some other people to think that
both expressions may be used in the
theoretical description with different meanings. Often, it is postulated that one
of the expressions, usually Abraham's, is the true mechanical momentum with the
other being some other kind of momentum, either pseudo-momentum  or quantum
mechanical momentum \cite{Gordon1973,Nelson1991,Barnett2010}.

In this paper we show that fancy interpretations, exotic forces and hidden
momentum are unnecessary hypotheses.  We contend
that the motivations to introduce them are wrong and that a totally consistent
description of electromagnetic momentum exists within the framework of
Maxwell-Lorentz electromagnetism. We show it is possible to
give a definite answer to the question of which is the energy-momentum tensor
of the electromagnetic field in matter; it is not necessary to take
different expressions to represent the classical and quantum electromagnetic
momentum in order to deal with the results  of the experiments.  The energy-momentum
tensor for the electromagnetic field is neither Minkowski's nor Abraham's and we
explain clearly why.  We construct by two different methods  an energy-momentum
tensor of the electromagnetic field in matter that is consistent with the microscopic
Lorentz force.  Our construction finally resolves the long dated Abraham-Minkowski
controversy and ratifies Minkowski's expression for the momentum density.  It is based
on the observation that, in opposition to what is claimed in a large part of the recent
literature, the matter and field contributions to the total energy-momentum tensor can
be rigorously distinguished by focusing on the mechanical behavior of matter and using
momentum conservation in the form of Newton's Third Law.

To resolve this important debate, it is essential to  describe very clearly the framework
in which the discussion will be pursued. So we begin
with a few remarks highlighting our suppositions and stating where
we disagree with other authors.

Classical electromagnetism in material media is a very subtle subject with
plenty of room for misunderstandings and confusion. To begin with,
the same notation is often used for microscopic and
macroscopic electromagnetic fields but they are very different objects.
The microscopic field, at the classical level,  allows for only monopolar couplings
and interacts with all the
microscopic charges and currents $\rho$ and ${\mathbf{j}}$. From the theoretical
point of view, it is accepted that the  description using the microscopic field
should be valid up to the scale where quantum effects begin to be relevant.
But in bulk matter this description cannot always be
experimentally explored making it necessary to introduce the macroscopic
field which is interpreted as an average field at some scale.
The macroscopic field has a  monopolar coupling  with the free charges and
currents $\rho_{\mathrm f}$ and ${\mathbf{j}_{\mathrm f}}$ and also has a dipolar
coupling with the polarization vector ${\mathbf{P}}$ and the magnetization
${\mathbf{M}}$. Of course $\rho_{\mathrm f}$, ${\mathbf{j}_{\mathrm f}}$,
${\mathbf{P}}$ and ${\mathbf{M}}$ are manifestations of quantum and classical
properties of the microscopic
quantities $\rho$ and ${\mathbf{j}}$.  The macroscopic fields may also have
quadrupolar and higher order couplings although this possibility  is usually not
investigated. In agreement with the experimental
evidence, we assume that  Maxwell's macroscopic equations are fulfilled.
So, our first remark is that it is necessary to focus
on the dipolar coupling of the macroscopic electromagnetic field to
${\mathbf{P}}$ and ${\mathbf{M}}$ instead of dwelling too much on the
constitutive relations between ${\mathbf{E}}$ and ${\mathbf{B}}$ and
${\mathbf{D}}$ and ${\mathbf{H}}$.  This approach is mandatory
when dealing with ferromagnetic materials.

Our second remark concerns relativistic invariance. The understanding of
the implications of relativistic invariance have evolved since the first
decade of the past  century when the controversy began. Putting things
simply, a relativistic description requires one to work with covariant objects.
Cavalier use of the electric and magnetic permeability constants $\epsilon$ and $\mu$
obscures the relativistic invariance of the equations.  Attempts
to regain relativistic invariance by introducing the velocity of the material
media and switching between the laboratory frame and the co-moving frame of
matter without choosing convenient covariant objects is a limiting procedure.
Such attempts  introduce the well known inconsistency of the concept
of rigid body and relativity. To guarantee the covariance of Maxwell's
equations in a material medium, it
is sufficient and necessary that the fields  ${\mathbf{P}}$ and ${\mathbf{M}}$
and, as a consequence, ${\mathbf{D}}$ and ${\mathbf{H}}$ transform in the same
 way as the electromagnetic fields ${\mathbf{E}}$ and ${\mathbf{B}}$.

Our third point concerns the force density that the field exerts on matter. The
force density and the total force  are observable quantities which can be measured in
a manifold of situations by observing the mechanical behavior of matter.
The energy-momentum tensor of matter is determined as the tensor whose four-divergence
is the force density. It includes all the energy parcels which contribute to inertia,
even those of electromagnetic origin. Momentum conservation, a
consequence of the homogeneity of space, then determines the energy-momentum tensor
of the field as the tensor whose four-divergence is the negative of the force density.
The force density is also subjected to theoretical constraints. It should be
relativistically covariant and should reduce to the Lorentz force when a microscopic
description of the electric and magnetic dipoles is re-introduced.  Einstein-Laub's,
Minkowski's, Abraham's and the Lorentz-Ampere force densities do not satisfy these criteria.
The force density $f^\mu_\mathrm{d}=\frac{1}{2}D_{\alpha\beta}\partial^\mu F^{\alpha\beta}$ 
does, as discussed below.

Our fourth introductory remark addresses the asymmetry of the energy-momentum tensor.
It is frequently assumed that the energy-momentum tensor of an isolated system should
be symmetric for one of the following three reasons. 1) To guarantee the conservation
of angular momentum. 2) Because there is a theorem
 \cite{BelF1939,Rosenfeld1940,Pauli1941,CvV1968} which states that it is always possible to
symmetrize the  tensor. 3) Because it is imposed by Einstein's equations of gravitation. None
of these reasons is valid. The energy-momentum tensor should be symmetric for conserving
the orbital angular momentum, not the total angular momentum.  There is no symmetrizing
theorem for the energy-momentum tensor. Belinfante-Rosenfeld's
construction \cite{BelF1939,Rosenfeld1940} yields a symmetric tensor, made up from the
energy-momentum tensor and derivatives of the spin density, which is not the mechanical
energy-momentum tensor. Einstein's equations impose that their source, not the mechanical
energy-momentum tensor, should be symmetric. The source of Einstein's equations is
physically a different object and there is no contradiction in relating it to the 
Belinfante-Rosenfeld tensor.

Our next observation is related to the hidden momentum approach and the CMMT.
In its usual form the CMMT is not valid in relativistic mechanics. The CMMT was invoked
repeatedly in discussions of  the electromagnetic momentum
to justify introduction of hidden momentum. It also was
used by Balazs to support Abraham's momentum arguing on  a mechanical
example. But the CMMT is not valid for systems with spin in general and
is not valid for the electromagnetic field interacting with polarizable matter in particular. 
What is  valid for systems which include spin and for which total angular momentum is
conserved, is a modified version of the theorem which states that for an isolated system a
quantity which we  call the center of mass and spin behaves inertially. This fact has been
overlooked in all previous discussions.

Our final point is about the distinction between the electromagnetic field and the
electromagnetic wave when
such a wave is propagating through a medium. In that case, traveling together with the
electromagnetic fields there are disturbances of  ${\mathbf{P}}$, ${\mathbf{M}}$ and the
internal stress. The energy and momentum contributions of these disturbances
should appear in the energy-momentum-stress tensor of matter.
Poynting's equation which is a consequence of Maxwell's equations for homogeneous media is
best suited to describe the wave as a whole. In particular Poynting's energy density, as we
will argue in this article,  is the energy of the wave, not the energy of just the field.
Many of the misunderstandings concerning the energy-momentum tensor of the field may be
traced back to this observation. This point is also very important to understand the meaning
of the von Laue-M\o{}ller \cite{vLaue1950,Moller} condition on the velocity of propagation
of the energy density.

Our paper is organized as follows. In section \ref{CMMT theorem}, we show the limitations of
CMMT for systems with spin, discuss the concept of center of mass and spin, and deduce
the modified version of the theorem which
holds for those systems. In section \ref{Poynting}, we discuss Poynting's equation and its
generalizations. In section \ref{AM},  a brief discussion of the Abraham-Minkowski
controversy is presented. The von Laue-M\o{}ller criterion and
the Balazs argument are considered. In section \ref{Force},  we obtain an
expression of the force which is covariant and reduces to the Lorentz force in the
microscopic description. Using this force in section \ref{theTensor}, we construct the true
energy-momentum tensor of the field. Although the energy density is not given by Poynting' s expression 
one recovers Minkowski's expression for the momentum density. This confirms our recent computation
\cite{MedinaandS2017} where in a simple setup we show that
if Maxwell's equations and Ohm's law are valid, only
Minkowski's momentum density is consistent with momentum conservation. On the other hand,
we recognize Poynting's energy density and Minkowski's
tensor as the objects which describe the whole electromagnetic wave thus satisfying the
von Laue-M\o{}ller criterion. In section \ref{Macroscopic}, we take an alternate path and
re-obtain our results by directly averaging the microscopic field equations and imposing
consistency with the results of a  macroscopic field with an arbitrary dipolar coupling.
 In the remaining sections we discuss some examples and consequences. In section
\ref{testing}, we compare the experimental consequences that predict the force density we
proposed with the other alternatives. We propose a method to elucidate
which is the true force density by measuring electro and magnetostriction.
  In section \ref{Wave} we consider the transmission of an electromagnetic wave in a medium,
 which corresponds to the setup of Balazs.
 We compute the force exerted by the fields on matter and show that the standard CMMT does
not hold, but that instead  the center of mass and spin behaves inertially, as we claim
it should.
In section \ref{Magnet-charge}, we consider an interesting magnet-charge system and show
that its description is fully consistent with our equations. We also show that, for this
system, the center of mass and spin behave inertially in a non trivial way.
In section \ref{Cylinder}, we consider a situation in which there is exchange of angular
momentum  between matter and field, which can be handled neatly  within our
approach. Section \ref{Conclusion} contains our conclusions.

\section{Angular momentum and the CM motion theorem}
\label{CMMT theorem}
\subsection{Orbital and spin angular momentum}

Let us consider a system with an energy-momentum tensor $T^{\mu\nu}$ upon which
a force density $f^\mu$ is exerted. A local version of Newton's Second Law holds
\begin{equation}
\label{EMT-eq}
\partial_\nu T^{\mu\nu}=f^\mu\ .
\end{equation}
For a localized system with $f^\mu=0$ the total energy
$U=\int T^{00}dV$ and the total momentum $p^i = c^{-1}\int T^{i0}dV$ are
conserved. The current density of the orbital angular momentum is defined as
$L^{\mu\nu\alpha} = x^\mu T^{\nu\alpha}-x^\nu T^{\mu\alpha}$ and it satisfies identically
\begin{equation}
\label{orbital-eq}
\partial_\alpha L^{\mu\nu\alpha} = T^{\nu\mu}-T^{\mu\nu}+x^\mu f^\nu-
x^\nu f^\mu\ .
\end{equation}
The last two terms of the right hand side of this equation represent the
torque of the force density. The asymmetry of the energy-momentum tensor
plays the role of a torque density exerted on the orbital angular momentum.
If the force density is zero and there are neither  a torque density  $\tau^{\mu\nu}$ nor a
spin density $S^{\mu\nu\alpha}$, the orbital angular momentum is conserved and the
energy-momentum tensor must be symmetric. This is not necessarily true when
these conditions are not fulfilled \cite{Papapetrou1949}. In presence of spin the relevant quantity is
the total angular momentum
 $J^{\mu\nu\alpha}= L^{\mu\nu\alpha}+S^{\mu\nu\alpha}$ whose equation of
motion 
\begin{equation}
\label{total-eq}
\partial_\alpha J^{\mu\nu\alpha} = \tau^{\mu\nu} + x^\mu f^\nu -x^\nu f^\mu\ ,
\end{equation}
allows for an external torque density $\tau^{\mu\nu}$.
Then the right hand side is the total torque density acting on the system. For isolated systems one assumes that the total 
angular momentum is conserved. 

Combining
(\ref{orbital-eq}) and (\ref{total-eq}) the spin equation is obtained
\begin{equation}
\label{spin-eq}
\partial_\alpha S^{\mu\nu\alpha} = T^{\mu\nu}-T^{\nu\mu} + \tau^{\mu\nu}\ .
\end{equation}
The internal torque density represented by the asymmetry of $T^{\mu\nu}$ 
couples spin and orbital angular momenta. If the system has no spin, but
there is a distributed torque then $T^{\mu\nu}-T^{\nu\mu} =- \tau^{\mu\nu}$.
Another interesting case is when the system is isolated; the total angular momentum is
conserved, but there may be an exchange between spin and orbital angular momentum
 as a result of microscopic interactions. That is what  happens in
a magnet when the magnetization changes. Since the magnetization is
proportional to the spatial part of the spin density, if a demagnetization occurs,  spin
becomes orbital angular momentum. This is the Einstein-de Haas \cite{Einstein-deHaas1915} 
effect which is used routinely to measure the gyromagnetic
ratio \cite{BarKen1952} of atoms and molecules and provides an example 
where the total energy-momentum tensor cannot be symmetric.
The statement, found some times in the literature  \cite{Abraham1910}, that conservation of angular momentum requires a symmetric energy-momentum
tensor, is not only false but in this and other cases  the opposite is
true: conservation of (total) angular momentum requires that the total
energy-momentum tensor should be non-symmetric. 

\subsection{The boost momentum}
The temporal component of the angular momentum equation plays an important role
in the treatment of the CMMT.
The total angular momentum of a system $J^{\mu\nu}=c^{-1}\int J^{\mu\nu 0}\,dV$
is an antisymmetric tensor. The spatial part of this tensor is the angular
momentum axial vector $\mathbf{J}$, $J^{ij} = \epsilon_{ijk}J^{k}$. The
temporal components $J^{0k}$ transform as a polar vector, which is a different
physical quantity, in the same sense that $\mathbf{E}$ is
different from $\mathbf{B}$. This quantity is seldom treated in the literature
and its physical meaning is not well appreciated. There is not even a generally
accepted name for it. A sensible name could be the boost momentum, since
$\{J^{0k}\}$ are the generators of the Lorentz boosts, or the arrow of angular
 momentum as we proposed elsewhere \cite{MedandSf}. To better understand the
significance of the boost momentum consider a non-relativistic
particle of mass $m$, position $\mathbf{x}$ and momentum $\mathbf{p}$. In a
different frame of reference moving with velocity $c\bm{\beta}$ the position
is incremented by $\Delta\mathbf{x}=-ct\bm{\beta}$, and the momentum is incremented by
 $\Delta\mathbf{p}=-mc\bm{\beta}$. The quantity which has the proper Poisson
brackets to generate this transformation is $\bm{\beta}\cdot\mathbf{g}$ with
$\mathbf{g}=ct\mathbf{p}-mc\mathbf{x}$.
For a system of particles of center of mass $\mathbf{X}$, total mass $M$ and
total momentum $\mathbf{P}$ the generator of the Galilean boost is
$\mathbf{g}=ct\mathbf{P}-Mc\mathbf{X}$. Conservation of this
quantity is equivalent to the CMMT. Note that $\mathbf{g}$ corresponds to the
orbital boost momentum $L^{0k}$ of the relativistic case. On the other hand,
the Galilean boost does not modify the angular velocitiy or the shape of
a body, that is, it does not modify the spin of the system (the angular momentum
with respect to the CM). In Newtonian physics the spin commutes with the
Galilean boost, and therefore  the existence of spin does not impair the CMMT.
This means that  spin boost momentum does not have a non-relativistic counterpart.

Like  boost momentum the temporal components of the torque also form a polar vector which is
different from the spatial torque. In some particular reference frame a system
may in principle have a non-zero boost momentum and a vanishing $\mathbf{J}$. This is also
true for the torque.
 
\subsection{The CM motion theorem}

Let us consider an isolated, localized system  with a non-symmetric
 energy-momentum tensor upon which no forces are exerted so that,
$\partial_\nu T^{\mu\nu}=0$.

To discuss the CMMT define the  center of mass by
\begin{equation}
\label{CenterMass}
X^i_T = \frac{1}{U}\int x^i T^{00}\,dV \  .
\end{equation}
Here the index $T$ refers to the fact that the CM depends on $T^{\mu\nu}$.
It is worth remarking that this definition of center of mass depends on the
frame of reference. If in some frame of
reference at time $t$ the center of mass is $\mathbf{X}$, and the event
$(t,\mathbf{X})$ is labeled by $(t^\prime,\mathbf{X}^\prime)$ in some other frame, and
if $\mathbf{Y}^\prime$ is the CM in the new frame at time $t^\prime$, then in
general $\mathbf{Y}^\prime\not=\mathbf{X}^\prime$.  Statements about centers
of mass should be done with caution. For example, the fact that the CM moves 
with constant velocity in some frame of reference
does not imply that the CM also moves with constant velocity in some other frame of
reference.

If there is no spin, $T^{\mu\nu}$ is symmetric and  orbital angular momentum
$L^{\mu\nu}=c^{-1}\int L^{\mu\nu 0}\,dV$ is conserved. Then, it is easy to see
that the center of mass moves with velocity $c^2 p^i/U$. Instead, for
non-symmetric $T^{\mu\nu}$ it is also easy to see that
\begin{equation}
\label{CenterMass-velocity}
\dot{X}^i_T = \frac{c}{U}\int T^{0i}\,dV \ .
\end{equation}
Note that what appears in (\ref{CenterMass-velocity}) is the energy current
density, not the momentum density. The
non-vanishing temporal spin of the system jeopardizes the validity of the CMMT.
To see this  define the spin matrix
$S^{\mu\nu}= c^{-1}\int S^{\mu\nu 0}\,dV$ and consider the quantity
\begin{equation}
\label{spincenter}
X^i_S=-\frac{c}{U}S^{0i}\ .
\end{equation}
From the conservation of  total angular momentum it follows that
\begin{eqnarray}
\label{CMSMT}
\frac{d}{dt}X^i_S&=&\frac{c}{U} \frac{d}{dt} L^{0i}= \frac{1}{U}\frac{d}{dt}\int \big[x^0 T^{i0}-x^iT^{00}\big]dV\nonumber\\
&=&\frac{c^2 p^i}{U}-\frac{d}{dt}X^i_T\ .
\end{eqnarray}
This shows that  the center of mass and spin defined by
\begin{equation}
\label{CenterMassSpin}
 X^i_\Theta=X^i_T+X^i_S
\end{equation}
moves with constant velocity $\dot{X}^i_\Theta = c^2 p^i/U $.

It is worth noting \cite{MedandSe} that $X^i_\Theta$ corresponds to the center
of mass computed using the symmetric Belinfante-Rosenfeld's tensor
$\Theta^{\mu\nu}$, which is a combination of the energy-momentum tensor and the
spin density. In the literature Belinfante-Rosenfeld's tensor is frequently
considered as an improved symmetrized energy-momentum tensor but we
consider that this interpretation is  wrong, at least from the mechanical point of view.

As a simple illustration of the ideas presented in this section, consider an isolated
uniformly magnetized sphere with a device included which heats the sphere respecting
the symmetry. In the rest frame the spin, which is proportional to the magnetization,
is purely spatial and both the center of mass and the center of mass and spin
are located at the center of the sphere. As the sphere heats the magnetization
diminishes and finally vanishes and by
conservation of total angular momentum the sphere  rotates. The center of mass
and the center of mass and spin remain in place. Looking at this system from a reference
frame where the sphere is moving in a direction which
is perpendicular to magnetization,  the center of mass is initially at the center of the
sphere. The temporal components of spin do not vanish and the center of mass and spin
is displaced from the center of the sphere in the direction perpendicular
to the magnetization and to the velocity. When the magnetization disappears the sphere is
rotating about the axis which is in the former  direction
of the magnetization. The velocity of matter in one side of the sphere is greater than
the velocity in the other side, and the center of mass is now displaced with
respect to the geometric center. Clearly, it has been accelerated. It may be shown
 \cite{MedandSf} that the center of mass and spin (\ref{CenterMassSpin}) which in the final situation
coincides with the center of mass behaves inertially along the whole process.

\section{Poynting's energy density and energy flux}
\label{Poynting}
\subsection{The dipolar density tensor}
In Gauss units and using the metric tensor
 $\eta^{\mu\nu}=\mathrm{diag}(-1,1,1,1)$, the electromagnetic field tensor
(either microscopic or macroscopic)
is $F_{\alpha\beta}=\partial_\alpha A_\beta-\partial_\beta A_\alpha$,
and is related to the electric and magnetic fields by $F^{0i}=-F^{i0}=E_i$ and
$F^{ij}=\epsilon_{ijk}B_k$.

The traditional way of treating polarizable matter is to assume that there are microscopic
electric and magnetic dipoles. The polarization $\mathbf{P}$ and magnetization $\mathbf{M}$
are defined so that an element of material of volume $dV$ has a total
electric dipole moment $\mathbf{P}dV$ due to the
electric dipoles and a total magnetic moment $\mathbf{M}dV$ due to the magnetic dipoles.
 In the non-relativistic
approximation this picture works fine but it is problematic in the general case, since
moving electric and magnetic dipoles have moments of both kinds. Because of this reason
in this paper we refrain from using those microscopic entities. In the following we
present a phenomenological and macroscopic definition of $\mathbf{P}$ and $\mathbf{M}$.

The fields and currents used in macroscopic physics are averages of
the microscopic fields and currents over small space-time regions. As
usual  we will assume, that at a microscopic scale the equations for the
vacuum hold. In a material sample at a macroscopic scale,
in addition to the averaged current density of free charges
 $j_{\mathrm f}^\mu= (c\rho_{\mathrm f},{\mathbf j}_{\mathrm f})$,
there are electric and magnetic dipole moments that couple directly to the EM tensor
$F^{\mu\nu}$. These dipole moments are due to bound charges. The total bound charge in
any piece of material is always zero. This is a kind of mechanical constraint.
The fields produced by the dipoles are described by the current density four-vector of bound
charges $j^\mu_\mathrm{b}$. Any piece of material may have a bound charge in the surface
 $Q_\mathrm{bS}$ and a bound charge in the bulk $Q_\mathrm{bV}$, but the
total bound charge should always be zero, $Q_\mathrm{bS}+ Q_\mathrm{bV} =0$. This
behavior is described in terms of $\mathbf{P}$, defined inside the material, if
the charge density of bound charges
is $\rho_\mathrm{b}=-\nabla\cdot\mathbf{P}$ and the surface charge density is
 $\sigma_\mathbf{b}=\mathbf{P}\cdot\hat{\mathbf{n}}$,
where $\hat{\mathbf{n}}$ is the normal unitary vector of the surface.  The continuity
equation for the bound charges is
\begin{equation}
\label{bound-continuity}
0=\frac{\partial \rho_\mathrm{b}}{\partial t}+\nabla\cdot\mathbf{j}_\mathrm{b}
=\nabla\cdot(\mathbf{j}_\mathrm{b} -\dot{\mathbf{P}}) \ .
\end{equation}

In general, in addition to the polarization
current density $\dot{\mathbf{P}}$, a divergence-less
magnetic current density
 $\mathbf{j}_\mathrm{m}=\mathbf{j}_\mathrm{b} -\dot{\mathbf{P}}$
also contributes  to the bound current density. This magnetic
current density is not related to any macroscopic motion of charge, and
 is expressed as the rotational of the magnetization field
$\mathbf{j}_\mathrm{m}=c\nabla\times\mathbf{M}$. $\mathbf{M}$ vanishes outside the
material. Since there
is no total magnetic current crossing a piece of material, in the surface of
any piece of material there is also a
magnetic surface current density $\mathbf{\Sigma}_\mathrm{m}=c\mathbf{M}\times
\hat{\mathbf{n}}$.

An element of material of volume $dV$ has an electrical dipole
moment $d\mathbf{d}=\mathbf{P}dV$ and a magnetic dipole moment
$d\mathbf{m}=\mathbf{M}dV$  due to the surface charges and surface
currents respectively. The bulk contributions which go like $r^3$ are negligible
with respect to the surface ones that go like $r^2$.

For a covariant description one should  define the  space-time dipolar density
$D_{\alpha\beta}$, whose  spatial part is the magnetization density
$D_{ij}=\epsilon_{ijk}M_k$
and whose temporal part is the polarization $D_{0k}=-D_{k0}=P_k$. The
bound current density is then
\begin{equation}
\label{bound-current}
j^\mu_\mathrm{b} = c\partial_\nu D^{\mu\nu} \ .
\end{equation}
As mentioned before, in order to have relativistic invariance, $D_{\alpha\beta}$
should be a Lorentz tensor. The fact that $D_{\alpha\beta}$ is antisymmetric implies
the conservation of the bound current
 $\partial_\mu j^\mu_\mathrm{b}=c\partial_\mu \partial_\nu D^{\mu\nu}=0$.

 \subsection{Maxwell's equations}
The displacement vector and the magnetizing field vector  are given in our notation by $\mathbf{D}=\mathbf{E}+4\pi\mathbf{P}$  
and  $\mathbf{H}=\mathbf{B}-4\pi\mathbf{M}$.  In what follows we assume that Maxwell's macroscopic equations 
\begin{eqnarray}
\label{Maxwell-mac}
\nabla\cdot{\mathbf{D}}&=&4\pi \rho_{\mathrm f}\ ,\qquad  \nabla\times {\mathbf E}+\frac{1}{c}\frac{\partial{\mathbf B}}{\partial t}=0\ ,\nonumber\\
\nabla\cdot{\mathbf{B}}&=&0\ ,\qquad  \nabla\times {\mathbf H}-\frac{1}{c}\frac{\partial{\mathbf D}}{\partial t}=\frac{4\pi}{c}{\mathbf j}_{\mathrm f}
\end{eqnarray}
are valid in any reference frame. To cast this in covariant notation one defines
\begin{equation}
\label{hmunu}
 H^{\mu\nu}=F^{\mu\nu}-4\pi D^{\mu\nu}
\end{equation}
whose components are related to the magnetizing field
$\mathbf{H}$ and to the electric displacement $\mathbf{D}$ by
$H^{ij} = \epsilon_{ijk} H_k$ and $H^{0i} = D_i$.
 
The field equations become
\begin{equation}
\label{Maxwell-covariant}
\partial_\beta H^{\alpha\beta}=\frac{4\pi}{c}j_{\mathrm f}^\alpha\ ,
\end{equation}
\begin{equation}
\label{Bianchi}
\partial_\mu F_{\nu\lambda}+\partial_\nu F_{\lambda\mu}
+\partial_\lambda F_{\mu\nu}=0 \ .
\end{equation}
The second of these equations is the Bianchi's identity. The first equation written
in terms of the macroscopic electromagnetic field is
\begin{equation}
\label{Maxwell-matter}
\partial_\nu F^{\mu\nu} = \frac{4\pi}{c}(j_{\mathrm f}^\mu+j_\mathrm{b}^\mu) \ .
\end{equation}
 
\subsection{The susceptibility and permeability tensors}
Isotropic linear materials are characterized by the familiar  relations  $\mathbf{D}=\epsilon\mathbf{E}$ and $\mathbf{H}=1/\mu\mathbf{B}$.
These relations are meant to be valid in a particular reference frame, which for homogeneous, isotropic media is the rest frame of matter. 
But much more complicated situations may be contemplated, and in order to obtain a relativistic invariant description it is necessary to 
introduce  generalized relations in terms of a four indexes susceptibility tensor. They take the form,
\begin{equation}
\label{Suscep}
 D_{\alpha\beta} =\frac{1}{2}
\chi_{\alpha\beta\gamma\delta}F^{\gamma\delta}\ .
\end{equation}
The susceptibilities form a four-tensor of four indexes with the following
symmetries
\begin{equation}
\chi_{\alpha\beta\gamma\delta}=\chi_{\gamma\delta\alpha\beta}=
-\chi_{\beta\alpha\gamma\delta}\ .
\end{equation}

The four-tensor $\chi_{\alpha\beta\gamma\delta}$ includes the susceptibilities
three-tensors, the  electric-electric $\chi^\mathrm{(e-e)}$,
the magnetic-magnetic $\chi^\mathrm{(m-m)}$, the electric-magnetic
$\chi^\mathrm{(e-m)}$ and the magnetic-electric $\chi^\mathrm{(m-e)}$.
The first two are symmetric tensors, and of the last two one is the transpose of
the other. In total there are $21$ independent components. The three-tensors
are defined by
\begin{equation}
\label{P-linear}
P_i = \chi^\mathrm{(e-e)}_{ij}E_j+\chi^\mathrm{(e-m)}_{ij}B_j
\end{equation}
and
\begin{equation}
\label{M-linear}
M_i = \chi^\mathrm{(m-m)}_{ij}B_j+\chi^\mathrm{(m-e)}_{ij}E_j\ .
\end{equation}
Note that $\chi^\mathrm{(m-m)}$ is different from the traditional magnetic
susceptibility tensor $\tilde{\chi}$, which is defined using $\mathbf{H}$
instead of $\mathbf{B}$,
 $\chi^\mathrm{(m-m)}=\tilde{\chi}(I+4\pi\tilde{\chi})^{-1}$.

 The relationship between the four-tensor and the three-tensors is
\begin{eqnarray}
\chi^\mathrm{(e-e)}_{ij}&=&\chi_{0i0j}\ ,\\
\chi^\mathrm{(m-m)}_{ij}&=&
\frac{1}{4}\epsilon_{ilm}\epsilon_{jfg}\chi_{lmfg}\ ,\\
\chi^\mathrm{(e-m)}_{ij}&=&\frac{1}{2}\epsilon_{jfg}\chi_{0ifg}\ ,\\
\chi^\mathrm{(m-e)}_{ij}&=&\frac{1}{2}\epsilon_{ifg}\chi_{fg0j} \ .
\end{eqnarray}

The proper way of treating the transformation of the susceptibilities when
the frame of reference is changed is by transforming the four-tensor.
When matter is at rest and the material is centrosymmetric the mixed
pseudo-tensors vanish, $\chi^\mathrm{(e-m)}=\chi^\mathrm{(m-e)}=0$. This
property is lost in other frames of reference.

Equivalent relations may be written in terms of the permeability tensor. Using
(\ref{hmunu}) one has 
\begin{equation}
\label{Permeab}
 H_{\alpha\beta} =\frac{1}{2}
\mu_{\alpha\beta\gamma\delta}F^{\gamma\delta}\ .
\end{equation}
with 
\begin{equation}
 \mu_{\alpha\beta\gamma\delta}=\delta_{\alpha\gamma}\delta_{\beta\delta}-\delta_{\alpha\delta}\delta_{\beta\gamma}+
 4\pi\chi_{\alpha\beta\gamma\delta}
\end{equation}
The permeability tensor has the same symmetry properties as the susceptibility
 tensor.  In particular one has,
\begin{equation}
\label{Displacement}
D_i =\epsilon_{ij}E_j+\xi_{ij}B_j
\end{equation}
and
\begin{equation}
\label{Magnetizing}
H_i =\zeta_{ij}B_j+\eta_{ij}E_j \ .
\end{equation}
The tensors $\epsilon_{ij}$ and $\zeta_{ij}$ must be symmetric and the
pseudo-tensors $\xi_{ij}$ and $\eta_{ij}$ are related by
$\eta_{ij}= -\xi_{ji}$. These conditions are
preserved by Lorentz transformations. At rest, for a centrosymmetric material,
the pseudo-tensors vanish, but that is not true in the general case.

In the particular  case of a homogeneous isotropic material with a dielectric constant $\epsilon$ and
magnetic permeability $\mu$, which in  the rest frame satisfy $\mathbf{D}^\prime=
\epsilon\mathbf{E}^\prime$ and $\mathbf{B}^\prime=\mu\mathbf{H}^\prime$ one may obtain the permeability tensor
by expressing the fields in the rest frame in terms of those in the moving frame. The familiar relations for 
the transformed fields are,
\begin{equation}
\label{rest-D}
\mathbf{D}+\bm{\beta}\times\mathbf{H}=\epsilon(\mathbf{E}+
\bm{\beta}\times\mathbf{B})
\end{equation}
and
\begin{equation}
\label{rest-H}
\mathbf{H}-\bm{\beta}\times\mathbf{D}=\frac{1}{\mu}(\mathbf{B}-
\bm{\beta}\times\mathbf{E})\ ,
\end{equation}
where $c\bm{\beta}$ is the velocity of the medium.

Solving these equations we get, $\gamma=1/\sqrt{1-\beta^2}$,
\begin{equation}
\epsilon_{jk}=\gamma^2[(\epsilon-\beta^2/\mu)\delta_{jk}-(\epsilon-1/\mu)
\beta_j\beta_k]\ ,
\end{equation} 
\begin{equation}
\zeta_{jk}=\gamma^2[(1/\mu-\epsilon\beta^2)\delta_{jk}-(\epsilon-1/\mu)
\beta_j\beta_k] \ ,
\end{equation} 
and
\begin{equation}
\xi_{jk}=-\eta_{jk}=\gamma^2(\epsilon-1/\mu)\epsilon_{jlk}\beta_l \ .
\end{equation} 

\subsection{Energy flux equation and Poynting's equation}

The flux of energy is determined by the following   relation which is 
a direct consequence of macroscopic  Maxwell's equations (\ref{Maxwell-mac}) 

\begin{eqnarray}
\label{energyflux2}
\frac{1}{4\pi}\Big(\mathbf{E}\cdot
\frac{\partial \mathbf{D}}{\partial t}&+&
\mathbf{H}\cdot\frac{\partial \mathbf{B}}{\partial t}\Big) \nonumber\\
&=&-\frac{c}{4\pi}\nabla\cdot({\mathbf E}\times{\mathbf H})-{\mathbf E}\cdot{\mathbf j}_{\mathrm f} \ .
\end{eqnarray}

This may be written in the two alternative forms
\begin{eqnarray}
\label{energyflux}
 \frac{1}{8\pi}\frac{\partial}{\partial t}(B^2&+&E^2)+\frac{\partial}{\partial t}({\mathbf E}\cdot{\mathbf P})+
 \frac{c}{4\pi}\nabla\cdot({\mathbf E}\times{\mathbf H})\nonumber\\&=&-{\mathbf E}\cdot{\mathbf j}_{\mathrm f}+
 {\mathbf P}\cdot\frac{\partial {\mathbf E}}{\partial t}+
{\mathbf M}\cdot\frac{\partial {\mathbf B}}{\partial t} 
\end{eqnarray}
and
\begin{eqnarray}
\label{energyflux1}
 \frac{1}{8\pi}\frac{\partial}{\partial t}(B^2&+&E^2)+\frac{c}{4\pi}\nabla\cdot({\mathbf E}\times{\mathbf B})\nonumber\\
 &=&-{\mathbf E}\cdot({\mathbf j}_{\mathrm f}+{\mathbf j}_{\mathrm b}) \ . 
\end{eqnarray}

The different terms in these equations should describe
the energy density and the energy current density of the
field and the power supplied to matter by the field.
 There is clearly more  that one way to establish these correspondences, but we will 
argue below that there is one way that should be preferred.
 
In 1884 Poynting \cite{Poynting} proposed his energy conservation equation using the 
particular case of these equations 
that applies for homogeneous isotropic materials with time-independent linear susceptibilities
 $\mathbf{D}=\epsilon\mathbf{E}$ and $\mathbf{H}=1/\mu\mathbf{B}$.  He postulates that the 
energy density is
\begin{equation}
\label{Poynting-energy}
u_\mathrm{P}=\frac{1}{8\pi}(\mathbf{E}\cdot\mathbf{D}+
\mathbf{H}\cdot\mathbf{B}) \ .
\end{equation}
With the mentioned restrictions on the medium it follows that 
\begin{equation}
\label{energy-deriv}
\frac{\partial u_\mathrm{P}}{\partial t}=\frac{1}{4\pi}\Big(\mathbf{E}\cdot
\frac{\partial \mathbf{D}}{\partial t}+
\mathbf{H}\cdot\frac{\partial \mathbf{B}}{\partial t}\Big) 
\end{equation}
and equation (\ref{energyflux2}) reduces to
\begin{equation}
\label{Poynting-eq}
 \frac{\partial u_\mathrm{P}}{\partial t} +\nabla\cdot\mathbf{S} =
-\mathbf{E}\cdot\mathbf{j}_{\mathrm f}\ ,
\end{equation}
where the energy current density $\mathbf{S}$ is  Poynting's vector,
\begin{equation}
\label{Poynting-vector}
\mathbf{S}=\frac{c}{4\pi}\mathbf{E}\times\mathbf{H}\ .
\end{equation} 
This led to the interpretation of $\mathbf{S}$ as the energy current density of the
electromagnetic disturbance, and $\mathbf{E}\cdot\mathbf{j}_{\mathrm f}$ as the time 
rate of work done by the field on the free charges. No work is done on the polarizable 
matter. At this point there is place to ask if $u_\mathrm{P}$ should necessarily be
interpreted as the energy density of the field and
if the work done on the free charges is all the work done by the field on matter. In
the following sections we argue that the answers to these questions are both negative.

The condition (\ref{energy-deriv}) is valid also for anisotropic linear
susceptibilities. Poynting's energy density for an  anisotropic material  in any reference frame 
is given by
\begin{equation}
u_\mathrm{P}=\frac{1}{8\pi}(\epsilon_{jk}E_j E_k+\zeta_{jk}B_j B_k) \ .
\end{equation}
From this expression it is readily shown that for
time-independent tensors the relationship (\ref{energy-deriv}) is
fulfilled, and that therefore Poynting's equation (\ref{Poynting-eq}) is valid.

Note that in order for $\epsilon_{jk}$ and $\zeta_{jk}$ be time independent in any reference frame, it is
required that the medium properties are not only time-independent but also homogeneous. Poynting's equation is valid
in any frame of reference only for material media which are homogeneous,
time-independent, and that are moving in a uniform translation.
 
\subsection{Electrostatic energy}
\label{electrostaticenergy}
The essential differences between non-dipolar and dipolar matter already appear
in electrostatics.  In this subsection we discuss the physical significance
of various ways of defining  electrostatic energy for dipolar matter. Let us
start by considering non-dipolar matter. The energy of a system of charges is
the sum of electrostatic interaction energy $U_\mathrm{e}$ and
$U_\mathrm{M}$ which represents all other kinds of energies. The electrostatic
 energy is
\begin{equation}
U_\mathrm{e}=\frac{1}{2}\int \rho\phi\,dV\ ,
\end{equation}
which can also be  expressed as
\begin{equation}
U_\mathrm{e}=\frac{1}{8\pi}\int E^2\,dV\ .
\end{equation}
If it is assumed that the matter energy is distributed in the volume
there would be a matter energy density $u_\mathrm{M}$ to be added to the
standard field energy density $u_\mathrm{S}$ which represents the electrostatic interaction
\begin{equation}
u_\mathrm{S}=\frac{1}{8\pi} E^2\ .
\end{equation}

For dipolar matter where there is a polarization $\mathbf{P}$ two facts have to be considered.
1) The energy density of matter becomes polarization-dependent and should be now denoted by
$u_\mathrm{b}(\mathbf{P})$ (here the label b stands for bare). 2) $u_\mathrm{S}$ still
represents the total electrostatic energy, but a part of this energy density, given by
\begin{equation}
 u_\mathrm{d}=-\mathbf{P}\cdot\mathbf{E}
\end{equation}
corresponds to the interaction energy of electric dipole moments. The total energy density is
given by $u_\mathrm{b}(\mathbf{P})+ u_\mathrm{S}$, but
$u_\mathrm{d}$ should be considered part of the energy of matter, since it is
attached to it  and contributes to its inertia.  The energy
density of (dressed) matter is then
\begin{equation}
 u_\mathrm{M}=u_\mathrm{b}(\mathbf{P})+u_\mathrm{d}\ .
\end{equation}
Correspondingly, the dipolar interaction energy must be subtracted from
the standard field energy density in order to obtain the actual field
energy density
\begin{equation}
\label{elecstaticenergy}
 u=u_\mathrm{S}-u_\mathrm{d}\ .
\end{equation}
This modifies the expression of power released to matter. The power density on the bare
dipoles is $\mathbf{E}\cdot\dot{\mathbf{P}}$, whereas for the dressed dipoles it is
$-\mathbf{P}\cdot\dot{\mathbf{E}}$.
In section \ref{theTensor}, we show that the generalization of
(\ref{elecstaticenergy}), which includes the magnetic field contribution should be regarded as
the energy density of the field.

This  field energy density is not only different from the standard one for
non-dipolar matter, it also differs from the Poynting density for
dipolar matter $\mathbf{E}\cdot\mathbf{D}/8\pi$. To compare with this last density consider
an isotropic medium  with linear
polarizability. The bare matter density $u_\mathrm{b}$ should have  a deformation
term quadratic in  $\mathbf{P}$
\begin{equation}
u_\mathrm{b}(\mathbf{P})=u_\mathrm{b}(0)+\frac{1}{2\chi}P^2\ .
\end{equation}
Here $\chi$  is the electric susceptibility. When an electric field is present the matter
polarizes itself by minimizing
$u_\mathrm{M}$ with the result $\mathbf{P}=\chi\mathbf{E}$. The dressed matter changes its
energy by an amount of $-\mathbf{P}\cdot\mathbf{E}/2$ which is the sum of the electrostatic
energy $-\mathbf{P}\cdot\mathbf{E}$ and the deformation energy
$\mathbf{P}\cdot\mathbf{E}/2$. The bare matter increases its energy by the
deformation energy. Poynting's density adds the deformation energy to the
standard energy density $u_\mathrm{S}$ of the field. This density  is
useful for determining the total energy of a system (field plus matter), for
example in a capacitor, but it cannot be used for determining the energy
exchange between matter and field. Poynting density is useless for
matter with non-linear polarizabilities.

\section{Abraham-Minkowski controversy}
\label{AM}
\subsection{Splitting the Electromagnetic tensor}
To discuss electromagnetic momentum it is necessary to have, not only an energy density and
an energy current density, but a complete energy-momentum tensor. For an isolated field-matter
 system the conserved  total
energy-momentum tensor $T^{\mu\nu}$ should be split in two terms,
\begin{equation}
\label{Tsplitting}
T^{\mu\nu}=T_{\mathrm{matter}}^{\mu\nu}+T^{\mu\nu}_{\mathrm{field}}
\end{equation}
corresponding to each part of the system. Contrary to what is sometimes stated
(see section \ref{testing}),
the matter contribution to the tensor is completely determined by its mechanical behavior
 through the equation
\begin{equation}
 \label{Tmatter}
  \partial_\nu T_{\mathrm{matter}}^{\mu\nu}=f^\mu \ ,
\end{equation}
where $f^\mu$ is the measurable force exerted by the field on matter. Correspondingly, all
the portions of the energy which contribute to inertia should be included in the matter
energy-momentum tensor, even those of electromagnetic origin.
On the other hand, conservation of momentum imposes that the field
energy-momentum tensor $ T^{\mu\nu}_{\mathrm{field}}$ should satisfy Newton's Third Law
\begin{equation}
\label{Tfield}
\partial_\nu T_{\mathrm{field}}^{\mu\nu}=-f^\mu \ .
\end{equation}
This gives a very stringent criterion for choosing the energy-momentum tensor
of the field, which may be only avoided at the expense of changing the description
of the mechanical behavior of matter relaxing (\ref{Tmatter}), or
by introducing other non standard elements like hidden momentum. Below we show
that none of these alternatives is necessary. Note that whereas in the temporal
component of (\ref{Tfield}) the energy density and the energy density
current are present, the spatial components are written in terms of the stress tensor
$T^{ij}$ and the momentum density $\mathbf{g}$,
\begin{equation}
\label{spatialforce}
 c\partial_0g^i+\partial_j T^{ij}=-f^i \ .
\end{equation}

For any  covariant  energy-momentum tensor introducing
Maxwell's equations in (\ref{Tfield}), the zero component yields an equation
equivalent to (\ref{energyflux}), with a specific choice of the
energy density, energy flux and power on matter. These quantities are given by definite
expressions of the fields, polarizations, and currents.

\subsection{Field energy-momentum tensor in vacuum}

Let us start from the well-known results of  microscopic EM field in vacuum,
which couples minimally to a microscopic conserved  electric current
density $j^\mu$,  $\partial_\mu j^\mu=0$ through the vector
potential $A_\mu$. The
four-vector Lorentz force density  acting on the currents is
\begin{equation}
\label{Lorentz}
f_{\mathrm L}^\mu = \frac{1}{c} F^{\mu\nu} j_\nu \ .
\end{equation}

 The electromagnetic tensor is determined by Maxwell's
equations and Newton's Third Law. Besides Bianchi's identity (\ref{Bianchi})
the vacuum field equations are
\begin{equation}
\label{Maxwell}
\partial_\nu F^{\mu\nu} = \frac{4\pi}{c}j^\mu \ .
\end{equation}

The standard energy-momentum tensor of the EM field is \cite{LanLL1994}
\begin{equation}
\label{TEI-S}
T^{\mu\nu}_\mathrm{S}=-\frac{1}{16\pi}F_{\alpha\beta}
 F^{\alpha\beta}\eta^{\mu\nu}
+\frac{1}{4\pi}F^\mu_{\ \alpha} F^{\nu\alpha} \ .
\end{equation}

Using (\ref{Lorentz}), (\ref{Maxwell}) and (\ref{Bianchi}) it is
straightforward to prove the identity
\begin{equation}
\label{FieldForce}
\partial_\nu T^{\mu\nu}_\mathrm{S} = -\frac{1}{c} F^{\mu\nu}j_\nu
= - f_{\mathrm L}^\mu \ ,
\end{equation}
valid for any solution of Maxwell's equations. Note that the
force density acting on the EM field is  the opposite of the force density
that the EM field exerts on the matter. That is, the action-reaction law
holds locally. As a consequence, if there are no other interactions, the
total energy and momentum of EM fields and matter are conserved.

To compare with our discussion below let us consider  the standard tensor (\ref{TEI-S})
when it is written in terms of the macroscopic field with dipolar coupling. In this case,
(\ref{TEI-S}) is referred in the literature as the Livens tensor \cite{Livens1916}.
Maxwell's equations take in this case  the general form (\ref{Maxwell-matter}) with
$j^\mu$ in (\ref{Maxwell}) replaced by $j_{\mathrm f}^\mu+j^\mu_\mathrm{b}$.
Since (\ref{FieldForce}) is an identity,  it is possible to write
directly the new identity
\begin{eqnarray}
\label{Identity}
\partial_\nu T_{\mathrm S}^{\mu\nu} &=&
 -\frac{1}{c} F^\mu_{\ \nu}(j_{\mathrm f}^\nu+j_\mathrm{b}^\mu) \nonumber\\
&=&-\frac{1}{c}F^\mu_{\ \nu}j_{\mathrm f}^\nu-F^\mu_{\ \nu }\partial_\alpha D^{\nu\alpha}\ .
\end{eqnarray}
The negative of the right-hand side of
this equation is what is known as the Amp\`ere-Lorentz force,
\begin{equation}
\label{Ampere-Lorentz}
 f_{\mathrm AL}^\mu=\frac{1}{c} F^\mu_{\ \nu}(j_{\mathrm f}^\nu+j_\mathrm{b}^\mu)\ .
\end{equation}

One should note that $T_{\mathrm S}^{\mu\nu}$ may not be identified with the energy-momentum
tensor of the field if the force density on the currents of bound charges which
contribute to the Amp\`ere-Lorentz force is not the same as the
true force density acting on the dipolar densities. We argue  in the next sections that they
are indeed different.

\subsection{Abraham and Minkowski momenta}

Many theoretical arguments and some experimental evidence have been used to try
to elucidate which form of the energy-momentum tensor
of the electromagnetic field in matter should be adopted. Some reviews written through the
years from different points of view  may be found in
\cite{Pauli1958,Skobel'tsyn1973,PeirlsB1973,Synge1974,Robinson1975,GinzburgU1976,
Dewar1977,Brevik1979,Novak1980,Loudon2004,Leonhart2006,Pfeifer2007,Obukhov2008,
MilonniBoyd2010, BarnettLoudon2010,Kemp2011,Cho2013}.

Equations (\ref{Poynting-energy}) and (\ref{Poynting-eq}) and the
requirement of Lorentz invariance were the main ingredients which
Minkowski \cite{Minkowski1908} used to propose  a non-symmetric energy-momentum tensor
 $T^{\mu\nu}_\mathrm{Min}$,
which he interpreted as the energy-momentum tensor of the field. In this tensor
the energy density and the energy  current density are Poynting's expressions
 (\ref{Poynting-energy}), (\ref{Poynting-vector}),
\begin{equation}
\label{MinPoynting}
T^{00}_\mathrm{Min}=u_\mathrm{P}\ ,\ \ \ c T^{0i}_\mathrm{Min} \hat{\mathbf{e}}_i =\mathbf{S}
\end{equation}
and  the density of  linear momentum is,
\begin{equation}
\label{Minmomentum-density}
\mathbf{g}_{\mathrm{Min}} = c^{-1} T^{i0}_\mathrm{Min} \hat{\mathbf{e}}_i =
\frac{1}{4\pi c}\mathbf{D}\times\mathbf{B}\ .
\end{equation}
For photons with energy $E$ this implies a momentum $nE/c$ with $n$ the
refraction index.
Minkowski's tensor is given by
\begin{equation}
\label{energy-momentum-tensor-Min}
T^{\mu\nu}_{\mathrm{Min}}=
-\frac{1}{16\pi}F_{\alpha\beta}H^{\alpha\beta}\eta^{\mu\nu}
+\frac{1}{4\pi}F^{\mu}_{\ \alpha}H^{\nu\alpha}
\end{equation}
and it is non symmetric . It is written here for the general case of arbitrary $D^{\mu\nu}$,
but was originally intended only for
linear materials.  In components, besides (\ref{MinPoynting}) and (\ref{Minmomentum-density})
it defines  Maxwell's tensor  by
\begin{equation}
\label{MaxwellMin}
 T^{ij}_{\mathrm{Min}}=\frac{1}{8\pi}(\mathbf{E}\cdot\mathbf{D}+\mathbf{B}\cdot\mathbf{H})\delta^{ij}-\frac{1}{4\pi}(E_iD_j+H_iB_j)\ .
\end{equation}
For linear isotropic materials this expression is symmetric, but in general it is not.
Using (\ref{Tfield}) the corresponding force density is
\begin{equation}
\label{force-density-Min}
f_{\mathrm{Min}}^\mu = \frac{1}{c}F^{\mu}_{\ \nu}j_{\mathrm{f}}^\nu+
 \frac{1}{4}(D_{\alpha\beta}\partial^\mu F^{\alpha\beta}
-F_{\alpha\beta}\partial^\mu D^{\alpha\beta})\ .
\end{equation}
Minkowski's force differs from  Amp\`ere-Lorentz's force (\ref{Ampere-Lorentz}).
In the rest frame, for linear
and homogeneous materials with no boundaries, with constant susceptibility tensor
(\ref{Suscep}), the second term above   vanishes, and  Minkowski's force density
reduces to  Lorentz's force density on free charges.
The equation for the flux of energy
(\ref{energyflux}) in Minkowski's approach is written as,
\begin{eqnarray}
\label{energyfluxMin}
\frac{\partial u_\mathrm{P}}{\partial t}+\nabla\cdot\mathbf{S} &=&
-\mathbf{E}\cdot\mathbf{j}_{\mathrm f}+\\
\frac{1}{2}\left[\mathbf{P}\cdot\frac{\partial \mathbf{E}}{\partial t}+
\mathbf{M}\cdot\frac{\partial \mathbf{B}}{\partial t}\right]&-&
\frac{1}{2}\left[\mathbf{E}\cdot\frac{\partial \mathbf{P}}{\partial t}+
\mathbf{B}\cdot\frac{\partial \mathbf{M}}{\partial t}\right]\ .\nonumber
\end{eqnarray}
The last two terms cancel each other
only for linear homogeneous materials,
allowing Poynting's (\ref{Poynting-eq}) equation to hold.

In 1909, Abraham \cite{Abraham1909}, requiring the energy-momentum tensor to be symmetric, gave
an alternative suggestion for the energy-momentum tensor of the field.
In the rest frame it agrees on the energy density and
the energy current density with Minkowski's tensor, but the linear momentum density
postulated is
\begin{equation}
\label{abrmomentum}
\mathbf{g}_{\mathrm{Abr}}=c^{-1} T^{i0}_\mathrm{Abr}=c^{-2}\mathbf{S}\ .
\end{equation}
This yields a symmetric energy-momentum object for homogeneous materials in the rest frame of
matter, and implies a momentum $E/nc$
for the photons. It also complicates the issue of relativistic invariance. The original
 Abraham's discussion was done partially from
a pre-relativistic point of view, and includes a comparison with Cohn's old fashioned
electrodynamics. The tensor expressed only
in terms of the fields (see the components below) cannot be written in covariant form
 (this was shown again recently in
\cite{VeselagoShchavlev2010} by an explicit computation).
Invariant versions of the tensor which reduce to the original tensor in the rest
frame, may be written
\cite{Abraham1914,Grammel1913,Pauli1958} for a homogeneous isotropic medium at the
expense of introducing a dependency on the velocity of the medium.
Pauli \cite{Pauli1958} gives the expression
\begin{equation}
\label{energy-momentum-tensor-AbrI}
T^{\mu\nu}_{\mathrm{AbrI}}=T^{\mu\nu}_{\mathrm{Min}}+\Delta_{\mathrm{AbrI}}^{\mu\nu}
\end{equation}
where
\begin{equation}
\Delta_{\mathrm{AbrI}}^{\mu\nu}=-\frac{(\epsilon\mu-1)}{4\pi c^4}F_{\beta\alpha}U^\alpha
 U_\gamma(U^\mu H^{\beta\gamma}-U^\gamma H^{\beta\mu})U^\nu
\end{equation}
with $U^\alpha$  the four-velocity of the medium. This tensor is considered only for a
homogeneous isotropic medium, in which case it is symmetric.
In the rest frame the only non zero components of $\Delta T_{\mathrm{AbrI}}^{\mu\nu}$ are
\begin{equation}
\Delta T_{\mathrm{AbrI}}^{i0}= -\frac{\epsilon\mu-1}{c}{S}_i
=\frac{1}{4\pi}(\mathbf{E}\times\mathbf{H}-\mathbf{D}\times\mathbf{B})_i\ .
\end{equation}
In this reference frame the components of the tensor are  given by (\ref{abrmomentum}) and by
\begin{equation}
\label{AbrPoynting}
 T^{00}_\mathrm{AbrI}=u_\mathrm{P}\ \ ,\ \ c T^{0i}_\mathrm{AbrI} \hat{\mathbf{e}}_i
 =\mathbf{S}
\end{equation}
and
\begin{eqnarray}
\label{MaxwellAbrI}
T^{ij}_{\mathrm{AbrI}}=\frac{1}{8\pi}(\mathbf{E}\cdot\mathbf{D}
+\mathbf{B}\cdot\mathbf{H})\delta^{ij}\nonumber\\
-\frac{1}{4\pi}(E_iD_j+H_iB_j)\ .
\end{eqnarray}
We note that the introduction of $U^\alpha$ in the formulation is a rather artificial
procedure, since this velocity is only defined if the
medium is taken to be a rigid body, and this enters in contradiction with special relativity.
 In consequence, there is  an intrinsic approximation in the approach.
The force density corresponding to this tensor is given by
\begin{equation}
\label{force-density-AbrI}
f_{\mathrm{AbrI}}^\mu = f_{\mathrm{Min}}^\mu-\partial_\nu\Delta T_{\mathrm{AbrI}}^{\mu\nu}\ .
\end{equation}
In the rest frame it reduces to
\begin{equation}
 \label{f0abrI}
 f_{\mathrm{AbrI}}^0=f_{\mathrm{Min}}^0
\end{equation}
\begin{equation}
\label{power-AbrI}
f_{\mathrm{AbrI}}^i = f_{\mathrm{Min}}^i+\frac{\epsilon\mu-1}{c^2}\partial_t{S}^i\ .
\end{equation}
The second term in this equation,
\begin{eqnarray}
\label{Abraham_term}
 \mathbf{f}_{\mathrm{Abraham}}&=&\frac{1}{4\pi}\partial_0[\mathbf{D}
\times\mathbf{B}-\mathbf{E}\times\mathbf{H}]\\
&=& \partial_0[\mathbf{P}\times\mathbf{B}+\mathbf{E}\times\mathbf{M}]
= \frac{\epsilon\mu-1}{c^2}\partial_t\mathbf{S}\nonumber
\end{eqnarray}
is what is usually called Abraham's force. The existence
of this force was apparently supported by some nice experiments
\cite{WalkerGL1975,WalkerGLW1975} in 1975, but in section \ref{testing} we show that
those experimental results are better explained within our approach.

For the forthcoming discussion it is convenient to introduce a second tensor,
which may be considered for non linear materials, and  reduces
to Abraham's proposal for isotropic linear materials. It is given by
\begin{equation}
\label{energy-momentum-tensor-AbrII}
T^{\mu\nu}_{\mathrm{AbrII}}=\frac{1}{2}(T^{\mu\nu}_{\mathrm{Min}}+T^{\nu\mu}_{\mathrm{Min}})+
\Delta T^{\mu\nu}_{\mathrm{AbrII}}\ ,
\end{equation}
\begin{eqnarray}
\Delta T^{\mu\nu}_{\mathrm{AbrII}} &=&\frac{1}{c^2 8\pi}
\Big[F_{\alpha\beta}U^\beta(H^{\alpha\mu}U^\nu+H^{\alpha\nu}U^\mu)\nonumber\\
&-& H_{\alpha\beta}U^\beta(F^{\alpha\mu}U^\nu+F^{\alpha\nu}U^\mu)\Big]\ .
\end{eqnarray}
This tensor is symmetric for arbitrary polarization and magnetization. Usually the
symmetrization of Minkowski's tensor in (\ref{energy-momentum-tensor-AbrII})
is not discussed because as already mentioned the spatial
part of this tensor is symmetric for isotropic homogeneous media, but if a symmetric tensor is
required for non linear materials this step is necessary. In the rest frame
the  equations corresponding to (\ref{abrmomentum}) and (\ref{AbrPoynting})
still hold but now  Maxwell's tensor appears explicitly symmetric
\begin{eqnarray}
\label{MaxwellAbrII}
T^{ij}_{\mathrm{AbrII}}&=&\frac{1}{8\pi}(\mathbf{E}\cdot\mathbf{D}
+\mathbf{B}\cdot\mathbf{H})\delta^{ij}\nonumber\\
&\ &-\frac{1}{8\pi}(E_iD_j+E_jD_i+B_iH_j+B_jH_i)\ .
\end{eqnarray}
Using (\ref{Tfield}), the corresponding force density is
\begin{equation}
\label{force-density-AbrII}
f_{\mathrm{AbrII}}^\mu = f_{\mathrm{Min}}^\mu+\Delta f_{\mathrm{AbrII}}^\mu
\end{equation}
where
\begin{equation}
 \Delta f_{\mathrm{AbrII}}^\mu=-\partial_\nu \Delta T^{\mu\nu}_{\mathrm{AbrII}}
+ \frac{1}{2}\partial_\nu(T^{\mu\nu}_{\mathrm{Min}}-T^{\nu\mu}_{\mathrm{Min}})\ .
\end{equation}
For the space time components in the rest frame we have,
\begin{equation}
\label{DeltafAbr0}
 \Delta f^{0}_{\mathrm{AbrII}}=0
\end{equation}
and
\begin{eqnarray}
\label{DeltafAbrII}
\Delta f^{i}_{\mathrm{AbrII}}&=&\frac{1}{4\pi}\partial_0[\mathbf{D}\times\mathbf{B}
-\mathbf{E}\times\mathbf{H}]^i\nonumber\\
&\ & + \frac{1}{2}\partial_j(T^{ij}_{\mathrm{Min}}-T^{ji}_{\mathrm{Min}})\ .
\end{eqnarray}
The first term  in this equation is again Abraham's force. The second term in
 (\ref{DeltafAbrII})
\begin{eqnarray}
\label{deltafsym}
 \Delta{f}^{i}_{\mathrm{Symm}}
&=&\frac{1}{2}\partial_j(T^{ij}_{\mathrm{Min}}-T^{ji}_{\mathrm{Min}})\\
 &=& \frac{1}{8\pi}\partial_j(E_iD_j-E_jD_i+B_jH_i-B_iH_j)\nonumber
\end{eqnarray}
vanishes for isotropic linear materials. For non linear materials it has various unpleasant
 characteristics.
For example, there is a contribution of the form
$\rho_bD_i $ which can be hardly justified on physical grounds.

In the rest frame the energy density and the energy density current for either of
the covariant Abraham tensors coincide with those of Minkowski's proposal.
Due to (\ref{f0abrI}) and (\ref{DeltafAbr0})
the energy flux equation for  both of them is given again
by (\ref{energyfluxMin}) in that reference frame. In other reference frames, additional
terms depending on the velocity should be added to the energy density and the energy
current density, and in consequence the power released to matter is also modified.

On the theoretical side the reasons used to justify Abraham's requirement of
symmetry of the tensor do not survive a careful analysis. They are explained for
example in \cite{Abraham1914}.  Here is the core
of Abraham's argument (the translation is ours):

``In the presentation of the Maxwell-Hertz theory usually the preference is given to
the Hertz's symmetrical stress tensor. The asymmetric Maxwell tensor implies that inside
a crystal torques are exerted by neighbor particles when placed in an electric field.
Those torques should be equilibrated by elastic torsion forces, unknown in the ordinary
 theory of elasticity.  An asymmetric electromagnetic tensor would be allowed only within
the framework of an appropriately extended elasticity theory.  As the example treated above
shows, the fact that torsional forces act on a crystal in an electric field 
in no way allows to infer the existence of an internal electrical torque.''

This argument is incorrect because it relies on the belief that in ordinary
elasticity theory the stress tensor must always be symmetric. Actually,
the stress tensor is proven to be symmetric under the hypothesis that there is
no internal torque or spin density \cite{AtanaT2000, Fung2001}. An asymmetry
of the stress tensor may be developed to compensate an internal torque. The existence of
an internal torque when $\mathbf{P}$ is not parallel to $\mathbf{E}$, or $\mathbf{M}$ is
not parallel to $\mathbf{B}$ follows just assuming that the torque on a particle
in the crystal depends only on the local  $\mathbf{E}$ and  $\mathbf{B}$ fields,
and is not affected by the state of the other particles. The reason also mentioned
by some authors that symmetry of the total energy-momentum tensor is required for the
conservation of angular momentum was long ago overruled by the discovery of spin.
But as discussed in section \ref{CMMT theorem}, this was not fully appreciated
in the literature and many people still use the argument in this context.

By the middle of last century, von Laue \cite{vLaue1950}
and M\o{}ller \cite{Moller} introduced a different argument which suggested that only
Minkowski's ansatz but not Abraham's is compatible with Poynting's equation and
relativity. That seemed for a while to have settled the issue. But soon apparently
independent arguments appeared to back Abraham's proposal, notably one based on
CMMT due to Balazs \cite{Balazs1953}. Let us briefly discuss both lines of reasoning.

\subsection{The von Laue-M\o{}ller argument}
Around 1950 von Laue \cite{vLaue1950} and M\o{}ller \cite{Moller}  
put forward a different criterion to pick the correct 
electromagnetic energy-momentum tensor. They basically argue that being an
observable quantity, the velocity of the electromagnetic disturbance in matter 
should behave in a way consistent with relativity.
Let us first discuss  the argument as stated by von Laue. He treats the case of
linearly polarized sinusoidal plane waves. He observed that there are two
four-vectors that may be related to the wave. One  is the
wave-four-vector $k^\mu$, $k^0=\omega/c$. The other should be constructed out of the 
velocity of energy propagation  $\mathbf{W}=(1/u)\mathbf{S}$. The  wave-four-vector is space-like,
while the beam four-velocity should be  time-like. Using the Lorentz transformations
(\ref{rest-D}) and (\ref{rest-H}) he shows that the beam velocity $\mathbf{W}$
satisfies Einstein's addition rule
when changing the frame of reference. From this he deduces that the energy current density associated to the system  
in any reference frame should be  Poynting's vector. He also shows that
$\mathbf{g}_\mathrm{Min}= (u/\omega)\mathbf{k}$. Abraham's energy current density 
(here meaning the one obtained from (\ref{energy-momentum-tensor-AbrI}) or (\ref{energy-momentum-tensor-AbrII})) in a moving frame is not 
Poynting's vector. von Laue took this result as an evidence that Abraham's tensor is not adequate to describe this system.
To illustrate his point von Laue considered
the reference frame in which $\mathbf{W}=0$.  If the momentum density were
 $c^{-2}\mathbf{S}$, the momentum of a wave-packet would be time-like
and the energy would be positive. What actually happens in
such a frame is that $\mathbf{E}=\mathbf{H}=u=\omega=0$ and
 $\mathbf{g}_\mathrm{Min}\not=0$ since neither $\mathbf{D}$ nor $\mathbf{B}$ vanish . We note
that this is a consequence of the mechanical difference between a particle and
a wave-packet. The particle can be stopped in any reference frame. The
wave-packet cannot be stopped; it is always moving with respect to the medium.
We also note that $U=pv$ should be expected for a non-dispersive wave and that
$p^\mu\propto k^\mu$ is what is consistent with Planck and De Broglie relations.

M\o{}ller discusses the same idea in a more abstract way. He also observes that if some
energy-momentum tensor $T^{\mu\nu}$ accounts for an observable energy density which
moves in a medium, then the quantity $W^i$
\begin{equation}
 W^i=c\frac{T^{0i}}{T^{00}}=\frac{S_\mathrm{T}^i}{T^{00}}
\end{equation}
which corresponds to von Laue beam velocity discussed above
should behave as a velocity under Lorentz
transformations. He proceed  to show that this implies that 
\begin{equation}
 W^\alpha=(c\gamma_w,\gamma_w w^i)=(\frac{cT^{00}}{\sqrt{-T^0_{\ \mu}T^{0\mu}}},\frac{cT^{0i}}{\sqrt{-T^0_{\ \mu}T^{0\mu}}})
\end{equation}
with $\gamma^{-1}_w=\sqrt{1-w^iw^i/c^2})$ is a Lorentz vector
although it is not written covariantly as a vector. For this to be true it is necessary
that the following relations are satisfied. 
\begin{eqnarray}
g^i_\mathrm{T}=(\frac{T^{00}}{S_\mathrm{T}^kS_{\mathrm{T}k}})T^{ij}S_{\mathrm{T}j}\ ,\quad 
T^{ij}=\frac{T^{ik}S_{\mathrm{T}k}S_\mathrm{T}^j}{S_\mathrm{T}^kS_{\mathrm{T}k}}
\end{eqnarray}
with $S^i_\mathrm{T}=cT^{0i}$ and $g^i_\mathrm{T}=c^{-1}T^{i0}$. In the case of an electromagnetic wave propagating in a medium,
Minkowski's tensor satisfies both these equations, but Abraham's does not. This implies 
that  Minkowski's tensor describes in fact an object which moves as a whole
in a relativistically invariant way. In section \ref{Wave} we show that it corresponds to the the electromagnetic
wave plus the polarizations of the medium.

\subsection{Balazs argument and hidden momentum}

In 1953, Balazs  \cite{Balazs1953} proposed a thought experiment with the aim to expose the inadequacy of
Minkowski's momentum density expression. He argued
that since Minkowski's momentum in matter is greater than in vacuum, photons
crossing a dielectric block will pull the block instead of pushing it and in consequence the
CMMT will be violated. As we discussed in section \ref{CMMT theorem} the CMMT is not valid in general. In particular,
it is never valid in the description of polarizable matter interacting with
the electromagnetic field because the polarization of matter induces a variation of the spin density current 
which contributes to (\ref{CMSMT}). This is illustrated in the examples of sections \ref{Wave} and \ref{Magnet-charge}. 
Nevertheless Balazs line of reasoning has shown to have a great
convincing power over some people and it has even been accepted on equal footing
with the experimental evidence in some review papers when discussing the status
of the controversy \cite{Pfeifer2007,Kemp2011}. The misunderstandings which introduces the
inadequate use of the CMMT when discussing electromagnetic momentum  also have a prominent
place in the line of arguments which led to the introduction
(and for the last fifty years the consideration) of the so called hidden momentum.
When hints appeared \cite{Cullwick1952,Taylor1965,SJ1967}, that in the framework of
Maxwell-Lorentz classical electrodynamics the CMMT may not hold for some systems,
they were dismissed in favor  of this very speculative hypothesis \cite{SJ1967}, which requires radical
modifications of the mechanical behavior of matter.
To illustrate how hidden momentum is
introduced consider a charged plane capacitor at rest in a transverse magnetic
field $\mathbf{B}$. The electromagnetic momentum of this system  
neglecting boundary effects is $p_{\mathrm em}=EBV$ and points in the direction
perpendicular to $\mathbf{B}$ and $\mathbf{E}$. Here, $V$ is the volume between
the shells of the capacitor, and $\mathbf{E}$ is the electric field. Hidden
momentum supporters argued that there is an opposite hidden momentum of equal
magnitude so that the total momentum vanishes and the CMMT holds. For discussions of the 
electromagnetic momentum using this concept see for example \cite{SJ1967,Furry1969,Abraham1909}. 
One of the results of this
paper is to show that the limitations of  the CMMT already discussed make of hidden momentum  an 
unnecessary hypothesis. 

\subsection{Alternative approaches}

As we mentioned in the introduction, other energy-momentum tensors have been considered besides
Abraham's and Minkowski's. 
The early pre-relativistic discussion on Maxwell's tensor already has implications on the 
definition of the force. As discussed by Pauli \cite{Pauli1958}, Maxwell and Heaviside
preferred a non symmetric expression  for the stress tensor whereas Hertz opted for the
symmetrized expression (\ref{MaxwellAbrII}). In a relativistic setup the force is given
by (\ref{spatialforce}) and 
without specifying the form of the momentum density, the force density is not uniquely defined.
This in particular means that  Abraham-Minkowski dilemma was already present in
the early development of the theory.

Einstein and Laub,  made a definite choice for Abraham's momentum expression. 
Based in the form of the force on elementary dipoles, proposed in the rest frame the stress 
tensor   
\begin{equation}
\label{MaxwellEinstein}
T^{ij}_{\mathrm{EL}}=\frac{1}{8\pi}(\mathbf{E}^2+\mathbf{H}^2)\delta^{ij}-\frac{1}{4\pi}(E_iD_j+H_iB_j)\ .
\end{equation}
This defines the Einstein Laub force,
\begin{eqnarray}
 \label{EinsteinLaubforce}
 f^i_{\mathrm{EL}}&=&-\partial_\nu T^{i\nu}_{\mathrm{EL}}=-\partial_0 T^{i0}_{\mathrm{EL}}-\partial_jT^{ij}_{\mathrm{EL}}\\
 &=&f^i_{\mathrm{Min}}+\frac{\epsilon\mu-1}{c^2}\partial_t S^i +\frac{1}{2}\partial^i (\mathbf{E}\cdot\mathbf{P}
 +\mathbf{H}\cdot\mathbf{M})\nonumber\ .
\end{eqnarray}
Since  Einstein and Laub gave no expression for the energy density the power released to matter  
cannot be computed in this case.

In a long series of papers summarized in Ref.~\cite{deGrootS}, de Groot and Suttorp 
approached the problem from the microscopic  point of view. Their construction depends on some
hypotheses about the matter distribution and the average procedure. For  the electromagnetic field
in a fluid (see Ref.~\cite{deGrootS}, Eq.V-92) they propose  the energy-momentum tensor,
\begin{eqnarray}
\label{deGrootEMT}
T^{\mu\nu}_{\mathrm{GS}}&=&
-\frac{1}{16\pi}F_{\alpha\beta}F^{\alpha\beta}\eta^{\mu\nu}
+\frac{1}{4\pi}F^{\mu}_{\ \alpha}H^{\nu\alpha}
+\frac{1}{c^2}[F^{\mu\gamma}D_{\gamma\alpha}V^\alpha\nonumber\\
&-& D^{\mu}_{\ \alpha}F^\alpha_{\ \beta}V^\beta-
\frac{1}{c^2}V^\mu V^\gamma D_{\gamma\alpha}F^\alpha_{\ \beta}V^\beta]V^\nu \ .
\end{eqnarray}
where $V^\mu$ is the local bulk velocity of the fluid. In the local rest frame this tensor takes the form
(see Ref.~\cite{deGrootS}, Eq.V-148)
\begin{eqnarray}
\label{dGEMTrest}
 T^{00}_{\mathrm{GS}}&=&\frac{1}{8\pi}(\mathbf{E}^2+\mathbf{B}^2)\ ,\ \ \ 
 T^{i0}_{\mathrm{GS}}=T^{0i}_{\mathrm{GS}}=\frac{1}{c}S^i\\
  T^{ij}_{\mathrm{GS}}&=&[\frac{1}{8\pi}(\mathbf{E}^2+\mathbf{B}^2)-
  \mathbf{B}\cdot\mathbf{M}]\delta^{ij}-\frac{1}{4\pi}(E_iD_j+H_iB_j)\nonumber\ .
\end{eqnarray}
The dependence of the last term in (\ref{deGrootEMT}) on the local velocity affects the form of the force density 
which cannot be computed directly from (\ref{dGEMTrest}). In absence of free charges it is given by (see Ref.~\cite{deGrootS}, Eq.V-156), 
\begin{eqnarray}
\label{GSforce}
 f_{\mathrm{GS}}^\mu &=&  \frac{1}{2}D_{\alpha\beta}\partial^\mu F^{\alpha\beta}-
 \frac{1}{c^2}\partial_\nu([F^{\mu\gamma}D_{\gamma\alpha}V^\alpha\nonumber\\
&-& D^{\mu}_{\ \alpha}F^\alpha_{\ \beta}V^\beta-
\frac{1}{c^2}V^\mu V^\gamma D_{\gamma\alpha}F^\alpha_{\ \beta}V^\beta]V^\nu) \ .
\end{eqnarray}
The first term in this equation is very interesting, and we will turn back to it later but, even for 
constant $V^\mu$, the second term  spoils the good properties of  the first term. 

In 1976, Peirls proposed yet another tensor based in an analysis of the microscopic momentum in matter 
for non magnetic media, and concluded that within his approximation scheme the energy-momentum tensor should be 
Minkowski's symmetrized  tensor. The force density for this  tensor is given by 
\begin{equation}
\label{Symmforce}
 f^i_{\mathrm{Symm}}=f^i_{\mathrm{Min}}+\frac{\epsilon\mu-1}{2c^2}\partial_t\mathbf{S}+ \Delta{f}^{i}_{\mathrm{Symm}}
\end{equation}
with $\Delta{f}^{i}_{\mathrm{Symm}}$ given by (\ref{deltafsym}). The contribution to the power is 
\begin{equation}
f^0_{\mathrm{Symm}}=\frac{1}{2}\partial_i[\mathbf{D}\times\mathbf{B}
-\mathbf{E}\times\mathbf{H}]^i\ .
\end{equation}
As already discussed above, this incorporates to the force density 
terms which are of difficult  interpretation. For the symmetrized 
tensor the energy current density clearly is not  Poynting's vector.

Besides those  discussed above, other alternative approaches have been proposed to discuss the energy-momentum 
tensor of the electromagnetic field. These include the tensors proposed by Beck  \cite{Beck1953} and 
Marx and Gy\"{o}rgyi \cite{MarxG1956}, 
the works of Haus and Penfield\cite{PenfieldH1967,Haus1969}, Gordon \cite{Gordon1973} and  Nelson 
\cite{Nelson1991}, which have been influential in  the discussions on the subject. 
There are also  constructions introduced for  specific situations, as for example
the work in \cite{KlimaP1975} for dielectric fluids, and reference \cite{Philbin2011} on dispersive media, 
and many others. There is no space in this paper to discuss properly the many details and different
points of view included in all these articles and many others that we are not mentioning. 
We only point out that in our opinion none of those 
approaches has  been widely accepted as a solution of the controversy.
 
For some authors \cite{TangM1961,Dewar1977,Kranys1979,Maugin1980,Pfeifer2007,Goto2011},
the original Abraham-Minkowski controversy
has been considered conceptually solved on the ground that the division of the
total energy-momentum tensor into electromagnetic and  material components
is arbitrary. Minkowski's, Abraham's, or any other electromagnetic
energy-momentum tensor has  a material counterpart (with or without hidden
momentum contributions), which adds to the same total energy-momentum tensor
and responds for any unwanted effect not accounted by the chosen tensor. 
This point of view denies the fact that the force
and torque densities are, within the experimental limitations, observable
quantities and that the field and matter contributions are constrained by
 (\ref{Tfield}) and (\ref{Tmatter}) as discussed in section \ref{AM}.

There is another approach to the problem which convinces some authors that the
controversy has been resolved. Those authors consider that both options
for the momentum of the electromagnetic field are correct, but that they are applicable to
different situations. For example, Gordon and Nelson try to identify Abraham's expression with
the mechanical momentum of light, and Minkowski's expression with the pseudo-momentum or crystal momentum. 
Novak \cite{Novak1980} talks 
about bare and dressed radiation quanta, distinguishing between the photon which would carry
Minkowski's momentum,  and the wave packet which would carry Abraham's. Similarly, in 
\cite{Barnett2010,BarnettLoudon2010},  Minkowski's  momentum is characterized as the canonical
momentum to be identified with the quantum mechanical momentum, and it is proposed to
identify Abraham's momentum as the mechanical momentum. We 
think that without obtaining a consistent and physically reasonable description at the
classical level such approaches 
which introduce the additional complications of quantum mechanics, including those related 
to the correspondence principle, cannot settle the question. 
Moreover, as we show in the rest of this article, Minkowski's expression is  the one which
gives a consistent description of the electromagnetic momentum at 
the classical level. This implies from our point of view that the mentioned distinction
proposed in \cite{Barnett2010}, and reviewed in \cite{ BarnettLoudon2010,MilonniBoyd2010},
should be basically unnecessary.

\subsection{Comparison of the tensors}
\label{Comparison}

Before entering in  the discussion of the adequacy of the force density expressions
discussed in the previous section to describe the outcome of the experiments,  let
us compare Minkowski's, Abraham's and the other tensors at the light 
of our remarks in the introduction.  First we note that most of the tensors with the
exception of  de Groot and Suttorp proposal focus 
on the constitutive relations between  ${\mathbf{E}}$ and ${\mathbf{B}}$ and
${\mathbf{D}}$ and ${\mathbf{H}}$, and use Poynting's energy, which is adequate only to
describe a  time-independent homogeneous medium. This is a general failure because none of
these tensors could be expected to describe materials with permanent polarization.

From the point of view of relativistic invariance, we note that 
$T^{00}_{\mathrm{Min}}=u_\mathrm{P}$ and $T^{0k}_{\mathrm{Min}}=c^{-1}S_k$ in any frame.
Minkowski's  tensor is written  in terms of the fields 
and by construction  transforms properly under Lorentz transformations. In consequence,
the symmetrized Minkowski's tensor related to Peirls approach also has
these two properties. All the other tensors depend on the velocity of matter. If this
velocity is taken to be constant and the same 
for all points of the medium as is usually done, one has to raise the question of the
compatibility with relativity. If the velocity is taken to be 
the local velocity of matter, the  correspondence with the intended expressions of the
energy density, the energy current density, and the 
momentum density are valid only in the local rest frame, and unwanted terms appear in the
force density. Even for constant velocity
the components of the energy-momentum tensor and  the force density gain additional
terms which depend on the velocity. In particular, the 
energy current density for any of Abraham's covariant tensors is
Poynting's vector only in the rest frame.

In our opinion the key point in the controversy is the proper identification of the true
expression of the force density consistent with the experiments. In the next section we
present our approach to this problem, and in section (\ref{testing}) we discuss some of the
experimental evidence on the force. We have already mentioned some problems
associated to the force density obtained from Minkowski's, Abraham's and the other tensors
but we postpone a detailed comparison up to these sections.

As explained in section \ref{CMMT theorem}, we do not regard symmetry as a necessary condition 
on the energy-momentum tensor of the electromagnetic
field but on the contrary we expect this tensor to be in some situations non symmetric. 
Nevertheless,  we note that the only truly symmetric tensors 
of the ones discussed are the symmetrized Minkowski tensor and the one we
  called $T^{\mu\nu}_{\mathrm{AbrII}}$.  The usual Abraham tensor is 
only symmetric for isotropic homogeneous media, and de Groot and Suttorp tensor is
non symmetric.  These last authors correctly addressed in their 
book the lack of justification for the symmetry constraint.

The detailed balance of the momentum exchange of the electromagnetic field with matter in 
Balzs thought experiment cannot be discussed  based only in the expressions of the momentum
density and the currents of momentum density of the fields. More input of the matter
configuration is needed.  In sections \ref{Wave} and 
\ref{Magnet-charge} we discuss two  specific examples for which it is possible to compare
the different approaches.

The importance of distinguishing between the electromagnetic wave and the electromagnetic
field acquires relevance in the light of  the argument  
used by von Laue  and M\o{}ller for rejecting Abraham's tensor \cite{vLaue1950,Moller}. When
applied to a wave in a linear medium, Abraham tensor does not describe an isolated
relativistic moving object. On the contrary, Minkowski's tensor does. Our interpretation 
of this fact is not that Minkowski's tensor is the correct energy-momentum tensor of the
field.  In our opinion, Poynting's energy as advanced in subsection 
\ref{electrostaticenergy} is not purely electromagnetic  but also includes the change of
the energy of matter due to polarization. Correspondingly 
the object described by Minkowski's tensor is the whole wave with the disturbances on matter
produced by the wave included. This point of view 
is confirmed by our results of section \ref{theTensor}. M\o{}ller-von Laue criterion 
although physically relevant does not allow by itself to identify the energy-momentum
tensor that should be ascribed to the electromagnetic field.

\section{Force and torque on dipolar systems}
\label{Force}
\subsection{Conditions on the force}
In this section we discuss the general conditions for the
splitting (\ref{Tsplitting}) of the energy-momentum tensor and the force on
dipolar. 

First, the separation between matter and field should be independent of the
frame of reference. This implies that each part of the energy-momentum tensor
must be a four-tensor and the force density $f^\mu$ a four-vector.

Second, the energy and momentum conservation means that equations
(\ref{Tmatter}) and (\ref{Tfield}) must be identities for any solution of
the dynamical equations. On the field sector, this implies that (\ref{Tfield})
must be an identity for any solution of of Maxwell equations and that therefore
the force density is completely determined by $T_\mathrm{field}^{\mu\nu}$ and vice versa.
For example, if the energy-momentum tensor of the field has the form of the
standard one in the vacuum $T_\mathrm{S}$ (Livens tensor), as it is some times
suggested \cite{Obukhov2003}, the force density should be the Amp\`ere-Lorentz
force on the total current density; see Eq.~(\ref{Identity}).

Third, for macroscopic matter the microscopic dynamical equations are
unknown, or in any case their solutions are unknown. So, (\ref{Tmatter}) becomes
a true dynamical equation useful for determining the motion. The splitting
should be such that the energy-momentum tensor of matter
keeps the standard form for the
different kinds of materials (for example for a continuum), excluding any
kind of exotic dynamics. This requirement indicates that there is only one
sensible splitting.

Fourth, since the macroscopic electromagnetic fields $F^{\mu\nu}$ are generated
by the free current density $j^\mu_\mathrm{f}$ and the polarizations
$D_{\mu\nu}$, a tensor that deserves the name of energy-momentum
tensor of the field should not depend
on matter variables, besides $j^\mu_\mathrm{f}$ and $D_{\mu\nu}$. Livens' and
Minkowski's tensors fulfill this requirement. Abraham's fulfills the requirement
only in the reference frame in which the matter is at rest. 
Velocity dependent terms have to be added for obtaining a tensor that
transforms properly \cite{Abraham1910, Grammel1913}. Abraham's tensor makes sense
only for non-rotating rigid matter. Maxwell's equations are valid in any
reference frame, but Abraham's tensor introduces a particular frame in which
matter is at rest.

Fifth, the problem is determined by the force density which is restricted 
by theoretical and experimental requirements. For
example, the force density should  be consistent with the fact that the
fields that produce the effects on charges and currents are $\mathbf{E}$ and
 $\mathbf{B}$, not
$\mathbf{D}$ or $\mathbf{H}$. Terms in which the free current density, the
polarization or the
magnetization are coupled to $\mathbf{D}$ or $\mathbf{H}$ are not acceptable.
For the electric sector there have never been any doubts on this. For the magnetic sector there
was a long dispute during XIX century starting with the molecular currents
proposed by Amp\`ere, but nowadays this issue has been
settled \cite{JacJ1998, Rasetti1944, Hughes-Burgy1951}.
Minkowski's symmetrized  tensor, which approximates one proposed by
Peirls \cite{Peirls1976}, results in a power term $-(\mathbf{E}+\mathbf{D})
\cdot\mathbf{j}_\mathrm{f}/2$ which is of course unacceptable. The Einstein-Laub
force has terms that couple $\mathbf{M}$ and $\mathbf{H}$
that are also unacceptable. 
 
\subsection{Dipolar systems}
To gain insight,  let us discuss first the microscopic description of a
localized system with dipolar moments. We consider a localized system of
charges, whose total charge is zero, observed from a distance much bigger than its size $R$. The
properties of the system, like forces felt, can be
expanded in powers of $R$, that correspond to the multipole expansion.
To fix ideas let us assume that the
system  is localized near the origin,
\begin{equation}
r>R \Longrightarrow \rho(\mathbf{r},t) = 0 \quad \hbox{and}\quad
\mathbf{j}(\mathbf{r},t)=0\ ,
\end{equation}
\begin{equation}
\label{zero-charge}
\int \rho\,dV=0\ .
\end{equation}

The electrical dipole moment of the system is 
\begin{equation}
\label{elec-dip}
\mathbf{d}=\int \mathbf{x}\rho(\mathbf{x},t)\,dV\ .
\end{equation}
The current density can be split in an electrical current that vanishes
when $\rho$ vanishes and a magnetic one that is independent of the charge
density, $\mathbf{j}=\mathbf{j}_\mathrm{e}+\mathbf{j}_\mathrm{m}$.
The corresponding continuity equations are
\begin{equation}
\frac{\partial\rho}{\partial t}+\nabla\cdot\mathbf{j}_\mathrm{e}=0
\end{equation} 
and
\begin{equation}
\nabla\cdot\mathbf{j}_\mathrm{m}=0\ .
\end{equation} 
From the definition (\ref{elec-dip}) it follows that
\begin{equation}
\label{d-dot}
\dot{\mathbf{d}}=\int \mathbf{j}_\mathrm{e}\, dV\ .
\end{equation}
On the other hand
\begin{equation}
\label{zero-curr}
\int \mathbf{j}_\mathrm{m}\, dV=0\ .
\end{equation}
The magnetic dipole moment is 
 \begin{equation}
\label{mag-dip}
\mathbf{m}=\frac{1}{2c}\int\mathbf{x}\times \mathbf{j}_\mathrm{m}\, dV\ .
\end{equation}
Conditions (\ref{zero-charge}) and (\ref{zero-curr}) imply that the dipole
moments are independent of the choice of the origin. Equation (\ref{d-dot})
implies that $\mathbf{j}_\mathrm{e}$ is of first order in $R$.

The definitions of the dipolar moments depend on the reference frame. It is
important to determine how they change under Lorentz transformations.
For each reference frame the system as a whole moves with some velocity
$\mathbf{v}$. Let us define the antisymmetric object
$d_{\alpha\beta}$, $d_{ij}=\gamma\epsilon_{ijk}m_k$ and
$d_{0i}=-d_{i0}=\gamma d_i$, where
$\gamma$ is the usual relativistic factor $\gamma^{-1}=\sqrt{1-v^2/c^2}$. This
object transforms as a four-tensor in the limit $R \to 0$. In fact
\begin{equation}
d^{\alpha\beta}=\frac{1}{c}\gamma\int (x^\alpha j^\beta - x^\beta j^\alpha)dV+
{\cal O}(R^2)\ .
\end{equation}
The second order term is due to $\mathbf{j}_\mathrm{e}$.

Let us study now what happens when the system is immersed in external electric $\mathbf{E}$
and magnetic $\mathbf{B}$ fields. The force on the system is given by the
Lorentz expression
\begin{equation}
\mathbf{F}=\int(\rho\mathbf{E}+\frac{1}{c}\mathbf{j}\times\mathbf{B})\,dV\ ,
\end{equation}
the power is
\begin{equation}
\dot{W}=\int\mathbf{j}\cdot\mathbf{E}\,dV\ ,
\end{equation}
and the torque is
\begin{equation}
\mathbf{t}=\int \mathbf{x}\times(\rho\mathbf{E}+
\frac{1}{c}\mathbf{j}\times\mathbf{B})\,dV\ .
\end{equation}
If the fields change slowly over distances of the
order of $R$ the system behaves as a dipole. We need also the multipole expansions obtained from the 
Taylor series of the fields
\begin{eqnarray}
\mathbf{E}&=&\mathbf{E}(0)+\mathbf{x}\cdot\nabla\mathbf{E}+\dots\ ,\\
\mathbf{B}&=&\mathbf{B}(0)+\mathbf{x}\cdot\nabla\mathbf{B}+\dots\ .
\end{eqnarray}

 \subsection{Forces and torques on dipoles}
\label{Forces and torques on dipoles}

The magnetic effects are those depending on
$\mathbf{j}_\mathrm{m}$. This field is divergence-less localized and
with closed field lines. Since the whole space can be subdivided into
infinitesimal closed current tubes, it is enough to treat the case of
current loops. The general case can be then obtained making the usual
substitution
$\mathbf{j}dV\leftrightarrow id\mathbf{l}$. For a closed curve $\cal C$ an
area pseudo vector $\mathbf{{\cal A}}$ is defined by
\begin{equation}
\oint_{\cal C} x_i\,dx_j=\epsilon_{ijk}{\cal A}_k
\end{equation} 
or equivalently by
\begin{equation}
\mathbf{{\cal A}} = \frac{1}{2}\oint_{\cal C}\mathbf{x}\times d\mathbf{l}\ .
\end{equation}
The magnetic moment of a loop  of current $i$ is
 $\mathbf{m}=i/c\mathbf{{\cal A}}$.

The torque on the loop $\cal C$ is then
\begin{eqnarray}
\label{mag-torque}
\mathbf{t}_\mathrm{m}&=&
\frac{i}{c}\oint_{\cal C} \mathbf{x}\times[d\mathbf{l}\times (\mathbf{B}(0)+
\mathbf{x}\cdot\nabla\mathbf{B})]\nonumber\\
&=&\frac{i}{c}\mathbf{{\cal A}}\times\mathbf{B}(0)\\
&=&\mathbf{m}\times\mathbf{B}(0)\nonumber\ .
\end{eqnarray}
The same expression is valid in the general case.

Analogously the force on the loop $\cal C$ is 
\begin{eqnarray}
\label{mag-force}
\mathbf{F}_\mathrm{m}&=&
\frac{i}{c}\oint_{\cal C} d\mathbf{l}\times (\mathbf{B}(0)+
\mathbf{x}\cdot\nabla\mathbf{B})\nonumber\\
&=&\frac{i}{c}[\nabla(\mathbf{B}\cdot\mathbf{{\cal A}}) 
-(\nabla\cdot\mathbf{B}) \mathbf{{\cal A}}] \nonumber\\
&=&(\nabla\mathbf{B})\cdot\mathbf{m}\ .
\end{eqnarray}
Also in this case the same expression is valid in the general case.

The power on the current loop is
\begin{eqnarray}
\label{mag-power}
\dot{W}_\mathrm{m}&=&i\oint_{\cal C}\mathbf{E}\cdot d\mathbf{l}
= i\int_S \nabla\times\mathbf{E}\cdot d\mathbf{S}\nonumber\\
&=& -\frac{i}{c}\int_S \frac{\partial\mathbf{B}}{\partial t}\cdot d\mathbf{S}
= -\frac{i}{c}\frac{\partial\mathbf{B}(0)}{\partial t}\cdot\mathbf{{\cal A}}+
\dots \nonumber\\
&=& - \frac{\partial\mathbf{B}}{\partial t}\cdot\mathbf{m} \ .
\end{eqnarray}
As before this result is valid in the general case.

The torque on the electric dipole is
\begin{eqnarray}
\label{elec-torque}
\mathbf{t}_\mathrm{e}&=&\int \mathbf{x}\times(\rho\mathbf{E}+
\frac{1}{c}\mathbf{j}_\mathrm{e}\times\mathbf{B})\,dV \nonumber\\
&=&\mathbf{d}\times\mathbf{E}(0)\ .
\end{eqnarray}
Note that the term with $\mathbf{j}_\mathrm{e}$ is quadrupolar.
The force on the electric dipole is
\begin{eqnarray}
\label{elec-force}
\mathbf{F}_\mathrm{e}&=&\int[\rho(\mathbf{E}(0)+\mathbf{x}\cdot\nabla\mathbf{E})
+ \frac{1}{c}\mathbf{j}_\mathrm{e}\times(\mathbf{B}(0)+\dots)]\,dV
\nonumber\\
&=& \mathbf{d}\cdot\nabla\mathbf{E}+\frac{1}{c}\dot{\mathbf{d}}\times\mathbf{B}
+\dots\nonumber\\
&=&-\mathbf{d}\times(\nabla\times\mathbf{E})+\nabla\mathbf{E}\cdot\mathbf{d}+
\frac{1}{c}\dot{\mathbf{d}}\times\mathbf{B}\nonumber\\
&=& (\nabla\mathbf{E})\cdot\mathbf{d}+\frac{1}{c}\frac{\partial}{\partial t}
(\mathbf{d}\times\mathbf{B})\ .
 \end{eqnarray}
The power on the electric dipole is readily obtained
 \begin{eqnarray}
\label{elec-power}
\dot{W}_\mathrm{e}&=&\int \mathbf{j}_\mathrm{e}\cdot(\mathbf{E}(0)+\dots)\,dV
= \dot{\mathbf{d}}\cdot\mathbf{E}\nonumber\\
&=&-\mathbf{d}\cdot\frac{\partial\mathbf{E}}
{\partial t}+\frac{\partial}{\partial t}(\mathbf{d}\cdot\mathbf{E})\ . 
\end{eqnarray}

Comparing (\ref{mag-force}) with (\ref{elec-force}), and (\ref{mag-power})
with (\ref{elec-power}) we note that the electric results have additional
terms that are time derivatives. Such terms would have the undesirable
consequence of spoiling the relativistic covariance of the four-force on the dipole.
This one would be expected to be $K^\mu = 2^{-1}\partial^\mu F^{\alpha\beta}d_{\alpha\beta}$.
Lorentz covariance is secured if one recognizes that those terms are actually
contributions to the momentum and energy of the dipolar system. Let us consider
first the potential energy, $\phi(\mathbf{x})\approx\phi(0)-
\mathbf{E}\cdot\mathbf{x}$,
\begin{equation}
\label{potential-dip}
U_\mathrm{i}=\int \phi\rho\,dV=-\mathbf{E}\cdot\mathbf{d}\ .
\end{equation}
If one defines the energy of the dressed dipole to include the interaction or
potential energy
\begin{equation}
U_\mathrm{dressed}=U_\mathrm{bare}+U_\mathrm{i}=
U_\mathrm{bare}-\mathbf{E}\cdot\mathbf{d}
\end{equation}
then the power on the dressed dipole is
\begin{equation}
\label{power-dip}
\frac{dU_\mathrm{dressed}}{dt}= -\frac{\partial\mathbf{E}}{\partial t}\cdot
\mathbf{d} - \frac{\partial\mathbf{B}}{\partial t}\cdot\mathbf{m} \ .
\end{equation}

The momentum can be treated in an analogous way. Recognizing that the
troublesome term in (\ref{elec-force}) represents a potential
momentum that has to be added to the bare momentum to get the momentum of the
dressed dipole
\begin{equation}
\mathbf{p}_\mathrm{dressed}=\mathbf{p}_\mathrm{bare}+\mathbf{p}_\mathrm{i}=
\mathbf{p}_\mathrm{bare}-\frac{1}{c}\mathbf{d}\times\mathbf{B}\ ,
\end{equation}
then the force on the dressed dipole is
\begin{equation}
\label{force-dip}
\frac{d\mathbf{p}_\mathrm{dressed}}{dt}=\nabla\mathbf{E}\cdot\mathbf{d}
+\nabla\mathbf{B}\cdot\mathbf{m}\ .
\end{equation}

Equations (\ref{potential-dip}) and (\ref{power-dip}) suggest a reinterpretation 
of equation (\ref{energyflux}), in which the second term on the left hand side is subtracted 
from the field energy because  it pertains to matter, and the last two terms on the right hand side are incorporated to 
the power transfered from  the field to matter. This agrees with the discussion
we have anticipated in section \ref{electrostaticenergy}. In section
 \ref{Poynting-Minkowski}, we show that such an interpretation 
emerges naturally after the covariant force density in matter is  identified.

In appendix \ref{dressed}, we show that these troublesome terms indeed may be seen as
the energy and momentum of the electromagnetic fields produced by the
dipoles themselves.

Before turning our attention to the continuum, let us stress that here we are considering only the 
electromagnetic properties of the dressed dipole. To construct a mechanical model, the internal stresses 
which also contribute to inertia should be included. For the treatment of related problem of a point charge and the resolution 
of the 4/3 factor problem  see reference \cite{MedR2006}.

\subsection{Torque, force and power densities in matter}

The force and  torque acting on the dipole moments of a microscopic piece of
polarizable matter give place to a torque density and a force density acting
locally on each element of the material. The fields $\mathbf{E}$ and
$\mathbf{B}$ discussed in the previous section do not include  the self fields produced by
the dipolar system itself. For an element of material the self fields are
negligible because they are proportional to the volume of the element. So the
EM fields  in the following discussion may be taken as the total macroscopic fields.
 The torque on an element of polarized material is calculated using
(\ref{mag-torque}) and (\ref{elec-torque}). The torque density  
is obtained after dividing by the volume of the element. In terms of $\mathbf{P}$ and
$\mathbf{M}$ it is written as
\begin{equation}
\label{torque-space}
\bm{\tau}_\mathrm{d}=\mathbf{P}\times\mathbf{E}
+\mathbf{M}\times\mathbf{B}\ .
\end{equation}
This is clearly what should be expected and the only physically sensible result. It is easy to see 
that this is the spatial part of the torque density four-tensor
\begin{equation}
\label{torque-density}
\tau^{\mu\nu}_\mathrm{d}=
D^{\mu\alpha}F^\nu_{\ \alpha}-D^{\nu\alpha}F^\mu_{\ \alpha}\ .
\end{equation}
The  temporal components are given by
\begin{equation}
\label{torque-time}
\tau^{0k}_\mathrm{d}=(-\mathbf{P}\times\mathbf{B}+\mathbf{M}\times\mathbf{E})_k \ .
\end{equation}
The temporal part of the torque plays
an important role in disentangling the apparent paradox discussed by Balazs
(see our treatment of a wave-packet in section \ref{Wave} ).

The force  and power densities on an element of polarized material should be
computed starting from (\ref{force-dip}) and (\ref{power-dip}). Dividing by the volume
one gets
\begin{equation}
\label{force-density-dip}
\mathbf{f}_\mathrm{d}=P_k\nabla E_k + M_k\nabla B_k
\end{equation}
and
\begin{equation}
\label{power-density-dip}
cf^0_\mathrm{d}= -\mathbf{P}\cdot\frac{\partial\mathbf{E}}{\partial t}-
\mathbf{M}\cdot\frac{\partial\mathbf{B}}{\partial t}\ .
\end{equation}
In covariant notation the force density four-vector is
\begin{equation}
 \label{dipolarforce}
 f^\mu_\mathrm{d}=\frac{1}{2}D_{\alpha\beta}\partial^\mu F^{\alpha\beta}\ .
\end{equation}

The Lorentz covariance is spoiled if the force on the bare
dipoles is used, as has been done in
some treatments of the problem \cite{Gordon1973,BarnettLoudon2010}. 
This shows the importance of imposing that 
the splitting of the energy-momentum tensor in matter and field components
should be Lorentz invariant.
Note that the derivative in (\ref{dipolarforce}) is applied on the 
electromagnetic field.
We note that (\ref{dipolarforce}) was one of the terms in the force density
(\ref{GSforce}) advocated by de Groot and Suttorp \cite{deGrootS}
for a neutral dielectric fluid.

The total force density on matter is obtained adding the force on the free
charges with the force on the dipoles
\begin{equation}
\label{force-density}
 f^\mu =f^\mu_\mathrm{f}+f^\mu_\mathrm{d}= \frac{1}{c}F^{\mu}_{\ \nu}j^\nu_\mathrm{f}+
 \frac{1}{2}D_{\alpha\beta}\partial^\mu F^{\alpha\beta}\ .
\end{equation}

It is easy to verify that (\ref{force-density}) is related to  Amp\`ere-Lorentz's 
(\ref{Ampere-Lorentz}) and  Minkowski's (\ref{force-density-Min}) forces by
\begin{eqnarray}
 f^\mu &=&f^\mu_\mathrm{AL}+\partial_\nu(F^{\mu\alpha}D^{\ \nu}_{\alpha})\ ,\nonumber\\
 f^\mu &=&f^\mu_\mathrm{Min}+\partial^\mu(\frac{1}{4}D_{\alpha\beta}
F^{\alpha\beta}) \ .
\end{eqnarray}
In consequence, the total force on a piece of material held in vacuum obtained from 
any of these expressions is the same. Abraham's total force is different in this situation.
Locally, the obtained $f^\mu$ is not the same as the
Amp\`ere-Lorentz's force on the bound charges and currents. If one considers a
piece of material subdivided into infinitesimal elements, the
dipolar moment and the force that  each element feels are determined by the surface charges 
and currents on the element. Only when considering the total force is that
the contributions  of adjacent elements cancel out; what remains are the contributions of
the bulk and of the external surface of the piece of material which correspond to the
Amp\`ere-Lorentz total force. To discriminate experimentally between Minkowski's,
Abraham-Lorentz's and our proposal the local force density should be
measured, not only the total force.

\subsection{Energy-momentum tensor of matter}
\label{EMT-of-matter}
Excluding the time-derivative terms which come from (\ref{elec-force}) and (\ref{elec-power}) in
(\ref{power-density-dip}) and (\ref{force-density-dip}) does not mean that
they do not have a mechanical effect. They have to  be incorporated to the
energy and momentum of matter, which become dependent on the fields. In the local rest
frame of matter the potential terms are added to the energy and momentum of bare matter
\begin{equation}
\label{matter-energy-density}
u_\mathrm{M}(\mathbf{P},\mathbf{M},\mathbf{E})=
u_\mathrm{b}(\mathbf{P},\mathbf{M})-\mathbf{P}\cdot\mathbf{E}
\end{equation}
and
\begin{equation}
\label{matter-momentum-density}
\mathbf{g}_\mathrm{M}(\mathbf{P},\mathbf{M},\mathbf{E},\mathbf{B}) = 
\mathbf{g}_\mathrm{b}(\mathbf{P},\mathbf{M},\mathbf{E})
-\frac{1}{c}\mathbf{P}\times\mathbf{B}\ .
\end{equation}
We note that in general the dynamics of matter implies that $\mathbf{P}$ and  $\mathbf{M}$
also depend on the field. This is the case for example for a material with linear polarization discussed in 
subsection \ref{electrostaticenergy}. 
The energy-momentum tensor of matter should have the following form
\begin{equation}
\label{EMT-matter}
T^{\mu\nu}_\mathrm{M}=c^{-2}\tilde{u}_\mathrm{M}u^\mu u^\nu + P^{\mu\nu}
-\Delta T^{\mu\nu}_\mathrm{FM}
\end{equation}
where $\tilde{u}_\mathrm{M}$ is the energy density of matter in the rest
frame, $u^\mu$ is the four-velocity and in the rest frame of matter,
\begin{equation}
 \Delta T^{k0}_\mathrm{FM}=(\mathbf{P}\times\mathbf{B})_k  \ .
\end{equation}
The first term in (\ref{EMT-matter}) describes the
energy-momentum flux due to the drift of matter. The last term is
the potential momentum density; when it is absent $P^{\mu\nu}$ is a four-tensor,
but in general only $T^{\mu\nu}_\mathrm{M}$ is a four-tensor. When there is
no other mechanism of energy-momentum transfer besides the drift of matter,
$P^{\mu\nu}$ is the stress. In this case, it is purely spatial in the rest frame. In
other frames it contributes to the inertia of the system
\cite{MedR2006,MedR2006b}. It is worth remarking that $P^{\mu\nu}$ is
obtained using Newton's second law (\ref{EMT-eq}) and the equations that
determine the mechanical behavior of the system. If there are other mechanisms
of energy-momentum transfer, for example if there are electrical
currents, heat flow or waves $P^{\mu\nu}$ may have temporal components even
in the rest frame.

In the next section, we  use our knowledge of the force to define
the energy-momentum tensor of the electromagnetic field  which is consistent
with Maxwell equations and momentum conservation.

\section{The energy-momentum tensor of the field}
\label{theTensor}

\subsection{The energy-momentum tensor of the field in matter}
Now we are in a position to determine the energy-momentum tensor of the electromagnetic field in matter. 
The matter contribution to the total energy-momentum tensor satisfies (\ref{Tmatter}) with $f^\mu$
given by (\ref{force-density}).  The force density on the dipoles is not
the same as the force on the current density of bound charges. The right hand side of
(\ref{Identity}) is not minus the total force on the matter
(\ref{force-density}) and the straightforward identification valid in vacuum, of
$T_{\mathrm S}^{\mu\nu}$ as the energy-momentum of the field does not hold. 
The true energy-momentum tensor of the electromagnetic field is given by
\begin{equation}
\label{energy-momentum-tensor}
T^{\mu\nu}_{\mathrm{F}}=
-\frac{1}{16\pi}F_{\alpha\beta}F^{\alpha\beta}\eta^{\mu\nu}
+\frac{1}{4\pi}F^{\mu}_{\ \alpha}H^{\nu\alpha}\ .
\end{equation}
because it satisfies
\begin{equation}
\label{energy-momentum-equation}
\partial_\nu T_{\mathrm{F}}^{\mu\nu} = -f^\mu\ .
\end{equation}
as can be seen after a simple manipulation of Eq. (\ref{Identity}) using Bianchi's identity.
Newton's third law between matter and field holds.   

For this tensor, the energy density is
\begin{equation}
\label{energy-density}
u = T_{\mathrm F}^{00} = \frac{1}{8\pi}(E^2+B^2)+\mathbf{E}\cdot\mathbf{P} \ .
\end{equation}
The energy current density $ c T_{\mathrm F}^{0i}=\mathbf{S} =c\mathbf{E}\times
\mathbf{H}/4\pi$ and the momentum density $ c^{-1}T_{\mathrm F}^{i0}=\mathbf{g} =\mathbf{D}\times
\mathbf{B}/4\pi c$ coincide with Minkowski's expressions (\ref{MinPoynting})
 and (\ref{Minmomentum-density}).  Maxwell's stress tensor is given by,
\begin{eqnarray}
\label{Maxwell-tensor}
 T_{\mathrm F}^{ij}  = \frac{1}{8\pi}(E^2&+&B^2)\delta_{ij} 
-\mathbf{B}\cdot\mathbf{M}\delta_{ij} \nonumber\\
 &-&\frac{1}{4\pi}(E_i D_j +H_i B_j) \ .
\end{eqnarray}
and also differs from Minkowski's expression.
The difference between our energy density and the energy density for the
vacuum, $u-u_\mathrm{S}=\mathbf{P}\cdot\mathbf{E}$, is the negative of the
electrostatic energy density of the polarization. As discussed
in the previous subsection, this energy is 
subtracted from the energy of the microscopic field and should be considered part of the energy of
matter. This makes physical sense because it contributes to the inertia of matter in 
exactly the same way  as nuclear interaction energy contributes to nuclei
mass.  Note that there is no similar magnetic term since no potential
magnetic energy exists.

The difference between the momentum density in matter and
vacuum $\mathbf{g} -\mathbf{g}_\mathrm{S}=\mathbf{P}\times\mathbf{B}/c$ is the
opposite of the interaction momentum of the dressed electrical dipoles.

\subsection{The meaning of Poynting's energy density and Minkowski's tensor}
\label{Poynting-Minkowski}

The  energy current density is  given in our approach by Poynting's vector
$\mathbf{S}=c\mathbf{E}\times\mathbf{H}/4\pi$. The energy density (\ref{energy-density})  
is  different from Poynting's expression (\ref{Poynting-energy}), 
$u_\mathrm{P}=(\mathbf{E}\cdot\mathbf{D}+\mathbf{B}\cdot\mathbf{H})/8\pi$. From (\ref{energyflux}) the energy
conservation equation in our formulation reads
\begin{equation}
\frac{\partial u}{\partial t}+\nabla\cdot\mathbf{S}=-\mathbf{E}\cdot
\mathbf{j}_\mathrm{f}+\mathbf{P}\cdot\frac{\partial\mathbf{E}}{\partial t } +
\mathbf{M}\cdot\frac{\partial\mathbf{B}}{\partial t }
\end{equation}
which differs from  Poynting's conservation equation (\ref{Poynting-eq}) and also from  
equation (\ref{energyfluxMin}) deduced for Minkowski's tensor
for non linear materials. Note that in general there are contributions to the power different
from  the Joule term on the free
charges. Our results are valid for any kind of material in any kind of condition:
ferromagnets, saturated paramagnets, electrets, matter moving or at rest,
solids, fluids, absorbing and dispersive media, etc. In each case the polarization tensor should be identified. 
Poynting's equation  (\ref{Poynting-eq}) may be reconstructed from our result in the particular
case of linear polarizabilities, that is, when
the polarization tensor $D_{\alpha\beta}$ is proportional to the field tensor
$F^{\mu\nu}$.

Let us return to Poynting's energy density. When the matter is immersed in electromagnetic fields the energy density
of matter changes. This energy difference can be calculated integrating our power expression
(\ref{power-density-dip}). For time-independent linear polarizabilities
(\ref{P-linear}) and (\ref{M-linear}) the work done on matter does not depend
on the way the fields change in time, it depends only on the final values of
the fields
\begin{eqnarray}
\Delta u_\mathrm{M} &=& u_\mathrm{M}(\mathbf{E},\mathbf{B})- u_\mathrm{M}(0,0)=
\int dw_\mathrm{d}\nonumber\\
 &=& -\int (d\mathbf{E}\cdot\mathbf{P}+
d\mathbf{B}\cdot\mathbf{M})\nonumber\\
&=& -\frac{1}{2}(\mathbf{E}\cdot\mathbf{P}+\mathbf{B}\cdot\mathbf{M}) \ .
\end{eqnarray} 
This includes the electrical potential energy
density $- \mathbf{E}\cdot\mathbf{P}$  
\begin{equation}
\Delta u_\mathrm{M}=-\mathbf{E}\cdot\mathbf{P}+
\frac{1}{2}(\mathbf{E}\cdot\mathbf{P}-\mathbf{B}\cdot\mathbf{M})\ .
\end{equation}
Poynting's energy density is
\begin{equation}
u_\mathrm{P}=u+\Delta u_\mathrm{M}=u_\mathrm{S}+
\frac{1}{2}(\mathbf{E}\cdot\mathbf{P}-\mathbf{B}\cdot\mathbf{M})\ .
\end{equation}
Therefore it results that Poynting's energy density corresponds to a mixed entity with contributions of ``fields
plus polarizations''. On the other hand Minkowski's is the only relativistic
tensor for which $T^{00}=u_\mathrm{P}$ and $cT^{0k}=S^k$ in any frame of reference. For an
electromagnetic wave propagating in a medium it makes sense to include the
polarization energy of matter as part of the wave energy. So as we suggested  in subsection \ref{Comparison}  Minkowski's
tensor properly represents the energy-momentum tensor of the EM wave in a
non-dispersive medium. It is
useful for calculating the theoretical force on the wave, but it cannot be
used for determining the force on matter, since the wave energy has a
component which belongs to  matter. One has to use (\ref{energy-momentum-tensor})
for doing that. Note that in general, it is the force on matter  the one that can be
measured.

Minkowski's energy-momentum tensor may be written as,
\begin{equation}
T^{\mu\nu}_{\mathrm{Min}}=T^{\mu\nu}_\mathrm{F}+\frac{1}{4}D_{\alpha\beta}
F^{\alpha\beta}\eta^{\mu\nu}\ .
\end{equation}
In this view the force density on the wave is
\begin{eqnarray}
\label{Minkowski-force}
\partial_\nu T^{\mu\nu}_{\mathrm{Min}}&=& -\frac{1}{c}F^{\mu}_{\ \alpha}
j^{\alpha}_\mathrm{f} -\frac{1}{2}\partial^\mu F^{\alpha\beta}
D_{\alpha\beta}
+\frac{1}{4}\partial^\mu(F^{\alpha\beta}D_{\alpha\beta})\nonumber\\
&=&  -\frac{1}{c} F^{\mu}_{\ \alpha}j^{\alpha}_\mathrm{f} +
\frac{1}{8}\partial^\mu\chi_{\alpha\beta\gamma\delta} F^{\alpha\beta}
F^{\gamma\delta}\ .
\end{eqnarray}
The second term is due to refraction.
In particular, the EM wave propagating in a time-independent homogeneous
medium conserves its energy and momentum. No magnetostriction or
electrostriction effects  are  described by Minkowski's tensor
\cite{Brevik1979}, but they appear when using (\ref{force-density}) and
(\ref{energy-momentum-tensor})  (see  section \ref{testing}).

Poynting's equation does not hold when the polarizabilities
are not linear as for example in the well known 
case of optical dispersive media \cite{JacJ1998,Furutsu1969,Philbin2011}. Our result should
be a good starting point for a fresh approach to study such cases. 
 
Further evidence of the inadequacy of Poynting's energy and
Minkowski's tensor for non-linear systems is obtained calculating  
the force on a piece of dipolar material in external fields as minus 
the gradient of the total energy 
(field and matter) when the piece is displaced. If the material
is linear this energy can be calculated using Poynting's energy density,
since the polarization terms in Poynting's energy density are equal to
the polarization-dependent terms in the energy density of bare matter.
This relation does not hold for non-linear systems; for example 
for an electret or a permanent magnet this procedure using Poynting's energy
yields  a force that is one half of the
correct result. On the other hand Minkowski's force, obtained from the
first equation of the set (\ref{Minkowski-force}),
gives the correct result even for this case. Therefore there is an
inconsistence if one pretends to calculate the total field-dependent energy of
non-linear systems using Minkowski's (Poynting's) energy density.
Our force density (\ref{force-density}) 
matches the result obtained by evaluating the energy using our field energy
density and the energy density of the dressed matter which includes the
dipolar potential energy density $-\mathbf{E}\cdot\mathbf{P}$.

\subsection{Orbital angular momentum and spin}

The identification of the spin density of the electromagnetic field, and in general
of gauge fields,
is an interesting problem, which  even in the vacuum has not yet been resolved satisfactorily. The canonical spin 
density which emerges from Noether's theorem  \cite{BelF1939} is 
gauge-dependent, and up to now there is no consistent and gauge-invariant
definition of the kinetic spin of the electromagnetic field. Nevertheless, from the kinematic point 
of view it is clear that whatever expression may it have,  equations
(\ref{total-eq}) and (\ref{spin-eq}) for the field quantities should be satisfied.

The non-symmetric part of  $T^{\mu\nu}_\mathrm{F}$ happens to be equal to the
torque density on the dipoles,
\begin{equation}
\label{dip-torque}
T^{\mu\nu}_\mathrm{F}-T^{\nu\mu}_\mathrm{F}=\tau^{\mu\nu}_\mathrm{d}\ .
\end{equation}
The orbital angular momentum equation (\ref{orbital-eq}) for the EM field is therefore
\begin{equation}
\label{orbital-eq-field}
\partial_\alpha L^{\mu\nu\alpha}_\mathrm{F} = -\tau^{\mu\nu}_\mathrm{d}-
x^\mu f^\nu +x^\nu f^\mu\ .
\end{equation}
The dipolar torque density is canceled out by  $T^{\mu\nu}_\mathrm{F}$'s
asymmetry in the spin equation (\ref{spin-eq}) of the field which reduces to
\begin{equation}
\label{spin-eq-field}
\partial_\alpha S^{\mu\nu\alpha}_\mathrm{F} = -\tau^{\mu\nu}_\mathrm{c}\ .
\end{equation}
where the tensor $\tau_\mathrm{c}$ represents an additional torque density acting on
the currents that could  exist.  The spin density is
not coupled to the torque density on the dipoles. There are phenomena, like the
absorption of circularly polarized light, in which the electromagnetic spin
is relevant. In those systems, the form of $\tau^{\mu\nu}_\mathrm{c}$ depends on the nature of the 
microscopic coupling responsible for the effect. As a result optical angular momentum may
be transferred to matter \cite{Beth1936,Carrara1949,Friese1996,Friese1998}.

As already said, the argument used in favor of Abraham's tensor  that the field  $T^{\mu\nu}$ should be
symmetric for angular momentum conservation  is
incorrect. The field exerts a torque density
(\ref{torque-space}) on  polarized matter, and the corresponding reaction
acts on the field. What is conserved is the total angular momentum of matter
and field (including spin), not the orbital angular momentum of the  field alone. It is precisely the
asymmetry of the energy-momentum tensor what couples in  (\ref{orbital-eq}) the orbital angular
momentum with the torque density. Our tensor has the right asymmetry to ensure angular momentum conservation.

\section{The average method}
\label{Macroscopic}

The traditional method to obtain the
phenomenological fields,  $\mathbf{P}$ and  $\mathbf{M}$ which appear in the
macroscopic Maxwell equations is  by making averages on
the microscopic electric and magnetic dipoles. In this approach, the
covariant treatment  presents some problems in how to define consistently the
dipole moments from microscopic charges and currents, particularly when the
dipoles are moving or changing in time \cite{JacJ1998,deGrootS}. To understand better 
the nature of the energy-momentum tensor obtained in the previous section, it is
convenient to look at its relation with this approach. In the present section, we re-obtain our result
in a form  which is independent of any microscopic model of the material and does not require to use the force on
the dipoles as an input  \cite{MedandSc}. We suppose that the macroscopic quantities may be obtained by
taking the average of microscopic fields and forces over small regions of
space-time, and that the macroscopic force density may be expanded  in
the macroscopic fields and their derivatives. In the dipolar approximation
only the first derivatives are relevant.  We define the free charge current
density as the four-vector that couples to the average field and the 
polarization density tensor as the quantities that couple to the first
derivatives in the expansion. Then we obtain
self-consistently the expressions for the free and bound (dipolar) charge and
current densities and for the energy-momentum tensor. Finally, we show that the
polarization tensor that was defined as the coupling parameters of the force
should indeed be interpreted as the polarization-magnetization tensor.

\subsection{Space-time average}
At a microscopic scale, we assume that for the electromagnetic field
$F^{\mu\nu}$ and the current density $j^\mu$, Maxwell's equations (\ref{Maxwell})
and Bianchi's identity (\ref{Bianchi}) hold
and that the force density is given by Lorentz expression (\ref{Lorentz}).
 Correspondingly, the
energy-momentum tensor for the microscopic electromagnetic field is the
standard tensor $T_\mathrm{S}$ (\ref{TEI-S}).
which satisfy identically $\partial_\nu T^{\mu\nu}_{\mathrm S}=-f_{\mathrm L}^\mu$ as a
 consequence of
(\ref{Maxwell}), (\ref{Bianchi}) and (\ref{Lorentz}). The
force density can alternatively be obtained from the energy-momentum tensor of
 matter by
$f_{\mathrm L}^\mu = \partial_\nu T^{\mu\nu}_{\mathrm M}$ if it is available.

The observable macroscopic quantities correspond to averages of the
microscopic quantities over small regions of space-time. Similar treatments
are usually done averaging only over space \cite{deGrootS}. We average also
over time for two reasons: 1) Lorentz covariance is preserved naturally,
2) average over time is required  to handle very fast processes as for
example electrons moving around nuclei. The average is done the following way
\begin{equation}
\label{average}
\langle A(x)\rangle = \int A(x^\prime)W(x-x^\prime)\,dx^\prime\ ,
\end{equation}
where $W(x)$ is a smooth function that is essentially constant inside a region
of size $R$ and  vanishes outside
\begin{equation}
W(x)\ge 0\ ,\qquad |x^\mu|>R \Longrightarrow W(x)=0\ ,
\end{equation}
\begin{equation}
 \int W(x)\,dx =1\ .
\end{equation}
Inside the region $W(x)\approx \langle W\rangle$.

At the scale $R$ the microscopic fluctuations are washed out,
and  therefore all the products of averages and fluctuations are
negligible. That is, if $\delta A= A-\langle A\rangle$ then
$\langle \langle B\rangle\delta A\rangle\approx 0$ and also
 $\langle\langle B\rangle\rangle \approx \langle B\rangle$.

With  these conditions it follows that
\begin{eqnarray}
\label{average_derivative}
 \partial_\nu \langle A\rangle &=& \int A(x^\prime)\partial_\nu W(x-x^\prime)
\, dx^\prime\nonumber\\ &=&
-\int A(x^\prime)\partial^\prime_\nu W(x-x^\prime)\,dx^\prime\nonumber\\ 
&=&-\int\partial^\prime_\nu[A(x^\prime)W(x-x^\prime)]\,dx^\prime\nonumber\\
&\ &\ + \int \partial^\prime_\nu A(x^\prime)W(x-x^\prime)\,dx^\prime\nonumber\\
&=& \langle \partial_\nu A\rangle \ .
\end{eqnarray}

In this section we denote the macroscopic fields obtained by averaging with a bar. From the discussion,
it emerges that they are equal to the corresponding macroscopic quantities used in the previous sections.
In particular we have,
$\bar{F}^{\mu\nu} =\langle F^{\mu\nu}\rangle$. Using (\ref{average_derivative})
one immediately obtains that Maxwell's equations (\ref{Maxwell}) are valid for the averaged
macroscopic fields and the averaged currents
\begin{equation}
\label{Maxwell_average}
\partial_\nu\bar{F}^{\mu\nu}=\frac{4\pi}{c}\langle j^\mu\rangle
\end{equation}
and that Bianchi's identity is valid for $\bar{F}^{\mu\nu}$ and also for
$\delta F^{\mu\nu}=F^{\mu\nu}-\bar{F}^{\mu\nu}$.

\subsection{Macroscopic force}
We avoid any controversy about the macroscopic force by making the sole hypothesis that the
microscopic force is given by the Lorentz expression. To obtain the average current 
density and the macroscopic
energy-momentum tensors of fields and matter we compute first the average of
the microscopic forces. Since the force (\ref{Lorentz}) is quadratic in the
microscopic quantities, the result in this case is not straightforward. We
assume that the macroscopic force density $\bar{f}^\mu$ on matter shares with
the microscopic force density 
the following four properties: 1) It is a
local functional of the average field.
 2) It is linear in the field. 
 3) The direction of the force is determined by the field. 4) It
transforms as a four-vector.
Therefore the force density can be expanded as a sum of terms in
which the derivatives of the field tensor are contracted with matter-dependent
tensors, and the free index belongs to the field derivative tensors.
 In the dipolar approximation,
 we neglect the coupling with second and higher order
derivatives which correspond to quadrupolar and higher order couplings.
 Taking into account Bianchi's identity, we prove that the most general
expression for the macroscopic force density  fulfilling the four  conditions 
stated  above takes the form
\begin{equation} 
\label{Mac-force}
\bar{f}^\mu = \frac{1}{c}\bar{F}^\mu_{\ \alpha}\bar{j}^\alpha_{\mathrm f}+
\frac{1}{2}\partial^\mu \bar{F}^{\alpha\beta} \bar{D}_{\alpha\beta}\ .
\end{equation}
It is written in terms of two phenomenological quantities related to matter,
a four-vector $\bar{j}^\alpha_{\mathrm f}$ and an antisymmetric four-tensor
$\bar{D}_{\alpha\beta}$ to be identified later with the current density of
free charges and the dipolar tensor density. Note that  the temporal component of 
$\bar{D}_{\alpha\beta}$ is then  identified with the polarization vector $\bar{\mathbf{P}}$ by $\bar{P}_k = \bar{D}_{0k}$, 
and the spatial components 
with the magnetization vector $\bar{\mathbf{M}}$  by $\bar{D}_{ij}=\epsilon_{ijk}\bar{M}_k$.

With the use  Bianchi's identity of and some simple algebra the force expression
takes the alternative form
\begin{equation}
\label{Mac-force-2}
\bar{f}^\mu = \frac{1}{c}\bar{F}^\mu_{\ \alpha}(\bar{j}^\alpha_{\mathrm f}+
c\partial_\beta \bar{D}^{\alpha\beta})+
\partial_\beta(\bar{F}^\mu_{\ \alpha} \bar{D}^{\beta\alpha})\ .
\end{equation}
In the microscopic formulation, the total energy-momentum tensor $T^{\mu\nu}$
splits in a contribution of the field given by (\ref{TEI-S}) and a
contribution $T^{\mu\nu}_{\mathrm M}$ of the matter. The average of the total
microscopic energy-momentum tensor is the macroscopic energy-momentum tensor.
How it splits in a macroscopic matter term $\bar{T}^{\mu\nu}_{\mathrm M}$,
and a macroscopic field term $\bar{T}^{\mu\nu}_{\mathrm F}$, should be
determined by Newton's third law. We write,
\begin{equation}
\bar{T}^{\mu\nu}_{\mathrm F} + \bar{T}^{\mu\nu}_{\mathrm M} =
 \langle  T^{\mu\nu}_{\mathrm S} + T^{\mu\nu}_{\mathrm M}\rangle \ .
\end{equation}
and impose,
\begin{equation}
\label{NewtonLaw}
\partial_\beta \bar{T}^{\mu\beta}_{\mathrm M}=\bar{f}^\mu = -\partial_\beta \bar{T}^{\mu\beta}_{\mathrm F}\ .
\end{equation} 
The average of the microscopic energy-momentum tensor of
electromagnetic fields is not the macroscopic energy-momentum of the field. It
may be expressed as, 
\begin{equation}
\langle T^{\mu\nu}_{\mathrm S}\rangle = \bar{T}^{\mu\nu}_{\mathrm S}+
\langle \delta T^{\mu\nu}_{\mathrm S}\rangle\ , 
\end{equation}
where $\bar{T}^{\mu\nu}_{\mathrm S}$ is the standard tensor built with the
macroscopic fields and $\delta T^{\mu\nu}_{\mathrm S}$ is the standard tensor
built with the fluctuation fields $\delta F^{\mu\nu}$. For the actual
macroscopic energy-momentum tensor of fields inside matter we write,
\begin{equation}
\label{mac-tensorF}
\bar{T}^{\mu\nu}_{\mathrm F} = \bar{T}^{\mu\nu}_{\mathrm S} + \Delta T^{\mu\nu}_{\mathrm F} \ ,
\end{equation} 
where $\Delta T^{\mu\nu}_{\mathrm F}$ is a possible polarization-dependent
correction to be determined self-consistently. The terms linear in the
fluctuations were neglected as explained above. To obtain the macroscopic
energy-momentum tensor of matter, the average fluctuation tensor of the field
has to be included and the dipolar correction must be subtracted,
\begin{equation}
\bar{T}^{\mu\nu}_{\mathrm M} = \langle T^{\mu\nu}_{\mathrm M}\rangle + 
\langle \delta T^{\mu\nu}_{\mathrm S}\rangle - \Delta T^{\mu\nu}_{\mathrm F}\ . 
\end{equation}
Note that the contribution of the fluctuation fields $\delta F$ should be considered part of the macroscopic matter.
This is the reason why $\bar{f}^\mu \not=\langle f^\mu\rangle$.

Using (\ref{NewtonLaw}), the macroscopic force density is
\begin{equation}
\label{Mac-force-3}
\bar{f}^\mu =\partial_\beta \bar{T}^{\mu\beta}_{\mathrm M}=\langle f^\mu\rangle+
\partial_\beta[\langle \delta T^{\mu\beta}_{\mathrm S}\rangle - \Delta T^{\mu\beta}_{\mathrm F}]\ .
\end{equation} 
Using (\ref{Maxwell}) and (\ref{Lorentz}) the averaged microscopic
force density is
\begin{equation}
\langle f^\mu\rangle =\frac{1}{4\pi}\langle F^\mu_{\ \alpha}\partial_\beta
F^{\alpha\beta}\rangle\ .
\end{equation}
Making the substitution $F^{\alpha\beta}=\bar{F}^{\alpha\beta}+
\delta F^{\alpha\beta}$ one gets
\begin{equation}
\langle f^\mu\rangle =\frac{1}{4\pi}\bar{F}^\mu_{\ \alpha}\partial_\beta
\bar{F}^{\alpha\beta}+ \frac{1}{4\pi}\langle\delta F^\mu_{\ \alpha}
\partial_\beta\delta F^{\alpha\beta}\rangle\ .
\end{equation}
Using (\ref{Maxwell_average}) in the first term of this last equation and
Bianchi's identity in the second, the average microscopic force density is then
\begin{equation}
\langle f^\mu\rangle =\frac{1}{c}\bar{F}^\mu_{\ \alpha}\langle j^\alpha\rangle
-\langle\partial_\beta\delta T^{\mu\beta}_{\mathrm S}\rangle\ .
\end{equation}
Substituting this result in (\ref{Mac-force-3}) and equating with
 (\ref{Mac-force-2}) one finally gets
\begin{eqnarray}
\label{condition}
\frac{1}{c}\bar{F}^\mu_{\ \alpha}(\bar{j}^\alpha_{\mathrm f}&+&c\partial_\beta
 \bar{D}^{\alpha\beta}-\langle j^\alpha\rangle)=\nonumber\\
&-&\partial_\beta(\bar{F}^\mu_{\ \alpha}
\bar{D}^{\beta\alpha}+\Delta T^{\mu\beta}_{\mathrm F})\ .
\end{eqnarray}
This last equation is satisfied identically by setting
\begin{equation}
\label{current_density}
\langle j^\alpha\rangle = \bar{j}^\alpha_{\mathrm f}+ c\partial_\beta \bar{D}^{\alpha\beta}
\end{equation}
and 
\begin{equation}
\label{correction}
\Delta T^{\mu\nu}_{\mathrm F} = -\bar{F}^{\mu}_{\ \alpha}\bar{D}^{\nu\alpha}\ .
\end{equation}
These solutions  are unique. Since (\ref{condition}) must be satisfied
for any field tensor, equation  (\ref{current_density}) follows. A tensor
$X^{\mu\nu}$ whose divergence
vanishes identically ($\partial_\nu X^{\mu\nu}\equiv 0$) could be added to 
$\Delta T^{\mu\nu}_{\mathrm F}$, but since $\bar{T}^{\mu\nu}_{\mathrm F}$ can
be at most quadratic in the fields, and should vanish when the fields vanish
such a tensor also vanishes.

Defining the current 
\begin{equation}
 \bar{j}^\alpha_{\mathrm b}=c\partial_\beta \bar{D}^{\alpha\beta}\ ,
\end{equation}
Maxwell's equations for the macroscopic fields (\ref{Maxwell_average}) are written in the familiar form,
\begin{equation}
\label{Maxwell_dipolar}
\partial_\nu\bar{F}^{\mu\nu}=
 \frac{4\pi}{c}(\bar{j}^\mu_{\mathrm f}+\bar{j}^\mu_{\mathrm b})\ .
\end{equation}
From here the identification of
$\bar{j}^\alpha_{\mathrm f}$ as the current of free charges and $ \bar{j}^\alpha_{\mathrm b}$ as 
the bound current density is evident. Noting also that in  terms of $\bar{\mathbf{P}}$ and $\bar{\mathbf{M}}$ defined above, the 
familiar expressions 
$\rho_{\mathrm b}= -\nabla\cdot\bar{\mathbf{P}}$  and $\mathbf{j}_{\mathrm b}=
 \dot{\bar{\mathbf{P}}}+ c\nabla\times\bar{\mathbf{M}}$ are obtained, the identification of $\bar{D}^{\nu\alpha}$ with
the dipolar density is almost completed.

The dipolar or bound charge is conserved identically, $\partial_\alpha
j^\alpha_{\mathrm b}=c\partial_\alpha\partial_\beta \bar{D}^{\alpha\beta}=0$. Since
the average charge is conserved, equation (\ref{current_density}) implies that the free charge 
is also conserved, $\partial_\alpha \bar{j}^\alpha_{\mathrm f}=0$.

It is worth noting that this whole scheme leaves out
processes, like ionization or capture of carriers, in which there is exchange
between free charges and bound charges.  In those cases (\ref{Mac-force}) does
not hold. 

Now we proceed to discuss the energy-momentum tensor. It is convenient to
define the tensor
\begin{equation}
\bar{H}^{\alpha\beta} = \bar{F}^{\alpha\beta} - 4\pi \bar{D}^{\alpha\beta}\ ,
\end{equation}
which is related to the electric displacement $\bar{\mathbf{D}}=\bar{\mathbf{E}}
+4\pi\bar{\mathbf{P}}$
and the magnetizing field $\bar{\mathbf{H}}=\bar{\mathbf{B}}-4\pi\bar{\mathbf{M}}$, $\bar{H}^{0k}=\bar{D}_k$
and $\bar{H}^{ij}=\epsilon_{ijk}\bar{H}_k$. With this notation, Maxwell's equations
 (\ref{Maxwell_dipolar}) become
\begin{equation}
\partial_\nu\bar{H}^{\mu\nu}=\frac{4\pi}{c}\bar{j}^\mu_{\mathrm f}
\end{equation}
and the macroscopic energy-momentum tensor of fields given by  (\ref{mac-tensorF}) and (\ref{correction}) is
\begin{eqnarray}
\bar{T}_{\mathrm F}^{\mu\nu}&=&\bar{T}_{\mathrm S}^{\mu\nu} +
 \Delta T^{\mu\nu}_{\mathrm F}\nonumber\\
&=& -\frac{1}{16\pi}\eta^{\mu\nu}\bar{F}^{\alpha\beta}\bar{F}_{\alpha\beta}+
\frac{1}{4\pi}\bar{F}^\mu_{\ \alpha}\bar{H}^{\nu\alpha}\ . 
\end{eqnarray}
This is exactly the result we obtained in  the previous section, with the notational difference 
of the bar used in this section to identify the macroscopic fields.

\subsection{Densities of dipole moment}
To  wholly recover the usual picture let us show that 
$\bar{\mathbf{P}}$ and $\bar{\mathbf{M}}$ are the densities of electric dipole moment and of
magnetic dipole moment respectively without relying in the model of microscopic dipoles 
(This is basically the reverse of the usual textbook computation). 

First we  find  the charges  at the surface of a piece of material. Near the
surface $\bar{\mathbf{P}}$, goes smoothly to zero in a distance of the order of
$R$.  The total dipolar charge is obtained by integrating
$\rho_{\mathrm b}=-\nabla\cdot\bar{\mathbf{P}}$ over the volume of the material.
Because of Gauss' theorem such integral is zero since $\bar{\mathbf{P}}=0$ at
the surface. Then the total dipolar charge is always zero. It is a bound
charge that cannot leave the material. At a macroscopic scale (much bigger
than $R$) there is a discontinuity at the surface. The dipolar charge at the
surface is always opposite to the charge in the bulk.  If $\sigma_{\mathrm b}$
is the surface density of polarization, then
\begin{equation}
0=\oint \sigma_{\mathrm b}\,dS-\int \nabla\cdot\bar{\mathbf{P}}\,dV=
\oint[\sigma_{\mathrm b}-\bar{\mathbf{P}}\cdot\hat{\mathbf{n}}\,]\,dS\ .
\end{equation}
This can happen for any surface if
\begin{equation}
\sigma_{\mathrm b} =\bar{\mathbf{P}}\cdot\hat{\mathbf{n}}\ ,
\end{equation} 
which of course is the usual expression given by the model of microscopic
dipoles.  The electrical dipole moment of a piece of material is computed as
always giving
\begin{eqnarray}
\mathbf{d}&=&\oint \mathbf{x}\sigma_{\mathrm b}\,dS +\int\mathbf{x}
\rho_{\mathrm b}\,dV=\int\bar{\mathbf{P}}\,dV\ . 
\end{eqnarray}
This completes the interpretation of $\bar{\mathbf{P}}$ as the density of dipole
moment.

The magnetic current density in the bulk is
$\bar{\mathbf{j}}_{\mathrm m}=c\nabla\times\bar{\mathbf{M}}$. In addition, there is also a
 surface current density $\mathbf{\Sigma}_{\mathrm m}$ that
can be obtained in a similar way  to that used for $\sigma_{\mathrm b}$. Let us
consider a closed
curve that is outside the piece of material and is the border of a surface
that cuts the material.  The total magnetic current that crosses
the surface is the flux of $\bar{\mathbf{j}}_{\mathrm m}$. Because of Stokes' theorem
such flux is the circulation of $c\bar{\mathbf{M}}$ in the curve, which is zero.
Therefore for any surface $S$ that cuts the material, the total magnetic current
that crosses the surface is zero. At a macroscopic scale the bulk current is
opposite to the surface current. Let ${\cal C}$ be the
intersection of $S$ with the surface of the piece of material, 
 $\hat{\mathbf{t}}$ the unitary vector tangent to the curve ${\cal C}$ and
$\hat{\mathbf{n}}$ the unitary vector orthogonal to the surface of the piece.
The unitary vector which is orthogonal to ${\cal C}$, and tangent to the surface
of the piece is $\hat{\mathbf{n}}\times\hat{\mathbf{t}}$. Then,
\begin{eqnarray}
0&=&\oint_{\cal C}\mathbf{\Sigma}_{\mathrm m}\cdot\hat{\mathbf{n}}\times
\hat{\mathbf{t}}\,dl
+\int_S c\nabla\times\bar{\mathbf{M}}\cdot d\mathbf{S}\nonumber\\
&=&\oint_{\cal C}[\mathbf{\Sigma}_{\mathrm m}\cdot\hat{\mathbf{n}}\times
\hat{\mathbf{t}}+ c\bar{\mathbf{M}}\cdot\hat{\mathbf{t}}\,]dl\nonumber\\
&=&\oint_{\cal C}[\mathbf{\Sigma}_{\mathrm m}\times\hat{\mathbf{n}}+
c\bar{\mathbf{M}}]\cdot\hat{\mathbf{t}}\,dl\ .
\end{eqnarray}
The expression that fulfills this equation for any $\hat{\mathbf{t}}$ is
\begin{equation}
\mathbf{\Sigma}_{\mathrm m}= c\bar{\mathbf{M}}\times\hat{\mathbf{n}}\ .
\end{equation}
The usual computation of the magnetic dipole moment of the magnetic currents, 
\begin{eqnarray}
\mathbf{m}&=&\frac{1}{2c}\oint \mathbf{x}\times\mathbf{\Sigma}_{\mathrm m}\,dS
+\frac{1}{2c}\int\mathbf{x}\times\bar{\mathbf{j}}_{\mathrm m}\,dV\nonumber\\
&=& \int\bar{\mathbf{M}}\,dV\ ,
\end{eqnarray}
completes the interpretation of $\bar{\mathbf{M}}$ as the density of
magnetic dipole moment.

\section{Testing the forces}
\label{testing}

\subsection{Experimental results}

The experimental researches deal with two related but different aspects of the controversy. 
One is the momentum of light in matter and the other is the force density that the
electromagnetic fields exert on matter.

It is the general view of most the authors which worked on the subject that the reported 
experiments 
\cite{JonesR1954,JonesL1978,Gibson1980,Campbell,Zuev2009} where  the momentum of
light  is observed, support Minkowski's expression for the momentum of the photons.
The analysis done  of Doppler and  Cherenkov effects
\cite{Ginzburg1960,Ginzburg1979} choose Minkowski's momentum too. This view is shared even
by some works which hold alternative positions on the subject as for example reference
\cite{ZhangLS2015}  which reports evidence  for
Abraham's pressure of light,  or reference \cite{Kemp2011} which argues in favor of
Barnett's proposal of different classical and quantum momenta. Since our momentum density
expression coincides with Minkowski's, those experimental results also support our approach.

In relation with the force measurements the opinions are much more divided. The interpretation 
of the experiments depends on the measurement of very
tiny quantities and it is not a surprise that it has been difficult to discard that other effects related to
the collective behavior of the samples may be responsible for the outcome of the experiments. 

Also the theoretical analysis of the force density is  confusing. 
In a recent paper \cite{JazayeriMehrany2014}, Jazayeri and Mehrany  presented a
comparative study of the most popular force densities that have been proposed,
Amp\`ere-Lorentz on the bound charges and currents, Einstein-Laub, Minkowski
and Abraham. Although we do not agree with their whole analysis, it is
interesting to consider their conclusions as a point of comparison. The article
is very critical about  Abraham's formula characterized as  
fanciful and devoid of physical meaning, Einstein-Laub tensor force which is
found to be incompatible with special relativity, and Minkowski's expression
for  excluding the interaction on the bound charges and currents inside
homogeneous media.  They spare Amp\`ere-Lorentz force stressing that it
corresponds to the assumption that the EM fields act 
on the total charges and currents.
The expression for the force density that we are proposing is different from
all of those. It considers the local force on the
dipoles and is compatible with relativity. Our force, Amp\`ere-Lorentz and
Minkowski's (when it is valid) each predicts the same
total force on a piece of material isolated in vacuum, but the local
distributions are different. Therefore they correspond to different stresses
inside the material.  Measuring the stress, that is magnetostriction and
electrostriction, gives an empirical method for determining which formula is the 
correct one. 

Most of the experiments that may be devised  to measure the force  which the
electromagnetic field exert on matter will measure the total force on a polar or
polarizable sample. As mentioned in section \ref{Force}, such experiments do not allow
to distinguish between Amp\`ere-Lorentz's, Minkowski's or our force density because in the 
three cases the total force is the same. Abraham's and Einstein-Laub's total forces are
different. On the other hand a determination of the force density depends more 
on the details of thermal and mechanical properties of the sample. 

For the specific case of an electromagnetic wave entering a medium with a higher
refraction index the question of weather the force is in the inward or outward
direction has a long history. Thomson and Poynting \cite{Poynting1905} suggested that
the force is in the outward direction, a result which was verified experimentally by
Barlow \cite{Barlow1912} for oblique incidence. This is
consistent with Minkowski's momentum expression while Abraham's momentum corresponds
to an inward force. The experiment of Ashkin and  Dziedzic \cite{AshkinD1973} tried to
elucidate this point by illuminating  a liquid with a laser beam and determining the
shape of the surface. Their observations were consistent with a raising of the liquid
level and in a first analysis with an outward force and Minkowski's momentum, but
issues related to the thermal and hydrodynamic response of the medium were invoked to
object this conclusion \cite{Gordon1973,LaiHY1976}. The same effect was observed in
\cite{Casner2001,Sakai2001,Astrath2014} but the opposite result has been recently
reported in \cite{ZhangLS2015}. Related experiments which also support Minkowski's
momentum are discussed in \cite{Kajorndejnukul2013,Qiu2015}.

In relation with Abraham's force term, there have been a few works which claimed
experimental evidence for it. These include the work in references
 \cite{WalkerGL1975,WalkerGLW1975,WalkerGW1977,SheWY2008,RikkenGT2012}.
Of them,  Walker, Ladhoz and Walker's experiment is the one which influenced most the opinion
of the researchers in favor of Abraham's point of view. In the next subsection we discuss the
experiment and show that its result is also completely consistent with our construction. 

As a further theoretical insight on the problem we recently show \cite{MedinaandS2017} with
a very simple computation that for an electromagnetic wave interacting with a conductor 
sheet in a dielectric medium if one supposes that Maxwell's equations and Ohm's law are valid, only
 Minkowski's momentum density is consistent with momentum conservation.

\subsection{Walker, Ladhoz and Walker experiment}

The experiment reported by Walker, Ladhoz and Walker
\cite{WalkerGL1975,WalkerGLW1975} was performed by suspending a non magnetic
disk of high dielectric constant as a torsional pendulum. The suspension was a 
tungsten fiber and the disk was located between the pole pieces of an
electromagnet. The disk had a central hole and the outer and inner surfaces
were coated with a layer of aluminum. The outer surface was 
connected to ground by means of a thin gold fiber and the inner one was connected
to the supporting fiber. A potential $V(t)=V_0\cos(\omega t+\phi)$ was
 applied between them. The movement of the
disk was observed by reflecting a laser beam
from a small mirror attached to the disk.

The electric  field in the disk is 
\begin{equation}
 \mathbf{E}(r)=\frac{V(t)}{r\ln(r_2/r_1)}\hat{\mathbf{e}}_r
\end{equation}
where $V(t)$ is the applied potential and $r_1$ and $r_2$ are the inner and
outer radii, respectively.
Abraham's force term  (\ref{Abraham_term}), would produce a torque in
 the vertical direction given by
\begin{eqnarray}
\label{TorqueAbraham}
\mathbf{t}_\mathrm{A}&=&\frac{\epsilon\mu-1}{4\pi c}\int\mathbf{r}\times
\frac{\partial}{\partial t}(\mathbf{E}\times\mathbf{H})\,dV\nonumber\\
&=&-\frac{\epsilon\mu-1}{4\pi c\mu\ln(r_2/r_1)}\frac{d V(t)}{d t}\int
\mathbf{B}\, dV\ .
\end{eqnarray}
Due to the symmetry of the system, the other contributions to Minkowski's and
Abraham's  forces do not produce any torque. Oscillations of the disk were
observed with the frequency of the potential adjusted to the resonance value
of the mechanical system. The authors of Ref.~\cite{WalkerGL1975,WalkerGLW1975}
report consistency of the outcome of the experiment with the results predicted
by (\ref{TorqueAbraham}) within the overall experimental error limit of about
$10\%$.

We can give an alternative explanation of this result, which makes no use of
Abraham's force term, and is fully consistent with our approach. As observed in
subsection \ref{EMT-of-matter} in a macroscopic description, the momentum of
matter acquires a term which depends on the field (see equation
 (\ref{matter-momentum-density})).
Correspondingly, the total angular momentum is given in our approach by
\begin{equation}
\mathbf{L}=\int\mathbf{r}\times\mathbf{g}_\mathrm{b}\, dV -\frac{1}{c}\int
\mathbf{r}\times(\mathbf{P}\times\mathbf{B})\, dV \ .
\end{equation}
Assuming that the only mechanism  of energy and
momentum  transport in the disk is the motion of mass, the matrix $P^{\mu\nu}$ that
appears in (\ref{EMT-matter}) is the stress which is purely spatial in the rest
frame, $P^{0\mu}=P^{\mu 0}=0$. In the non-relativistic limit the bare
momentum density is just
 $\mathbf{g}_\mathrm{b}=\tilde{u}_\mathrm{M}c^{-2}\mathbf{v}$ 
and the bare angular momentum $\mathbf{L}_\mathrm{b}$ is the usual classical
 angular momentum. 
The time derivative of the total orbital angular momentum $\mathbf{L}$ is
equal to the total torque $\mathbf{t}_\mathrm{t}$ applied to the disk. The
opposite of time derivative of the potential angular momentum acts as an
 additional torque on the bare angular momentum
\begin{equation}
\frac{d\mathbf{L}_\mathrm{b}}{dt}=\mathbf{t}_\mathrm{t} +
\int \mathbf{r}\times\partial_0(\mathbf{P}\times\mathbf{B})\, dV\ .
\end{equation}

For the conditions of the experiment
\begin{equation}
\label{TorquePotential}
\frac{d\mathbf{L}_\mathrm{b}}{dt}-\mathbf{t}_\mathrm{t}=
-\frac{\epsilon-1}{4\pi c\ln(r_2/r_1)}\frac{dV}{dt}\int \mathbf{B}\, dV\ .
\end{equation}

This is only slightly different from the expected effect of the Abraham term
in the general case and
exactly the same in the conditions of the experiment of Walker, Ladholz and
Walker in which the disk was made of a nonmagnetic material ($\mu=1$). 

It is not easy, using linear materials, to take advantage of the differences between
(\ref{TorquePotential}) and (\ref{TorqueAbraham}) to distinguish experimentally
between the two alternatives because, to the effect to be measurable
$\epsilon$ should be large. But we can  see from (\ref{Abraham_term}) that the
difference between the Abraham term and the effective force in our approach is
the presence in the former of a term that depends on the magnetization. Using a
ferromagnet (or a paramagnet close to the Curie point) with small electric
permittivity it could be possible to rule out Abraham's force. 

\subsection{Electrostriction}
Consider a material with an
isotropic dielectric constant $\epsilon$.  Our force density is
\begin{equation}
\label{M-S-force}
\mathbf{f} = P_k\nabla E_k =\frac{(\epsilon-1)}{4\pi}E_k\nabla E_k=
\frac{(\epsilon-1)}{8\pi}\nabla E^2\ ,
\end{equation}
Amp\`ere-Lorentz's is
\begin{equation}
\label{A-L-force}
\mathbf{f} = -(\nabla\cdot\mathbf{P})\,\mathbf{E}\ ,
\end{equation}
and Minkowski's is
\begin{equation}
\label{Min-force}
\mathbf{f} = -\frac{1}{8\pi}(\nabla\epsilon) E^2\ .
\end{equation}

In the textbook example of a capacitor of parallel plates partially filled
with a slab of dielectric material, all three forces give the same total force
on the slab. This force can also be obtained starting from the energy of the capacitor as a
function of the penetration of the slab. What is different in each case
is the interpretation. Our force (\ref{M-S-force}) is exerted on the
dipoles where the magnitude of the electric field is varying. This happens at the border of the
capacitor that is filled by the slab. Amp\`ere-Lorentz's (\ref{A-L-force})
force would appear due to
the interaction  between the free charges in the plates and the
polarization charges on the surface of the slab. For a homogeneous material
Minkowski's force (\ref{Min-force}) vanishes in the volume, but there is a
surface force on the slab because of the discontinuity of $\epsilon$; the total
force is determined by the force on the surface of the slab that is inside the
capacitor since the forces on the two surfaces parallel to  the plates cancel.  
Our force is parallel to the plates, while
in the other two cases there is also an inward tension perpendicular to the
plates.

It is interesting to consider  a variant of this example where the dielectric is
a liquid. The system is then made of two communicating vessels, where one is a
capacitor of parallel plates. In this case \cite{Brevik1979},  the surface of the 
liquid inside the capacitor rises above the level of the liquid outside.  
The level difference may be computed by an elementary method using the energy of
the capacitor and corresponds to a pressure difference
($\epsilon=1$ for the gas)
\begin{equation}
\label{pressure-diff}
\Delta p = \frac{\epsilon-1}{8\pi}E^2 \ .
\end{equation}
Brevik \cite{Brevik1979} cites the work of Hakim and Higham \cite{Hakim1962} as
supporting a different experimental result consistent with  a force density proposed by
Helmholtz \cite{Stratton1941}. The work in Ref. \cite{Hakim1962} 
uses a different geometry and a very 
intense electric field which may introduce nonlinear phenomena.

The rise of the liquid in this experiment  is enough to exclude  Amp\`ere-Lorentz's 
force since it cannot be achieved by applying a  force density  at
the lateral surfaces. Both (\ref{M-S-force}) and (\ref{Min-force})
agree with (\ref{pressure-diff}), the difference being that Minkowski's force
predicts that the pressure difference appears at the liquid-gas interface
inside the capacitor (the liquid is pulled from the top), while for our force
it appears in the submerged border of the capacitor (the liquid is pushed from
the bottom). Therefore by measuring the local pressure inside the liquid
it would be possible to discriminate between the two forces.

Another experiment which may be used to establish the correct force density is to consider 
a cylindrical electret with  polarization along its axis. Suppose the electret is 
immersed in a uniform electric field parallel to 
its polarization. For ours (\ref{M-S-force}), Minkowski's
(\ref{Min-force}) and Amp\`ere-Lorentz's (\ref{A-L-force}) force densities the total force
on the electret vanishes. Our force density itself vanishes and produces no stress on the electret. 
The bound charges for this system are located at the circular surfaces 
of the electret.  Amp\`ere-Lorentz's force predicts a  stretching in the direction of the axis.
Minkowski's force is applied at all the pieces of the material's surface  because there the 
dielectric constant varies discontinuously. Correspondingly, besides a longitudinal stretching it also
predicts a radial stretching. It is possible to discriminate between the alternatives if the electret 
is sufficiently elastic or if the internal stress is detected by optical means.

 \subsection{Magnetostriction}
The experiment discussed above has a magnetic analogous. Consider a cylindrical permanent
magnet magnetized along its axis in an intense magnetic field oriented in the same direction. 
Our force density vanishes again and predicts no additional stress on the magnet. 
Amp\`ere-Lorentz's (\ref{A-L-force}) force density acts on the magnetization current in the cylindrical 
surface and produces a radial stress. Minkowski's force density acts again at the whole boundary and
produces a longitudinal and a radial stress. It is possible to discriminate between the alternatives 
if the magnet  is sufficiently elastic or if the internal stress is detected by optical means.

Consider now magnetostriction in a ferromagnet in the
case where there is no magnetizing field, $\mathbf{H}=0$.
Then, as in any case with permanent magnets,
Minkowski's tensor is useless; it  predicts no effect since
 $T^{ij}_\mathrm{Min}=0$.  Instead Maxwell's tensor  for our formulation reduces to
\begin{equation}
T^{ij}_\mathrm{F}= -\frac{1}{2}\mathbf{M}\cdot\mathbf{B}\,\delta_{ij}\ .
\end{equation}

Outside the material both matter and field tensors vanish. Inside it
must be $T^{ij}_\mathrm{M}+T^{ij}_\mathrm{F}=0$. Therefore there is an
isotropic pressure inside the material $p=1/2\mathbf{M}\cdot\mathbf{B}$.
This pressure can  be calculated directly from the forces between magnetic
dipoles. For example, if the magnet is a long rod uniformly magnetized along
its axis, the stress along the axis corresponds to
the force of attraction between the two pieces of the rod divided by a plane
perpendicular to the axis.

 The stress along directions perpendicular  to the axis may also be calculated.
 The force density (\ref{force-density-dip})
vanishes in the bulk. It is felt only in the narrow superficial region in which
 the magnetization decreases from the bulk
value to zero.  This force density is perpendicular to the  external surface and
directed inward;  when it is integrated over the thickness of the surface
region the pressure is obtained.

The pressure contributes to the spontaneous volume magnetostriction of the
ferromagnet. It results in a contraction of the ferromagnet during the magnetization process.
Besides the purely magnetic effect there are also contributions due to the
Heisenberg interactions that align the spins.

The purely magnetic volume magnetostriction can be calculated dividing the
pressure by the bulk elastic modulus $K$
\begin{equation}
\label{V-magnetostriction}
\omega=\frac{\delta V}{V}= -\frac{1}{2K}\mathbf{M}\cdot\mathbf{B}\ .
\end{equation}

In the following table there is a comparison of the measured volume
magnetization $\omega_\mathrm{exp}$ of various ferromagnets
 \cite{Richter-Lotter1969,Turek-et-al2004}
with the values calculated with (\ref{V-magnetostriction}) using the saturation
magnetization.

\bigskip
\centerline{
\begin{tabular}{lcccc}\hline\hline
&\ $M$\ (MA/m)&\ \ $K$\ (GPa)&\ \ $\omega\times 10^6$&
\ \ \ $\omega_\mathrm{exp}\times 10^6$\\\hline
Fe&1.75&170&-11.3&\ \ 400\\
Co&1.43&180&\ -7.1&\ -250\\
Ni&0.51&180&\ -0.9&\ -270\\
Gd&2.14&\ 38&-75.8&-5000\\
\hline\hline\\
\end{tabular}
}

\bigskip
The purely magnetic effect is usually not taken into account in
the theoretical papers on the topic \cite{James-Kinderlehrer-1993, Wu-2002};
although it is only a few percent of the total volume change it should not
be completely neglected.

\section{Electromagnetic waves in matter}
\label{Wave}
\subsection{Wave-matter interaction}
A good test for the energy-momentum tensor and the force density presented in
this paper is to compute the momentum and energy exchange between a packet of
electromagnetic waves and a dielectric medium  \cite{MedandSb}. Here the solution 
of Maxwell's equations is an independent input which depends only on the properties of the
medium which is implicitly supposed to be homogeneous, isotropic and rigid.
Suppose that the region  $x>0$ is filled
by a non-dispersive medium with dielectric constant $\epsilon$
and magnetic permeability $\mu$. A packet of linearly polarized
plane waves approaches the $yz$ surface traveling in the $x$ direction.

As the wave enters into matter the electromagnetic field exerts a force on the
medium sharing its energy and momentum with it. The medium becomes polarized
and correspondingly changes its energy and momentum. The dipolar
energy in (\ref{matter-energy-density}) goes along with the electromagnetic
fields at the light velocity of the medium. For the idealized model of a
semi-infinite medium used in this example, the momentum exchange comprises two
different processes. First, the pulling (see below) that the field exerts on
the medium sets in movement a mechanical wave which travels at the sound
velocity, absorbs the impulse transferred to the matter and does not interact
with the field anymore. Second, as discussed in subsection \ref{EMT-of-matter},
part of the electromagnetic momentum (\ref{matter-momentum-density}) is
transferred to the medium. As we show below the medium response is to adjust
the internal stress components of $P^{\mu\nu}$ to compensate this gain of
momentum. The resulting perturbation which has zero total momentum also travels
with the transmitted wave packet at the light velocity.  

Let us turn to the solution of Maxwell's equations. The electric field is 
\begin{equation}
\mathbf{E}_1(x,y,z,t) = E_1 g(t-x/c)\theta(-x)\hat{\mathbf{y}} \ .
\end{equation}
where $E_1$ is the amplitude, $\theta$ is the Heaviside step function and $g(t)$ is
a dimension-less well-behaved function (otherwise arbitrary) that vanishes
for $t<0$ and $t>T$. At the surface of
the material ($x=0$) the packet is reflected and transmitted. The reflected and
transmitted packets  are
\begin{eqnarray}
\mathbf{E}_2(x,y,z,t) &=& E_2 g(t+x/c)\theta(-x)\hat{\mathbf{y}} \ ,\\
\mathbf{E}_3(x,y,z,t) &=& E_3 g(t-x/v)\theta(x)\hat{\mathbf{y}} \ ,
\end{eqnarray}
where the speed of light in the material is $v=c/n$ with $n=\sqrt{\epsilon\mu}$.
For $t<0$ only the incident packet is present, for $t>T$  the
reflected one is in the region  $x<0$ and the transmitted 
one is in the region $x>0$. For $0<t<T$ the
three packets are touching the surface $x=0$.
The corresponding magnetic fields of the three packets are
\begin{eqnarray}
\mathbf{B}_1 &=& B_1 g(t-x/c)\theta(-x)\hat{\mathbf{z}}\ ,\\
\mathbf{B}_2 &=& B_2 g(t+x/c)\theta(-x)\hat{\mathbf{z}}\ ,\\
\mathbf{B}_3 &=& B_3 g(t-x/v)\theta(x)\hat{\mathbf{z}}\ .
\end{eqnarray}
Using Maxwell's equations the magnetic amplitudes are 
\begin{equation}
B_1 = E_1\ ,\qquad B_2 = -E_2\ , \qquad B_3 = \sqrt{\epsilon\mu}E_3\ .
\end{equation}
By the continuity conditions at $x=0$
\begin{eqnarray}
E_2 = \frac{1-\sqrt{\epsilon/\mu}}{1+\sqrt{\epsilon/\mu}}E_1\ ,\ \  
E_3 = \frac{2}{1+\sqrt{\epsilon/\mu}}E_1 \ .
\end{eqnarray}
\subsection{Energy and momentum conservation}
For $t<0$ the energy  of a cylindrical piece of
the incident packet with axis parallel to $x$ and cross section  $A$ is,
\begin{eqnarray}
U_1 &=& \int T^{00}_{\mathrm{S}}(1)\,dV\nonumber\\
 &=& \frac{AE^2_1}{4\pi}\int_\infty^0 g(t-x/c)^2\,dx
=\frac{Ac\bar{T}}{4\pi}E_1^2
\end{eqnarray}
with
\begin{equation}
\bar{T} = \int_0^T g(t)^2\,dt \ .
\end{equation}
The momentum of the incident wave-packet is
\begin{equation}
\mathbf{p}_1=\int\mathbf{g}(1)\,dV=
\int c^{-1} T^{i0}_\mathrm{S}(1)\hat{\mathbf{e}}_i\,dV  =
 \frac{U_1}{c}\hat{\mathbf{x}} \ .
\end{equation}
For the reflected packet  ($t>T$) the energy and momentum are
\begin{equation}
U_2 = \frac{Ac\bar{T}}{4\pi}E^2_2\ ,\quad  
\mathbf{p}_2 = \int \mathbf{g}(2)\,dV = -\frac{U_2}{c}\hat{\mathbf{x}}\ .
\end{equation}
The energy and momentum  transferred to the $x>0$ side of the
space are
\begin{eqnarray}
U_1-U_2 &=& \frac{Ac\bar{T}}{4\pi}(E^2_1-E^2_2)=
\frac{Ac\bar{T}}{4\pi}E^2_3\sqrt{\epsilon/\mu}\\
\mathbf{p}_1-\mathbf{p}_2&=&\frac{A\bar{T}}{4\pi}(E^2_1+E^2_2)\hat{\mathbf{x}}
=\frac{A\bar{T}}{8\pi}E^2_3(1+\epsilon/\mu)\hat{\mathbf{x}}\;.\ \ \ \;
\end{eqnarray}
The EM energy and momentum of the transmitted packet are
\begin{eqnarray}
U_3 &=& \int T^{00}_{\mathrm{F}}(3)\,dV =\frac{Ac\bar{T}}{8\pi\sqrt{\epsilon\mu}}
E^2_3(\epsilon\mu+2\epsilon-1) \ ,\ \\
\mathbf{p}_3 &=& \int \mathbf{g}(3)\,dV=
 \frac{A\bar{T}v}{4\pi c}E^2_3\epsilon\sqrt{\epsilon
\mu}\hat{\mathbf{x}} \ .
\end{eqnarray}
Using (\ref{force-density}), the power on the matter at time $t$ is
\begin{eqnarray}
\dot{W}&=&c\int f^0 dV=-\int(\mathbf{P}\cdot\dot{\mathbf{E}}+\mathbf{M}\cdot\dot{\mathbf{B}})dV\nonumber\\
&=& -\frac{AE^2_3}{8\pi}[\epsilon -1 + (\mu-1)\epsilon]
\int_0^\infty \frac{\partial g(t-x/v)^2}{\partial t}dx\nonumber\\
&=& - \frac{Ac}{8\pi\sqrt{\epsilon\mu}}E^2_3(\epsilon\mu -1)g(t)^2\ .
\end{eqnarray}
Integrating in time the work done on matter is
\begin{equation}
W = -\frac{Ac\bar{T}}{8\pi\sqrt{\epsilon\mu}}E^2_3(\epsilon\mu-1) \ .
\end{equation}
This work changes the energy of the matter where the wave-packet is located, 
and includes the integral of the term $-\mathbf{E}\cdot\mathbf{P}$ of electrostatic dipolar energy.
It has to be added to the EM energy to obtain the total transmitted energy. One gets,  
\begin{equation}
 U^\prime_3 = U_3+W =U_1-U_2\ . 
\end{equation}
Energy is conserved.  $U^\prime_3$ is equal to
the integral of Poynting's
energy density (\ref{Poynting-energy}) of the transmitted packet. 
This reinforces our interpretation of  Minkowski's
tensor as the one which describes the whole wave in matter. Note also that
 $\mathbf{p}_3=c^{-1}U^\prime_3 n\hat{\mathbf{x}}$,
as would be expected for a non-dispersive wave, not for a particle-like excitation.

To verify momentum conservation one computes
the impulse on matter. The force on matter has a volume component
given by (\ref{force-density}) and a surface component due
to the discontinuity at $x=0$. The volume component is 
\begin{eqnarray}
\label{force-V}
\mathbf{F}_{\mathrm V} &=&\int (P_i\nabla E_i+M_i\nabla B_i)dV\nonumber\\
&=& \frac{A}{8\pi}\int_0^\infty[(\epsilon-1)\partial_x E^2+(1-1/\mu)
\partial_x B^2]dx \,\hat{\mathbf{x}}\nonumber\\
&=&-\frac{AE^2_3}{8\pi}(\epsilon\mu-1)g(t)^2\hat{\mathbf{x}} \ .
\end{eqnarray}
The surface component of the force at $x=0$ is equal to the momentum flux
exiting the vacuum side minus the momentum flux entering the matter side.
That is
\begin{equation}
\mathbf{F}_{\mathrm S}=
A(T^{11}_{\mathrm{S}}(-)-T^{11}_{\mathrm{F}}(+))\hat{\mathbf{x}} \ .
\end{equation}
Using (\ref{TEI-S}) and (\ref{energy-momentum-tensor}),
\begin{eqnarray}
T^{11}_{\mathrm{S}}(-)-T^{11}_{\mathrm{F}}(+)&=&\nonumber\\
\frac{1}{8\pi}[B^2(-)&-&B^2(+)+2(1-1/\mu)B^2(+)]\nonumber\\
&=&\frac{\epsilon E^2_3}{8\pi}(1/\mu+\mu-2)g(t)^2 \ .
\end{eqnarray}
Therefore the total force is
\begin{equation}
\label{force-VS}
\mathbf{F}=\mathbf{F}_{\mathrm V}+\mathbf{F}_{\mathrm S}=
\frac{AE^2_3}{8\pi}(1+\epsilon/\mu- 2\epsilon)g(t)^2\hat{\mathbf{x}} \ .
\end{equation}
We note that if diamagnetism does not prevail the wave packet pulls the dielectric. The impulse is
\begin{equation}
\label{Impulse}
\mathbf{I} = \int \mathbf{F}\,dt=
\frac{A\bar{T}E^2_3}{8\pi}(1+\epsilon/\mu-2\epsilon)
\hat{\mathbf{x}} \ .
\end{equation}
The total momentum transferred to the region $x>0$ for $t>T$ is
\begin{equation}
\mathbf{I}+\mathbf{p}_3=\frac{A\bar{T}E^2_3}{8\pi}(1+\epsilon/\mu)\hat{\mathbf{x}}
=\mathbf{p}_1-\mathbf{p}_2
\end{equation}
in a consistent way with momentum conservation.

\subsection{The center of mass and spin}
We now treat the motion of the center of mass.Consider first
the transmitted electromagnetic wave. For $t>T$, the position of the center
of mass of the field wave and of the whole wave including the polarization
energy of matter is the same. This is expressed by the following equation:
\begin{eqnarray}
\label{CenterOfMass}
\mathbf{X}_\mathrm{F3}(t)&=&
\frac{1}{U_3}\int xu\,dV\,\hat{\mathbf{x}}\\
&=&\frac{1}{v\bar{T}}\int x g(t- x/v)^2 \,dx\,\hat{\mathbf{x}}\nonumber\\
&=&\frac{1}{U^\prime_3}\int xu_\mathrm{P}\,dV\,\hat{\mathbf{x}}
=\mathbf{X}_\mathrm{W3}\nonumber\ ,
\end{eqnarray}
where $\mathbf{X}_\mathrm{F3}$ is the center of mass of the field and
$\mathbf{X}_\mathrm{W3}$ that of the whole wave.
It immediately follows that
 $\mathbf{X}_\mathrm{F3}(t)=\mathbf{X}_\mathrm{F3}(T) + (t-T)v\hat{\mathbf{x}}$, and that
$\dot{\mathbf{X}}_\mathrm{F3}=v\hat{\mathbf{x}}=\dot{\mathbf{X}}_\mathrm{W3}$.
This velocity is indeed constant. It can be  written also as
\begin{equation}
\label{CMvelocity}
\dot{\mathbf{X}}_\mathrm{W3}=\frac{1}{U^\prime_3}\int \mathbf{S}\,dV=
\frac{1}{U^\prime_3}\int T^{0i}_\mathrm{F}(3)\hat{\mathbf{e}}_i\,dV \ .
\end{equation}
For the strong CMMT to hold for this object it would be necessary to take
Abraham's value $c^{-2}U^\prime_3\dot{\mathbf{X}}$ as its momentum. But then
overall momentum conservation would be lost.

Let us turn to the movement of the center of mass of the whole system. Suppose
that the medium is initially at rest, has a total mass $M$ and is
free to move. To be consistent with the assumption of a rigid medium at rest
used in solving Maxwell's equations, let us  consider a situation in which
$u_\mathrm{M}\gg u_\mathrm{P}$ and the sound speed is much smaller than light
speed, so that the energy of the mechanical wave that appears as the
wave-packet enters the medium would be negligible in comparison with $U_1$.

Initially the total energy of
the system is $U=U_1+Mc^2$. When the wave is moving towards the dielectric there is
no spin contribution to the center of mass and spin and we may write 
\begin{equation}
\label{CMinitial}
 \dot{\mathbf{X}}_\Theta= \dot{\mathbf{X}}_\mathrm{T}=
 \frac{U_1}{U} \dot{\mathbf{X}}_\mathrm{W1}=\frac{c^2p_1}{U}\hat{\mathbf{x}}\ ,
\qquad  t<0\ .
\end{equation}

After the wave has penetrated the dielectric, it is convenient to separate the
field and matter contributions to the center of mass. For $t>T$ one has,
\begin{eqnarray}
U\mathbf{X}_\mathrm{T}(t)=
U_2\mathbf{X}_\mathrm{F2}(t)&+&U_3\mathbf{X}_\mathrm{F3}(t)\nonumber\\
&+&(W+Mc^2)\mathbf{X}_\mathrm{M}(t)\ ,
\end{eqnarray}
$(W+Mc^2)$ being the total energy of matter and $\mathbf{X}_\mathrm{M}$ the center of mass of matter.
The polarization energy travels with speed $v$ along with the electromagnetic wave. Then
\begin{equation}
 \mathbf{X}_\mathrm{M}(t)=\frac{W\mathbf{X}_\mathrm{F3}(t)
+Mc^2\mathbf{X}_\mathrm{b}(t)}{W+Mc^2}
\end{equation}
with $\mathbf{X}_\mathrm{b}(t)$ the center of mass of the bare matter (the
medium perturbed by the mechanical wave). The energy-momentum of matter has
two components, one which is symmetric, that is related to the mechanical
wave, and the other related to the polarization energy. For the mechanical
wave the CMMT is valid, so that
$M\dot{\mathbf{X}}_\mathrm{b}=\mathbf{I}$, where
 $\mathbf{I}$ is the impulse computed in (\ref{Impulse}). Assembling the parts
we have,
\begin{eqnarray}
 \dot{X}^1_\mathrm{T}&=&\frac{-U_2c+(U_3+W)v+Ic^2}{U}\nonumber\\
&=&\frac{Ac^2\bar{T}E^2_3}{16\pi U}[1+\epsilon/\mu+2\sqrt{\epsilon/\mu}+4/\mu-
4\epsilon]\ \ \;
\end{eqnarray}
which is not equal to the right hand side of (\ref{CMinitial}). 
The CMMT does not hold in this case.

Let us  consider now the spin contribution. The total
spin density of system which has contributions from
matter and field should be used. The identification of these separate contributions in the
general case is a difficult  problem which we will not discuss
here. Since the wave is linearly polarized, we do not expect any 
field spin contribution in this case, but this supposition is unnecessary to reach our
conclusion in what follows. The matter-field system is isolated,
suppose for a while that the energy-momentum tensor of matter
is symmetric. The symmetry question arises only for the contribution
related to the transmitted packet.
 Using (\ref{dip-torque}) we have
 \begin{equation}
\label{torque-spin}
\partial_\alpha S^{\mu\nu\alpha} = \tau^{\mu\nu}_{\mathrm{d}} \ .
\end{equation}
with $\tau^{\mu\nu}_{\mathrm{d}}$ given by (\ref{torque-space}) and (\ref{torque-time}).
Focusing in the temporal components, which are the ones that contribute  
to the center of mass and spin definition (\ref{spincenter}),
we have,
\begin{eqnarray}
\label{spintorque}
\partial_\alpha S^{0k\alpha}&=&(-\mathbf{P}\times\mathbf{B}
+\mathbf{M}\times\mathbf{E})_k \nonumber\\
&=&-\frac{\mu\epsilon-1}{4\pi\mu}(\mathbf{E}\times\mathbf{B})_k
\end{eqnarray}
where we have used the constitutive equations
\begin{equation}
\mathbf{P}=\frac{\epsilon-1}{4\pi}\mathbf{E}\ ,\qquad  \mathbf{M}=\frac{\mu-1}{4\pi\mu}\mathbf{B}\ .
\end{equation}
Spin transport in this system is due to the drift, $S^{0ki}=S^{0k0}v^i_m$ with $v^i_m$ the
matter velocity. In this case it vanishes because the mechanical wave is much slower.
Then $\partial_iS^{0ki}=0$ and using that the right hand side of
(\ref{spintorque}) points in the $x$ direction we have
\begin{equation}
 \partial_0S^{010}=-\frac{(\mu\epsilon-1)\sqrt{\epsilon\mu}}{4\pi\mu}g^2(t-x/v)E^2_3
\end{equation}
Integrating in space the spin term which appears in equation (\ref{CMSMT})
for $t>T$ is 
\begin{equation}
\frac{\partial}{\partial t}S^{01}=
-\frac{AcE^2_3\bar{T}(\mu\epsilon-1)}{4\pi\mu}=
-\frac{U}{c}\dot{X}^1_\mathrm{S}\ .
\end{equation}
Taking all together we verify  that for $t>T$,
\begin{equation}
 \dot{X}^1_\Theta=\dot{X}^1_\mathrm{T}+ \dot{X}^1_\mathrm{S}
=\frac{Ac^2\bar{T}E_1^2}{4\pi U}=\frac{c^2p_1}{U}
\end{equation}
as requested by the improved  theorem (\ref{CMSMT}). This justifies our
supposition that the matter energy-momentum tensor is symmetric.
We treat this point with more detail in the following paragraphs. 

The electromagnetic field modifies the matter when it penetrates. 
According to (\ref{matter-energy-density}) and 
(\ref{matter-momentum-density}) part of the energy and momentum of the 
microscopic field are kidnapped by the matter during the polarization process.
This  is accompanied by a mechanical response which tunes the internal stress to
the value necessary to guarantee the right properties of the medium and
relativistic invariance. To  explore this view consider the transport of the
energy of matter in the packet. The energy density of the matter, which is still at
rest when the wave is passing, is
\begin{equation}
\label{en-den-lin-mat}
u_\mathrm{M}=u_\mathrm{b}(0,0)+\frac{2\pi}{\epsilon-1}P^2
-\frac{2\pi\mu}{\mu-1}M^2 -\mathbf{P}\cdot\mathbf{E}
\end{equation}
whereas the equation of transport of energy in the matter is
\begin{equation}
\partial_0 u_\mathrm{M}+\partial_k P^{0k}= -\mathbf{P}\cdot\partial_0\mathbf{E}
   -\mathbf{M}\cdot\partial_0\mathbf{B}\ . 
\end{equation}
By taking the time derivative in (\ref{en-den-lin-mat}), it is obtained
that $\partial_k P^{0k}=0$ and, given the symmetry of the problem,
it follows that $P^{0k}=0$ everywhere since $P^{\mu\nu}$ vanishes outside the packet.  
That is, Poynting's vector is the whole energy current density
of the wave-packet. 
If $T^{\mu\nu}_\mathrm{M}$ is symmetric then $P^{k0}$ compensates the
potential momentum density appearing in (\ref{matter-momentum-density})
and vice versa,
\begin{equation}
0=T^{0k}_\mathrm{M}=T^{k0}_\mathrm{M}=P^{k0}-(\mathbf{P}\times\mathbf{B})_k\ .
\end{equation}
The spatial stress $P^{jk}$ is symmetric because the spatial torque on matter
vanishes.

On the other hand, M{\o}ller-von Laue criterion allows to recognize
$u_\mathrm{P}$ and $\mathbf{g}_{\mathrm{Min}}$ as the total energy and momentum
traveling with the wave. Since $\mathbf{g}_{\mathrm{Min}}$ is the momentum of the electromagnetic field
no additional momentum attached to the matter is following the
electromagnetic wave. The energy-momentum tensor of matter is symmetric 
and the conditions for the validity of the improved theorem (\ref{CMSMT}) are met.
In our picture the electromagnetic field exchanges energy with matter and
exerts locally a force on it, although the total force vanishes once the
wave-packet is completely into the matter.

\section{The magnet-charge system}

\label{Magnet-charge}
\begin{figure}
\centering
\includegraphics[scale=0.8]{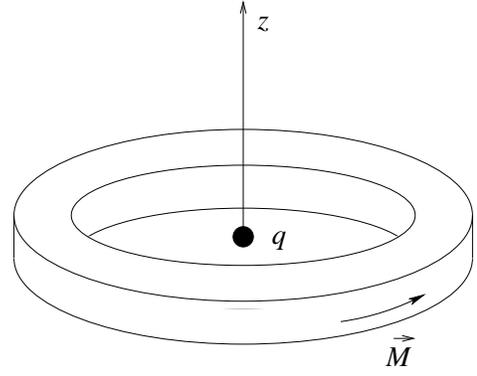}
\caption{Magnetized ring}
\label{washer}
\end{figure}

\subsection{Violation of the CMMT}
As a second illustration of our discussion  \cite{MedandSa}, consider a ring with the
shape of a washer made of a ferromagnetic insulator
(see Fig.~\ref{washer}). The axis of the ring is along the $z$-axis and the
center is at the origin. For simplicity we take $\epsilon=1$. The ring is magnetized  around its axis with  magnetization 
$\mathbf{M}$ of uniform magnitude.
In this situation $\mathbf{B}=4\pi\mathbf{M}$ inside the ring whereas
 $\mathbf{H}=0$ everywhere.
Suppose that at the center of the ring there is a particle of  charge $q$. Since the ring is an insulator the electrostatic 
field of the charge penetrates the magnet.  The total momentum defined by $T_{\mathrm{F}}^{\mu\nu}$ does not vanish but 
points in the $z$ direction.  Using cylindrical
coordinates $(\rho, \theta, z)$, $\mathbf{M}=M\hat{\mathbf{{\theta}}}$ and
 $\mathbf{r}=\rho\hat{\bm{{\rho}}}+ z\hat{\mathbf{z}}$ we have
\begin{equation}
\mathbf{g}=\frac{1}{4\pi c}\mathbf{D}\times\mathbf{B}=
\frac{q M}{c r^3}(-z\hat{\bm{\rho}}+\rho \hat{\mathbf{z}})\ .
\end{equation}
and integrating over the volume
\begin{equation}
\label{momentum}
\mathbf{p}= \frac{q M}{c}\int dV\, \frac{\rho}{r^3}\,\hat{\mathbf{z}} \ .
\end{equation}

The center of mass of the system (including the electromagnetic contribution to the energy) is at rest at the center of 
the ring. Then  the usual CMMT is violated. Note that Abraham's momentum density vanishes since $\mathbf{H}=0$, so that if 
one takes it to define the momentum of the electromagnetic field,  the total momentum of the system including the 
electromagnetic field is zero. 

Let us now examine what happens when the fields  change by considering the scenario in
which the  washer demagnetizes ({\it e.~g.~}  by the action of a device included in the magnet
which allows  its temperature to rise above the Curie point). The magnetic
induction field $\mathbf{B}$ changes and therefore an induced electric field
appears. It is always possible to split the electric field in a static field,
$\mathbf{E}_S$, $\nabla\cdot\mathbf{E}_S=4\pi\rho$, $\nabla\times\mathbf{E}_S=0$
and an induced field $\mathbf{E}_{\mathrm{I}}$, $\nabla\cdot\mathbf{E}_{\mathrm{I}}=0$,
$\nabla\times\mathbf{E}_{\mathrm{I}}=-c^{-1}\dot{\mathbf{B}}$.  It is easily  shown that
\begin{equation}
\mathbf{E}_{\mathrm{I}} = -\frac{1}{4\pi c}\int dV\,\frac{\partial\mathbf{B}}{\partial t}
\times \frac{\hat{\mathbf{r}}}{r^2}\ .
\end{equation}
Neglecting the radiation field $\mathbf{B}=4\pi\mathbf{M}$. Then at the origin
the induced field is
\begin{equation}
\mathbf{E}_{\mathrm{I}}(0)=-\frac{\dot{M}}{c}\int dV\,\frac{\rho}{r^3}\,\hat{\mathbf{z}}\ .
\end{equation}
If the demagnetization process is fast enough, the force on the charge
is $\mathbf{F}=q\mathbf{E}_{\mathrm{I}}(0)$, and the impulse  on the (dressed) particle may also be computed
\begin{equation}
\label{charge-impulse}
\mathbf{I}_{\mathrm{Charge}}=\int dt\,\mathbf{F}=-\frac{q}{c}\int dt\,\dot{M}\,
\int dV\,\frac{\rho}{r^3}\,\hat{\mathbf{z}}\ .
\end{equation}
Note that  no $\mathbf{B}$ field is produced during the demagnetization process by
the charge and there is no force acting on the magnet which therefore remains at rest.
The mechanical momentum of the system is now different from zero. If one uses Abraham's
momentum density, the unavoidable conclusion is that momentum is not conserved. But on the
other hand, (\ref{charge-impulse}) is exactly 
the momentum of the initial configuration calculated (\ref{momentum}) using Minkowski's
momentum density, which  is then shown to be consistent with momentum conservation. 
 Since the center of
mass of the magnet does not move, the center of mass of the system 
moves in the direction in which the particle moves with velocity
\begin{equation}
\label{MagnetCM}
  \dot{X}_\mathrm{T}^i=\frac{c^2}{U}p^i
\end{equation}
where $U$ is the total rest energy of the system. The CMMT is again violated.

\subsection{The motion of the center of mass and spin}

The result obtained is nevertheless consistent with our center of mass and spin motion theorem (\ref{CMSMT}). To see this we note 
that in the initial situation the total  spin density satisfies again (\ref{torque-spin}).  For the temporal components we have,
\begin{equation}
 \partial_\alpha S^{0i\alpha}=\tau^{0i}=(\mathbf{M}\times\mathbf{E})^i  \ ,
\end{equation}
and using again that matter is at rest we are left with
\begin{equation}
 S^{0i0}=c(\mathbf{M}\times\mathbf{E})^it +C_1
\end{equation}
with $C_1$ a constant. Integrating over space and using $\mathbf{B}=4\pi\mathbf{M}$ we obtain
\begin{equation}
 \dot{X}_\theta^i=\dot{X}_\mathrm{S}^i=-\frac{d}{dt}\frac{c}{U}S^{0i}=\frac{c^2}{U}p^i \ ,
\end{equation}
in agreement with (\ref{CMSMT}) and (\ref{MagnetCM}). There is a
short period when the magnetization varies for which both
$\dot{\mathbf{X}}_\mathrm{T}$ and $\dot{\mathbf{X}}_\mathrm{S}$ are
variable.

\subsection{Storing momentum in the magnet}
To complete the picture, let us investigate  the origin of the momentum that in ours and Minkowski's 
formulation is stored in the electromagnetic field. Let us start with  the charge  far 
away and analyze how, as it is brought to the center of the ring, the field momentum builds up. 
The magnet does not produce an electromagnetic
field outside itself. Hence there is no force acting upon the charge. There is, though, a
force acting on the magnet produced by the magnetic field $\mathbf{B}$ generated as
the charge moves. To keep the magnet in place an opposite force must be
applied to it. The impulse generated by this force is precisely the stored momentum in 
the electromagnetic field. For a charge moving along the $z$-axis the calculation can be done easily.

Let us call $\mathbf{x}$ the position of the charge, $\mathbf{x}=\zeta\hat{\mathbf{z}}$. The velocity
is $\mathbf{v}=\dot{\zeta}\hat{\mathbf{z}}$.  For $v\ll c$ the magnetic field produced by
the moving charge is
\begin{equation}
\mathbf{B}=\frac{q}{c}\frac{\mathbf{v}\times\mathbf{r}^\prime}{{r^\prime}^3}=
\frac{q\dot{\zeta}\rho\hat{\bm{\theta}}}{c[\rho^2+(z-\zeta)^2]^{3/2}} \ ,
\end{equation}
where ${\mathbf{r}}^\prime=\mathbf{r}-\mathbf{x}=\rho\hat{\bm{\rho}}+(z-\zeta)\hat{\mathbf{z}}$.
Using (\ref{force-density}), the force density on the magnet is
\begin{equation}
\mathbf{f}=M_k\nabla B_k=\frac{q}{c}\dot{\zeta}M\left[\hat{\bm{\rho}}
\frac{\partial}{\partial\rho}+\hat{\mathbf{z}}\frac{\partial}{\partial z}\right]
\frac{\rho}{{r^\prime}^3}\ .
\end{equation}
The force is obtained by integrating over the volume
\begin{eqnarray}
\mathbf{F} = \int dV\,\mathbf{f}&=& \frac{q}{c}\dot{\zeta}M\int dV\,\frac{\partial
}{\partial z}\frac{\rho}{{r^\prime}^3}\,\hat{\mathbf{z}}\\
&=& -\frac{q}{c}M\frac{d}{dt}\int dV\,\frac{\rho}{{r^\prime}^3}\,\hat{\mathbf{z}}\ .
\end{eqnarray}
We have used $\partial_z =-\partial_\zeta$. The impulse on the magnet is then
\begin{equation}
\mathbf{I}_{\mathrm{Magnet}}=\int dt\,\mathbf{F}=
-\frac{q}{c}M\int dV\,\frac{\rho}{r^3}\,\hat{\mathbf{z}} \ .
\end{equation}
The momentum stored by  the field is the opposite of this quantity and again
agrees with (\ref{momentum}). Also in this situation CMMT does not hold. If no
external force is applied to keep the magnet in place, the charge would move
with constant velocity but the magnet would be accelerated in the negative
direction of the $z$ axis, as the center of mass would be too.

\subsection{Toroidal currents}
\begin{figure}
\centering
\includegraphics[scale=0.8]{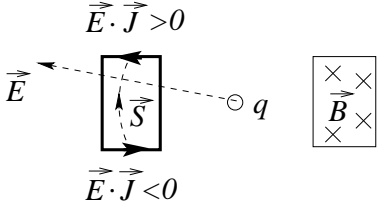}
\caption{Toroidal currents}
\label{currents}
\end{figure}

Taylor \cite{Taylor1965} has treated a related example, with toroidal currents
instead of a magnet. It is interesting to have a view on the differences with
our discussion in the starting situation with the charge at rest in the center. 
In this  case the proper energy-momentum tensor is known to be the standard
tensor defined in the vacuum, which is symmetric, $\mathbf{S}=c^2\mathbf{g}=
c\mathbf{E}\times\mathbf{B}/4\pi$, and therefore CMMT holds. If one considers
a free current distribution of value $\mathbf{j}=c\nabla\times\mathbf{M}$, with
$\mathbf{M}$ the magnetization of our previous example, the magnetic induction
field and the induced electric field are  the same as in that case and
the electromagnetic momentum density is equal to the one obtained using
Minkowski's expression in that situation. But now unlike in the magnet
configuration the Poynting vector does not vanish. The force exerted on the
currents by $\mathbf{E}$ is still zero, but the power density
$\mathbf{E}\cdot\mathbf{j}$ is not, see Fig.~\ref{currents}. In the equivalent
situation to the one discussed above, inside the structure that supports the currents, the
energy is increasing in the top,  and decreasing in the base. The total power
does vanish. Since the force is zero, the velocity of the supporting structure
maintains its null value; nevertheless, its center of mass moves because 
the internal energy is transported by Poynting's vector from the bottom to the
top. The CMMT imposes that the velocity of the center of mass should equal the stored momentum
divided by the total energy.  This makes this configuration ephemeral. Even
disregarding Joule's effect it could last only until the energy storage in the
base is depleted. Regretfully, the analysis presented in Taylor's paper is not
correct because it does not take into account adequately the electromagnetic
momentum. The energy transport mechanism just described is in contrast to what
occurs with the magnet where the power density and the Poynting vector are zero.
For this reason, the center of mass of the magnet may remain at rest.
  
\section{Angular momentum of a magnetized cylinder}
\label{Cylinder}
\begin{figure}
\centering
\includegraphics{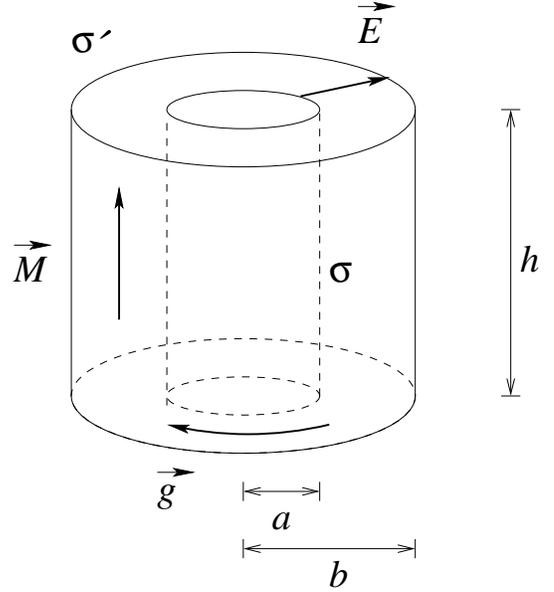}
\caption{Magnetized hollow cylinder}
\label{MagCyl}
\end{figure}

In our last illustration we consider the implications of the momentum density 
expression to orbital angular momentum equilibrium. Let us consider an 
infinite hollow cylinder made of
a ferromagnetic insulating material with $\epsilon = 1$ (see
 figure \ref{MagCyl}).
The radius of the concentric  hole is $a$ and the
external radius is $b$. The cylinder is magnetized along its axis with
a uniform magnetization $\mathbf{M}$.
In the surface of the hole there is a uniform surface charge density
$\sigma$ and in the external surface of the cylinder there is a
cylindrical shell of insulator
charged with a surface density $\sigma^{\prime} = -\sigma a/b$. In this
condition inside the wall of the cylinder there is a uniform magnetic
field $\mathbf{B}=4\pi\mathbf{M}$ and a radial electrostatic field 
\begin{equation}
\mathbf{E}=4\pi\sigma\frac{a}{r}\hat{\mathbf{r}}\quad \hbox{for}\quad a<r<b
\end{equation}
and $\mathbf{E}=0$ otherwise. The momentum density is
\begin{equation}
\mathbf{g}=\frac{1}{4\pi c}\mathbf{D}\times\mathbf{B}=
-\frac{4\pi M\sigma a}{cr}\hat{\bm{\theta}}\ ,\qquad a<r<b \ .
\end{equation}
Since $\mathbf{g}$ circulates around the axis there is a net orbital angular momentum
$\mathbf{L}$  stored in the electromagnetic field. For a  section
of cylinder of  height $h$ the angular momentum is
\begin{equation}
\mathbf{L}=\int \mathbf{r}\times\mathbf{g}\,dV=
-\frac{(2\pi)^2}{c}M\sigma a (b^2-a^2)h\hat{\mathbf{z}}\ .
\end{equation}
Note that the EM fields store angular momentum  although macroscopically there
is nothing rotating.

If the cylinder demagnetizes an induced electric field is generated. The
field produces a torque on the charged shell and  the angular momentum is
transferred to the shell\footnote{In this example there is also a spin
proportional to the magnetization which will also be transferred to the angular
momentum of the magnet (Einstein-de Haas effect). The coupling is due to the
asymmetry of the stress tensor of the magnet.}.
 For $a<r\le b$ the induced electric field  is
\begin{equation}
\mathbf{E}_{\mathrm in}=-\frac{2\pi}{cr}(r^2-a^2)\dot{M}\hat{\bm{\theta}} 
\end{equation}
and the torque on the shell is
\begin{equation}
\bm{\tau} = \int \mathbf{r}\times\mathbf{E}_{\mathrm in}\sigma^\prime dS
= \frac{(2\pi)^2}{c} \sigma a (b^2-a^2)h\dot{M}\hat{\mathbf{z}} \ .
\end{equation}
The angular impulse is indeed equal to the angular momentum
of the electromagnetic fields,
\begin{equation}
\int \bm{\tau}\,dt=\mathbf{L} \ .
\end{equation}

In this situation only Minkowski's momentum density is consistent with 
angular momentum conservation. Abraham's momentum density  yields a 
null angular momentum in the initial situation since $\mathbf{H}=0$.

\section{Conclusion}
\label{Conclusion}

In this paper we attain a complete resolution of the Abraham-Minkowski
controversy on the energy-momentum tensor of the electromagnetic field in
matter by introducing a new energy-momentum tensor which is different from
those previously proposed.  Our approach emphasizes the local character of
the force and  the torque exerted by the field on matter and requires that the
energy-momentum tensor  be locally  well determined. It cannot be
arbitrarily modified by the addition of a divergence-less term, 
even if by doing so the total energy and momentum do not change. 

Our discussion departs from the main lines of argumentation 
prevailing in the recent literature on the topic. It is based on four main
points: First, the force density on matter is in principle a measurable
quantity and should be computable as the divergence of the matter tensor.
Momentum conservation determines in an unambiguous way the
energy-momentum tensor of the electromagnetic field. Second, 
all  terms which contribute to inertia, including those of
electromagnetic origin, should be included in the matter tensor
and consequently excluded from the electromagnetic one. Third, the force
density should be compatible with the microscopic Lorentz force.
Fourth,  magnetization is a manifestation of spin and  EM fields
produce torques on polarized matter, hence a complete
understanding of the mechanical behavior of polarizable matter 
in an electromagnetic field requires the dynamics of spin  too.

In this work  the necessity of a relativistic invariant description of
the matter-field interaction 
and the understanding of the true consequences of such a description play
a key role.
  Throughout the discussion we insisted on the idea that the 
polarizations tensor which includes the electric polarization and 
the magnetization in a relativistic covariant object, should be used to  
describe the polarization of matter. We avoid arguments that
rely on the properties of the electric and magnetic permeabilities which are defined 
only in the rest frame of matter (which not always
exist). As an element which has more general consequences
we discuss the serious limitations of the CMMT, which does not take into account 
the contribution of spin to the dynamics. The uncritical 
use of the CMMT in this context has been the source of many confusions and even
mistakes in previous papers on this subject.

The conditions imposed on the form of the force density on matter by the
relativity principle, the validity of Maxwell's 
microscopic and macroscopic equations and the consistency with Lorentz
microscopic force allow us to determine uniquely an 
expression of the force density on polarizable matter. This expression is
similar but different to some of the expressions discussed in the past. 
The total force that the electromagnetic field exerts on an isolated sample of matter 
computed with our expression is the same as the total Lorentz force
computed on the bound charges and currents of the sample. It is also equal to the
total force predicted by the use of Minkowski's tensor but it is different
to the one consistent with Abraham's tensor. The term known as 
Abraham's force is not present in our formulation. Our analysis of the force
density uncovers also the portion of the energy momentum tensor of matter which depends
on the field. This includes the electrostatic potential energy of the dipoles,
a momentum contribution required for the polarization of matter and, depending on
the external conditions on matter, a field dependent mechanical response of matter
which modifies  its stress tensor.

The energy-momentum tensor of the field is fixed by the form of
the force density and we are able to get its correct expression. The energy density of this 
tensor is not Poynting's classic formula. It is shown to be the energy
density of the electromagnetic field in vacuum minus 
the electrostatic energy of the dipoles. This term represents a
electromagnetic contribution to inertia. Poynting's vector maintains 
its interpretation as the energy density current once the correct expression of
the  power transmitted by the field to matter is taken into account. 
The momentum density is shown to be Minkowski's expression. It is shown to
be the momentum in vacuum minus the field dependent momentum of matter. This puts an
end to the long standing controversy. 

The energy-momentum tensor of the field is asymmetric. This has important
consequences. This asymmetry is shown to be necessary 
for the coupling of the total spin density  (magnetization) and the orbital
angular momentum of matter.  The asymmetry 
cancels exactly the effect of the torque density in the dynamical equation of
the field spin density with the general result that the 
spin of the field is decoupled from the polarization of matter.

We presented a second approach to deduce the form of the electromagnetic
energy-momentum tensor in matter. For this we perform a space-time 
average starting from the microscopic equations and postulate a monopolar and a
dipolar coupling of matter with  the macroscopic field. 
By imposing consistency with Maxwell's equations we identify the monopolar 
current with the free charges, and show that the dipolar coupling 
is described by the polarization tensor in such a way that we recover exactly
the formulation of the first approach. Although it may be foreseeable, 
this result has a deep meaning: Relativistic invariance and the physical
hypotheses discussed in section \ref{Force} uniquely determine the equations
of the macroscopic field up to the dipolar approximation. It will be
interesting to investigate if the quadrupolar approximation maintains 
such robustness.
 
In our presentation, we also note that although in a general situation
Minkowski's tensor is not particularly useful, for a material with
non-dispersive linear polarizabilities
($D_{\alpha\beta}=2^{-1}\chi_{\alpha\beta\mu\nu}F^{\mu\nu}$) it may be
interpreted as the energy-momentum tensor
of the electromagnetic field plus the fraction of the energy of matter arising from
the polarization. Working with care it can be used in this case.
Nevertheless its divergence is not the reaction of the force acting on
matter.  This explains the absence of magnetostriction and electrostriction in
Minkowski's approach. When an electromagnetic wave propagates on  such a
medium Minkowski's tensor describes the whole wave and hence satisfies
von Laue-M{\o}ller criterion. 

The arguments presented in this paper are of theoretical nature. Of
course, the ultimate test of the validity of our approach 
should be the direct measurement of the force density in different situations.
For some of the experiments
that have been done in the past, it is not completely straightforward to
conciliate the measurements obtained with a definite choice of the force 
density or momentum expressions. In consequence there have been confronting
opinions on how to interpret  correctly the
results of those experiments. Nevertheless there are some situations in which
in our opinion our analysis  proves to be clearly better.
The applications and examples presented in the text were devised with this in
mind.

In section \ref{Force} we discuss some of the implications of our approach for
the analysis of  the experiments which have been performed to determine the
electromagnetic force density in matter. In particular we show that the
result of the Walker, Ladhoz and Walker experiment, which is usually presented
as an evidence for the existence of Abraham's force term, may be better
understood as a manifestation of the field dependent terms of the matter
energy-momentum tensor. Our discussion of the electromagnetic forces on solid
and liquid dielectrics strongly supports the validity of the force density that
we are proposing.

We use our force density, the energy and momentum definitions obtained
from $T^{\mu\nu}_{\mathrm{F}}$ and Maxwell's equations to verify energy
and momentum conservation in the
interaction of a  packet of electromagnetic waves with a dielectric medium.
 In opposition to the statement of the influential  argument first
advanced by Balazs \cite{Balazs1953}, that for sixty years has been considered
a strong support in favor of Abraham's point of view,
we  show that for $n>1$  the wave packet effectively pulls the material
when it enters a medium (See Eq.(\ref{force-VS})).
This in fact is a most comfortable  result after observing that it has a 
simple physical explanation. Dielectric and paramagnetic materials are
attracted while diamagnetic materials are repelled in the direction to high
field regions. When the wave packet is entering the medium it pulls the
material unless diamagnetism prevails.  For the same reason when the wave
leaves, it drags the block forward. We complete our picture of this process by
computing the evolution of 
the spin density. This calculation shows that the boost momentum
gives the precise contribution necessary for our center of mass and
spin to behave inertially.

The example of the charge-magnet system presented in section \ref{Magnet-charge},
gives in a very simple setup an illustration of the consequences of the main
ideas behind our formulation. In the same spirit, for the magnetized cylinder
analyzed in section \ref{Cylinder} we show how our approach
gives an appealing and consistent description of a system where the orbital angular
momentum stored in the EM fields is transferred to matter.

As a by-product of  the  analysis presented in this paper the weakness of the
CMMT was exposed in specific physical situations, and the validity
of the improved theorem which includes the spin contributions was tested. These
results also show that the hidden momentum hypothesis
is unnecessary. Although we have not made a detailed check in each case (which
would take a disproportionate effort), we think that the many 
analysis found in the literature which are based in this hypothesis should  prove
wrong  or at least superfluous.

Our results open a range of theoretical and experimental possibilities of
research in the study of the interaction of the electromagnetic fields 
with polarizable matter. Our construction of the force density suggests to make
a methodological shift
in the approach to the many applications in which it plays a role  and concentrate 
in the determination of the physically acceptable  polarization tensor of the
system under study, preferably in the laboratory 
reference frame. This may be applied to  a variety of systems, which include
ionized gases, dielectric liquids, and non linear magnetic matter, to give a few examples.
In the past these were either ignored or studied in terms of the permeabilities. We
note that the polarization tensor of some of these systems may depend 
on the local velocity of matter, so that velocity dependent effects are not 
banned from our formulation. It is only  Abraham's like terms which are
found unnecessary. On the experimental side the determination of force 
density instead of the total force advocated by some authors and in particular
by Brevik \cite{Brevik1979} in the past acquires new importance.
A critical reanalysis an even a remaking of the experiments which perform a
direct measurement of the momentum of photon should be useful for establishing 
without further doubts Minkowski's momentum as the classical and quantum
momentum of light. Experiments exploring the inertiality of the center of mass 
and spin in this context which may include the real versions of the idealized
situations discussed in sections \ref{Wave}, \ref{Magnet-charge} 
and \ref{Cylinder} will also contribute to the complete experimental
clarification of this subject.

\appendix 
\section{Energy and momentum of dressed dipoles}
\label{dressed}
The time derivative terms that spoil the relativistic covariance in
(\ref{elec-force}) and (\ref{elec-power}) actually  correspond to part of
the momentum and energy of the microscopic  electromagnetic interaction. In 
addition to the external fields we  considered in section
\ref{Forces and torques on dipoles} there are also the
electromagnetic fields produced by the  dipolar system itself
\begin{eqnarray}
\mathbf{E}_\mathrm{t}&=&\mathbf{E}+\mathbf{E}_\mathrm{d}\ ,\\
\mathbf{B}_\mathrm{t}&=&\mathbf{B}+\mathbf{B}_\mathrm{d}\ .
\end{eqnarray}

The energy density of the total electromagnetic field is
\begin{equation}
u_\mathrm{t}=u+u_\mathrm{d}+\frac{1}{4\pi}(\mathbf{E}\cdot\mathbf{E}_\mathrm{d}
+\mathbf{B}\cdot\mathbf{B}_\mathrm{d})
\end{equation}
and the momentum density is
\begin{eqnarray}
\mathbf{g}_\mathrm{t}&=&\frac{1}{4\pi c}\mathbf{E}_\mathrm{t}\times
\mathbf{B}_\mathrm{t}\nonumber\\
&=& \mathbf{g}+ \mathbf{g}_\mathrm{d}+\frac{1}{4\pi c}(\mathbf{E}\times
\mathbf{B}_\mathrm{d}+\mathbf{E}_\mathrm{d}\times\mathbf{B})\ .
\end{eqnarray}

The integrals of the energy density $u_\mathrm{d}$ and of the momentum density
$\mathbf{g}_\mathrm{d}$ that only depend on the dipolar fields
$\mathbf{E}_\mathrm{d}$ and $\mathbf{B}_\mathrm{d}$ are
part of the energy and momentum of the dressed dipole.

The interaction momentum and energy are the integrals over the whole space
of products of an external field and a dipolar field. We  assume that the
the external fields are essentially
constant where the dipolar system is located and that the fields of the dipolar
system are well-approximated by the fields of dipolar moments where the external
charges and currents are located (dipolar approximation). The integrals can be split 
in  a local term corresponding to the region around the dipolar system and
an external term  outside this region,
\begin{equation}
\int_\mathrm{whole-space} {\dots}\,dV=\int_\mathrm{local} {\dots}\,dV +
\int_\mathrm{external} {\dots}\,dV\ .
\end{equation}

The local  electromagnetic interaction energy and interaction momentum
also contribute to  the dressed dipole energy and momentum.
The local interaction energy is
\begin{equation}
U_\mathrm{i}=\frac{1}{4\pi}\int_\mathrm{local}
(\mathbf{E}\cdot\mathbf{E}_\mathrm{d}+
 \mathbf{B}\cdot\mathbf{B}_\mathrm{d})\, dV
\end{equation}
and the local interaction momentum is
\begin{equation}
\mathbf{p}_\mathrm{i}=\frac{1}{4\pi c}\int_\mathrm{local}
(\mathbf{E}\times\mathbf{B}_\mathrm{d}
+\mathbf{E}_\mathrm{d}\times\mathbf{B})\,dV\ .
\end{equation}

To evaluate the contribution of the local interaction terms we use 
the fact that the external fields are essentially constant. Then
\begin{equation}
U_\mathrm{i}\approx\frac{1}{4\pi}\left(\mathbf{E}\cdot\int
\mathbf{E}_\mathrm{d}\,dV+
 \mathbf{B}\cdot\int\mathbf{B}_\mathrm{d}\, dV\right)\ ,
\end{equation}
and
\begin{equation}
\mathbf{p}_\mathrm{i}\approx\frac{1}{4\pi c}\left(
\mathbf{E}\times\int\mathbf{B}_\mathrm{d}\,dV
+\int\mathbf{E}_\mathrm{d}\,dV \times\mathbf{B}\right)\ .
\end{equation}

These approximations have a pitfall because the whole-space
integrals of the dipolar fields are non-convergent improper integrals.
Nevertheless, in the next appendix we argue that the physically consistent
values of the integrals are
\begin{equation}
\label{B-int}
\int \mathbf{B}_\mathrm{d}\,dV=0
\end{equation}
and 
\begin{equation}
\label{E-int}
\int \mathbf{E}_\mathrm{d}\,dV=-4\pi \mathbf{d}\ .
\end{equation}
Therefore
\begin{equation}
\label{int-energy}
U_\mathrm{i}= -\mathbf{d}\cdot\mathbf{E}
\end{equation}
and
\begin{equation}
\label{int-momentum}
\mathbf{p}_\mathrm{i}= -\frac{1}{c}\mathbf{d}\times\mathbf{B}\ .
\end{equation}

As we have shown in subsection \ref{Forces and torques on dipoles}, these contributions
to the interaction, (\ref{int-momentum}) and (\ref{int-energy}),
exactly cancel the additional terms in (\ref{elec-force}) and
(\ref{elec-power}). In other words (\ref{elec-force}) and (\ref{elec-power})
correspond to the force and power on the bare dipole while for the dressed
dipole the offending terms are not present.

\section{Integrals of self fields of dipolar systems}

In this section we prove that the whole-space volume integrals of the dipolar
fields that appear in appendix \ref{dressed}
are non-convergent improper integrals. We show that the physically
consistent values of these integrals are those given by (\ref{B-int}) and
(\ref{E-int}). We assume that the external fields are $\mathbf{E}=-\nabla\phi$
and $\mathbf{B}=\nabla\times\mathbf{A}$ with the potentials vanishing at
infinity as $1/r$. Near the dipolar system the fields are constant, then
$\phi\approx \phi(0)-\mathbf{E}(0)\cdot\mathbf{x}$ and
$\mathbf{A} \approx\mathbf{A}(0)+ \mathbf{B}(0)\times\mathbf{x}/2$.

The integral of the self magnetic field on a finite volume $V$ is 
\begin{eqnarray}
\int_V \mathbf{B}_\mathrm{d} \,dV &=&
 \int_V \mathbf{B}_\mathrm{d}\cdot\nabla\mathbf{x}\,dV\nonumber\\
&=&\int_V [\nabla(\cdot\mathbf{B}_\mathrm{d}\mathbf{x})-
\nabla\cdot\mathbf{B}_\mathrm{d}\,\mathbf{x}]\,dV\nonumber\\
&=&\oint_{\partial V} \mathbf{x}\,\mathbf{B}_\mathrm{d}\cdot d\mathbf{S}
-\int_V\nabla\cdot\mathbf{B}_\mathrm{d}\,\mathbf{x}\,dV\,.\;
\end{eqnarray}

The surface integral is not convergent as the surface $\partial V$ goes to
infinity. For example, for a sphere the surface integral is
$(8\pi/3)\mathbf{m}$ regardless of the radius. Instead if the volume is
taken as the doughnut-shaped figure whose surface is made of field lines
then the integral vanishes.

Let us assume that there is some way of defining the limit that gives the
``correct'' physical value. Given the symmetry of the dipolar field the
limit of the surface integral must be proportional to the dipole moment
\begin{equation}
\oint_{\partial V}\mathbf{x}\,\mathbf{B}_\mathrm{d}\cdot d\mathbf{S}
\to \lambda\mathbf{m}\ .
\end{equation}

Since $\nabla\cdot\mathbf{B}_\mathrm{d}=0$ the total integral is
\begin{equation}
\int \mathbf{B}_\mathrm{d} \,dV = \lambda\mathbf{m}\ .
\end{equation}

The electrical case is quite similar.
Since the magnetic and electric fields have the same form far away from the
dipole, the surface integral must be the same for both fields.
\begin{eqnarray} 
\int \mathbf{E}_\mathrm{d} \,dV &=& \lambda\mathbf{d} - 
\int\nabla\cdot\mathbf{E}_\mathrm{d}\,\mathbf{x}\,dV\nonumber\\
&=& \lambda\mathbf{d} -4\pi\int\rho\mathbf{x}\,dV\nonumber\\
&=& \lambda\mathbf{d}-4\pi\mathbf{d}\ .
\end{eqnarray}

For determining the value of $\lambda$ let us calculate the whole
interaction electric energy
\begin{eqnarray}
\frac{1}{4\pi}\int \mathbf{E}\cdot\mathbf{E}_\mathrm{d} \,dV &=&
-\frac{1}{4\pi}\int \nabla\phi\cdot\mathbf{E}_\mathrm{d} \,dV\nonumber\\
&=& -\frac{1}{4\pi}\int [\nabla\cdot(\phi\mathbf{E}_\mathrm{d})
-\phi\nabla\cdot\mathbf{E}_\mathrm{d}]\,dV\nonumber\\
&=&\int\phi\rho\,dV\nonumber\\
&=&-\mathbf{E}(0)\cdot\mathbf{d}\ ,
\end{eqnarray}
which of course is the electrostatic energy of  the dipole and should
be considered part of the energy of the dressed dipolar system. Therefore, it
must be that $\lambda=0$. The integral of the magnetic field is then zero and
there are not magnetic contributions to the momentum or energy of the
dressed dipole. Nevertheless, the whole interaction magnetic energy is not
zero,
\begin{eqnarray}
\frac{1}{4\pi}\int \mathbf{B}\cdot\mathbf{B}_\mathrm{d} \,dV &=&
\frac{1}{4\pi}\int \nabla\times\mathbf{A}\cdot\mathbf{B}_\mathrm{d} \,dV
\nonumber\\
&=& \frac{1}{4\pi}\int\mathbf{A}\cdot\nabla\times\mathbf{B}_\mathrm{d} \,dV
\nonumber\\
&=& \frac{1}{c}\int\mathbf{A}\cdot\mathbf{j}_\mathrm{m} \,dV
\nonumber\\
&=&\mathbf{B}(0)\cdot\mathbf{m}\ .
\end{eqnarray}
This energy must be assigned to the external system. If an external field
$\mathbf{B}$ is turned on in a region where there is a constant magnetic dipole, a
work $-\mathbf{B}\cdot\mathbf{m}$ is done on the bare dipole. The opposite
is the work done on the external system.

\vfill



\begin{thebibliography}{long}
\bibitem{Minkowski1908} H.~Minkowski, Nachr.~Ges.~Wiss.~G\"ottingen, 53, (1908).
\bibitem{Abraham1909} M.~Abraham, Rend.~Circ.~Mat.~Palermo \textbf{28}, 1--28, (1909).
\bibitem{Abraham1910} M.~Abraham, Rend.~Circ.~Mat.~Palermo \textbf{30}, 33--46, (1910).

\bibitem{EinsteinL1908} A.~Einstein and J.~Laub, Ann.~Phys.~(Leipzig) \textbf{26}, 532--540 (1908)
\bibitem{EinsteinL1908b} A.~Einstein and J.~Laub, Ann.~Phys.~(Leipzig) \textbf{26}, 541--550 (1908)
\bibitem{Beck1953} F.~Beck, Z.~Phys. \textbf{134}, 136--155 (1953).
\bibitem{MarxG1956} G.~Marx and G.~Gy\"{o}rgyi, Ann.~Phys. (Leipzig) \textbf{16}, 241--256 (1956).
\bibitem{deGrootS}S.~R.~de Groot and L.~G.~Suttorp, {\it Foundations of
 Electrodynamics}, North Holland, Amsterdam, 1972.
\bibitem{Peirls1976} R.~Peirls, Proc.~Royal.~Soc. \textbf{A347}, 475--491, (1976).
\bibitem{Peirls1977} R.~Peirls, Proc.~Royal.~Soc. \textbf{A355}, 141--151, (1977).

\bibitem{SJ1967} W.~Shockley and R.~P.~James, Phys.~Rev.~Lett. \textbf{18}, 876--879 (1967).
\bibitem{Pryce1948} M.~L.~H.~Pryce, Proc.~Roy.~Soc. \textbf{195}, 62--81 (1948).
\bibitem{Hill1951} E.~L.~Hill, Rev.~Mod.~Phys. \textbf{23}, 253--260 (1951).

\bibitem{TangM1961} C.~L.~Tang and J.~Meixner, Phys.~of ~Fluids \textbf{4}, 148--154 (1961).
\bibitem{Dewar1977} R.~L.~Dewar, Aus.~J.~Phys. \textbf{30}, 533--575 (1977).
\bibitem{Israel1977} W.~Israel, Phys.~Lett. \textbf{B67}, 125--128 (1977).
\bibitem{Kranys1979} M.~Krany\v{s}, Can.~J.~Phys. \textbf{57}, 1022--1026 (1979).
\bibitem{Maugin1980} G.~Maugin, Can.~J.~Phys. \textbf{58}, 1163--1170 (1978).
\bibitem{Pfeifer2007} Robert N.~C.~Pfeifer {\it et al}, Rev.~Mod.~Phys.
\textbf{79}, 1197--1216 (2007).
\bibitem{Goto2011} Shin-tiro~Goto, Robin~W.~Tucker and
Timothy J.~W.~Alton, Proc.~Royal.~Soc.  \textbf{A467}, 59--78, (2011)

\bibitem{Gordon1973} James~P.~Gordon, Phys.~Rev.\textbf{A8}, 14--21 (1973).
\bibitem{Nelson1991} D.~F.~Nelson, Phys.~Rev.\textbf{A44}, 3985--3996 (1991).
\bibitem{Barnett2010} Stephen~M.~Barnett, Phys.~Rev.~Lett.  \textbf{104}, 070401 (2010).

\bibitem{BelF1939} F.~J.~Belinfante, Physica \textbf{VI}, 887--898 (1939).
\bibitem{Rosenfeld1940} L.~Rosenfeld, Mem.~Acad.~Roy.~Belg. (Cl. Sciences) \textbf{18},fasc. 6, 2--30 (1940).
\bibitem{Pauli1941} W.~Pauli, Rev.~Mod.~Phys. \textbf{13}, 203--232 (1941).
\bibitem{CvV1968} Sidney~Coleman and J.~H.~Van Vleck Phys.~Rev. \textbf{171}, 1370--1375 (1968).

\bibitem{vLaue1950} M.~v.~Laue, Z.~Physik \textbf{128}, 387--394 (1950).
\bibitem{Moller} C.~M\o{}ller, {\it The theory of relativity}, Oxford, (1952).
\bibitem{MedinaandS2017} Rodrigo~Medina and J.~Stephany,  Eur.~J.~Phys. {\bf 38}, { 015208 (8pp), (2017)}.

\bibitem{Papapetrou1949} A.~Papapetrou, Phil.~Mag. \textbf{40}, 937--946 (1949).
\bibitem{Einstein-deHaas1915} A.~Einstein, W.~J.~de Haas, Deutsche Physikalische
Gesellschaft, Verhandlungen \textbf{17}, 152--170 (1915).
\bibitem{BarKen1952} S.~J.~Barnett and G.S.~Kenny, Phys.~Rev. \textbf{91}, 408--411 (1953).

\bibitem{MedandSf} Rodrigo~Medina and J.~Stephany, {\it An improved inertia principle}, arXive:1404.1590.
\bibitem{MedandSe} Rodrigo Medina and J.~Stephany, {\it Belinfante-Rosenfeld tensor and the inertia principle}, arXiv:1404.3334.

\bibitem{Poynting}J.~H.~Poynting, Phil.~Trans.~R.~Soc.~ \textbf{175}, 343--361 (1884).
\bibitem{LanLL1994} L.~Landau and E.~Lifchitz, {\it Classical theory of fields},
Butterworth-Heinemann, Oxford, 1994, p. 87.
\bibitem{Livens1916} G.~H.~Livens, Phil.~Mag. \textbf{32}, 162--171 (1916).

\bibitem{Pauli1958} W.~Pauli, The theory of relativity, Pergamon, (1958).
\bibitem{Skobel'tsyn1973}  D.~V.~Skobel'tsyn, Sov.~Phys.~Usp. \textbf{16}, 381--401 (1973).
\bibitem{PeirlsB1973} M.~G.~Burt and R.~Peirls, Proc.~Royal.~Soc.  \textbf{A333}, 149--156, (1973).
\bibitem{Synge1974} J.~L.~Synge, Hermathena, \textbf{117}, 80--84, (1974).
\bibitem{Robinson1975} F.~N.~H.~Robinson, Phys.~Rep. \textbf{16}, 313--354 (1975).

\bibitem{GinzburgU1976} V.~L.~Ginzburg and V.~A.~Ugarov, Sov.Phys.Usp. \textbf{19}, 94-101 (1976).
\bibitem{Brevik1979} I.~Brevik, Phys.~Rep. \textbf{52}, 133--201 (1979).
\bibitem{Loudon2004} R.~Loudon, Fortschr.~Phys. \textbf{52}, 1134--1140 (1980).
\bibitem{Leonhart2006} U.~Leonhart, Nature \textbf{444}, 823--824 (2006).
\bibitem{Obukhov2008} Y.~N.~Obukhov, Ann.~Phys.(Berlin) \textbf{17},830--851 (2008).
\bibitem{Novak1980} M.~N.~Novac, Fortschr.~Phys. \textbf{28}, 285--353 (1980).
\bibitem{MilonniBoyd2010} Peter W.~Milonni and Robert W.~Boyd, Advances in
Optics and Photonics \textbf{2}, 519--553 (2010).
\bibitem{BarnettLoudon2010} Stephen~M.~Barnett and Rodney~Loudon, Phil.~
Trans.~R.~Soc. A \textbf{368}, 927--939 (2010).
\bibitem{Kemp2011} B.~A.~Kemp, J.~Appl.~Phys. \textbf{109}, 111101 (2011).
\bibitem{Cho2013} Adrian~Cho, Science \textbf{327} 1067 (2013).

\bibitem{Abraham1914} M.~Abraham, Ann.~Phys.(Leipzig) \textbf{44},537--543 (1914).
\bibitem{Grammel1913} R.~Grammel, Ann.~Phys.(Leipzig) \textbf{41},570--580 (1913). 

\bibitem{VeselagoShchavlev2010} V.~G.~Veselago, V.~V.~Shchavlev,
Physics--Uspekhi \textbf{53}, 317--318 (2010).


\bibitem{AtanaT2000} Teodor M.~Atanackovic and Ard\'eshir Guran,
{\it Theory of Elasticity for Scientists and Engineers}, Birkh\"auser, Boston, 2000, p. 11. 
\bibitem{Fung2001} Yuan-cheng Fung and Pin Tong, {\it Classical and
 Computational Solid Mechanics}, World Scientific Publishing Co., Sigapore,
 2001, pp. 75, 77--78. 
\bibitem{Balazs1953} N.~L.~Balazs, Phys.~Rev. \textbf{91}, 408--411 (1953).

\bibitem{Cullwick1952} E.~G.~Cullwick, Nature, \textbf{170}, 425 (1952).
\bibitem{Taylor1965} T.~T.~Taylor, Phys.~Rev.~\textbf{137}, B467--B471 (1965).
\bibitem{Furry1969} W.~H.~Furry, Am.~J.~Phys. \textbf{37}, 621--636 (1969).
\bibitem{Babson2009} D.~Babson, S.~P.~Reynolds, R.~Bjorkquist and D.~J.~Griffiths, Am.~J.~Phys. \textbf{77}, 826--833 (2009).
\bibitem{PenfieldH1967} P.~Penfield~Jr. and H.~A.~Haus, {\it Electrodynamics of
Moving Media} MIT, Cambridge, MA, 1967.
\bibitem{Haus1969} H.~A.~Haus, Physica \textbf{43}, 77-91 (1969).

\bibitem{KlimaP1975} R.~Kl\'{\i}ma and V.~A.Petr\v{z}\'{\i}lka, Ann.Phys.(NY) \textbf{92}, 395--405 (1975).
\bibitem{Furutsu1969} K.~Furutsu, Phys.~Rev. \textbf{185}, 257--272 (1969).
\bibitem{Philbin2011} T.~G.~Philbin,  Phys.~Rev. \textbf{A83}, 013823 (2011).
\bibitem{Obukhov2003} Yuri N.~Obukhov, Friedrich W.~Hehl, Phys.~Lett. 
\textbf{A311}, 277--284, (2003).
\bibitem{JacJ1998} J.~D.~Jackson, {\it Classical Electrodynamics 3rd ed.}, John Wiley\&Sons, New York, 1998, p. 189, p. 213.
\bibitem{Rasetti1944} F.~Rasetti, Phys.~Rev.~\textbf{66}, 1--5, (1944).
\bibitem{Hughes-Burgy1951} D.~J.~Hughes and M.~T.~Burgy, Phys.~Rev.~\textbf{81},
498--506, (1951).
\bibitem{MedR2006} Rodrigo~Medina, J.~Phys.~A: Math Gen \textbf{39}, 3801--3816
(2006).
\bibitem{MedR2006b} Rodrigo~Medina, Am.~J.~Phys. \textbf{74}, 1031--1034 (2006).
\bibitem{Beth1936} R.~A.~Beth, Phys.~Rev. \textbf{50}, 115--125 (1936). 
\bibitem{Carrara1949} N.~Carrara, Nature \textbf{164}, 882-883 (1949). 
\bibitem{Friese1996} M.~E.~J.~Friese, J.~Enger, H.~Rubinsztein-Dunlop,
and N.~R.~Heckenberg, Phys.~Rev. \textbf{A54}, 1593--1596 (1996). 
\bibitem{Friese1998} M.~E.~J.~Friese, T.A.~Nieminen, H.~Rubinsztein-Dunlop,
and N.~R.~Heckenberg, Nature \textbf{394}, 348--350 (1998). 
\bibitem{MedandSc} Rodrigo~Medina and J.~Stephany, {\it Electromagnetic fields in matter revisited } arXiv:1404.5189
\bibitem{JonesR1954} R.~V.~Jones and J.~C.~S.~Richards, Proc.~Roy.~Soc. \textbf{A221},
480--498 (1954).
\bibitem{JonesL1978} R.~V.~Jones and B.~Leslie, Proc.~Roy.~Soc. \textbf{A360},
347--363 (1978).
\bibitem{Gibson1980} A.~F.~Gibson, M.~F.~Kimmitt, A.~O.~Koohian, D.~E.~Evans and G.~F.~D.~Levy, 
Proc.~Royal.~Soc. \textbf{A370}, 303--311, (1980)
\bibitem{Campbell} G.~K.~Campbell, A.~E.~Leanhardt, Jongchul Mun, M.~Boyd,
E.~W.~Streed, W.~Ketterle and D.~E.~Pritchard, Phys.~Rev.~Lett. \textbf{94},
170403 (2005).
\bibitem{Zuev2009} V.~S.~Zuev and G.~Ya.~Zueva,Opt.~and Spect. \textbf{106}, 248--251. (2009).
\bibitem{Ginzburg1960} V.~L.~Ginzburg, Sov.~Phys.~Usp. \textbf{2}, 874--893 (1960). 
\bibitem{Ginzburg1979} V.~L.~Ginzburg, {\it Theoretical physics and Astrophysics}, Pergamon Press (1979). 
\bibitem{ZhangLS2015} Li~Zhang, Weilong~She, Nan~Peng and Ulf~Leonhardt, New J.~Phys. \textbf{17},
053035 (2015).

\bibitem{JazayeriMehrany2014} Amir M.~Jazayeri and Khashayar~Mehrany,
Phys.~Rev. \textbf{A89}, 043845 (2014).

\bibitem{Poynting1905} J.~H.~Poynting, Phil.~Mag. \textbf{9}, 393--361 (1905).
\bibitem{Barlow1912} G.~Barlow, Proc.~Roy.~Soc. \textbf{A87}, 1--16 (1912).

\bibitem{AshkinD1973} A.~Ashkin and J.~M.~Dziedzic, Phys.~Rev.~Lett. \textbf{30}, 139--142 (1973).
\bibitem{LaiHY1976} H.~M.~Lai and K.~Young, Phys.~Rev. \textbf{A14}, 2329-2333 (1976).
\bibitem{Casner2001} A.~Casner and J.~P.~Delville, Phys.~Rev.~Lett. \textbf{87} 054503 (2001).
\bibitem{Sakai2001} K.~Sakai,D.~Mizuno and K.~Takagi,Phys.~Rev. \textbf{E63} 046302 (2001).
\bibitem{Astrath2014} N.~G.~C.~Astrath, L.~C.~Malacarne, M.~L.~Baesso, G.~V.~B.Lukasievicz
and S.~E.~Bialkowski,~Nat.~Commun.~\textbf{5}, 4363 (2014).
\bibitem{Kajorndejnukul2013} V.~Kajorndejnukul, Weiqiang Ding, S.~Sukhov, Cheng-Wei~Qiu and
A.~Dogariu, Nature Photon. \textbf{5}, 1--4 (2011).
\bibitem{Qiu2015} Cheng-Wei~Qiu, Weiqiang~Ding, M.~R.~C.~Mahdy, Dongliang~Gao, Tianhang~Zhang, Fook~Chiong~Cheong,
A.~Dogariu, Zheng~Wang and Chwee~Teck~Lim, Light: Sci.~Appl. \textbf{4}, e278, (2015).
\bibitem{WalkerGL1975} G.~B.~Walker and D.~G.~Lahoz, Nature \textbf{253},
339--340 (1975).
\bibitem{WalkerGLW1975} G.~B.~Walker, D.~G.~Lahoz and G.~Walker, Can.Jour.Phys.
 \textbf{53}, 2577--2586 (1975).
\bibitem{WalkerGW1977} G.~B.~Walker and G.~Walker, Can.Jour.Phys. \textbf{55},
2121--340 (1977).
\bibitem{SheWY2008} Weilong~She, Jianhui~Yu and Raohui~Feng, Phys.~Rev.~Lett. \textbf{101},
243601 (2008).
\bibitem{RikkenGT2012} G.~L.~J.~A.~Rikken and B.~A.~van Tiggelen, Phys.~Rev.~Lett. \textbf{108},
230402 (2012).
\bibitem{Hakim1962} S.~S.~Hakim and J.~B.~Higham, Proc.~Phys.~Soc. \textbf{80},
190 (1962).
\bibitem{Stratton1941} J.~A.~Stratton, {\it Electromagnetic Theory}, Mc Graw-Hill, New York, 1941, p. 140.
\bibitem{Richter-Lotter1969} F.~Richter and U.~Lotter, Phys.~Stat.~\textbf{34}
 K149--K152 (1969).
\bibitem{Turek-et-al2004} I.~Turek, J.~Rusz, and M.~Divi\v{s},
 Czech.~J.~Phys.~\textbf{54}, D279 (2004).
\bibitem{James-Kinderlehrer-1993} M.~D.~James and D.~Kinderlehrer, Phil.~Mag.~B
\textbf{68}, 237--274 (1993).
\bibitem{Wu-2002} Ruqian~Wu, J.~Appl.~Phys.~\textbf{91}, 7358--7360 (2002) 
\bibitem{MedandSb} Rodrigo~Medina and J.~Stephany, {\it The force density and the kinetic
energy-momentum tensor of electromagnetic fields in matter} arXiv:1404.5250
\bibitem{MedandSa} Rodrigo~Medina and J.~Stephany, {\it Violation of the center of mass
theorem for systems with electromagnetic interaction} arXiv:1404.5251

\end{thebibliography}
\end{document}